\documentclass[notitlepage,superscriptaddress,twocolumn,pra]{revtex4-2}

\expandafter\let\csname equation*\endcsname\relax
\expandafter\let\csname endequation*\endcsname\relax
\expandafter\let\csname eqnarray*\endcsname\relax
\expandafter\let\csname endeqnarray*\endcsname\relax
\usepackage{array}
\usepackage{amsmath}
\usepackage{amssymb}
\usepackage{graphicx}
\usepackage{braket}
\usepackage{color}
\usepackage{float}
\usepackage{soul}
\usepackage{enumitem}
\usepackage[dvipsnames]{xcolor}
\usepackage{comment}

\newcommand{\RN}[1]{\textup{\uppercase\expandafter{\romannumeral#1}}}

\renewcommand{\rm}{\mathrm}

\begin{document}
\setlength{\parindent}{0mm}

\title{A Quantum Klystron - \\Controlling Quantum Systems with Modulated Electron Beams}

\author{Dennis R\"atzel}
\email{dennis.raetzel@physik.hu-berlin.de}
\affiliation{Institut f\"ur Physik, Humboldt-Universit\"at zu Berlin, Newtonstra{\ss}e 15, 12489 Berlin, Germany}

\author{Daniel Hartley}
\address{Vienna Center for Quantum Science and Technology, Atominstitut, TU Wien, Stadionallee 2, 1020 Vienna, Austria}

\author{Osip Schwartz}
\affiliation{Dept. of Physics of Complex Systems, Weizmann Institute of Science, Rehovot, Israel}

\author{Philipp Haslinger}
\email{philipp.haslinger@tuwien.ac.at}
\address{Vienna Center for Quantum Science and Technology, Atominstitut, TU Wien, Stadionallee 2, 1020 Vienna, Austria}

\begin{abstract}

Coherent control of quantum transitions  -- indispensable in quantum technology -- generally relies on the interaction of quantum systems with electromagnetic radiation.
Here, we theoretically demonstrate that the non-radiative electromagnetic near-field of a temporally modulated free-space electron beam can be utilized for coherent control of quantum systems. We show that such manipulation can be performed with only classical control over the electron beam itself, and is readily realizable with current technology. This approach may provide a pathway towards spectrally selective quantum control with nano-scale spatial resolution, harnessing the small de Broglie wavelength of electrons.

\end{abstract}

\maketitle

\section{Introduction}

Coherent manipulation of quantum systems with precisely controlled electromagnetic (EM) fields, such as laser or microwave pulses, is a ubiquitous tool of quantum science from the search for new physics \cite{parker2018measurement, safronova2018search} to quantum information processing \cite{monroe2002quantum,imamog1999quantum}. 
\begin{figure}[hbt!]
\hspace*{-3mm}
\includegraphics[width=8.8cm,angle=0]{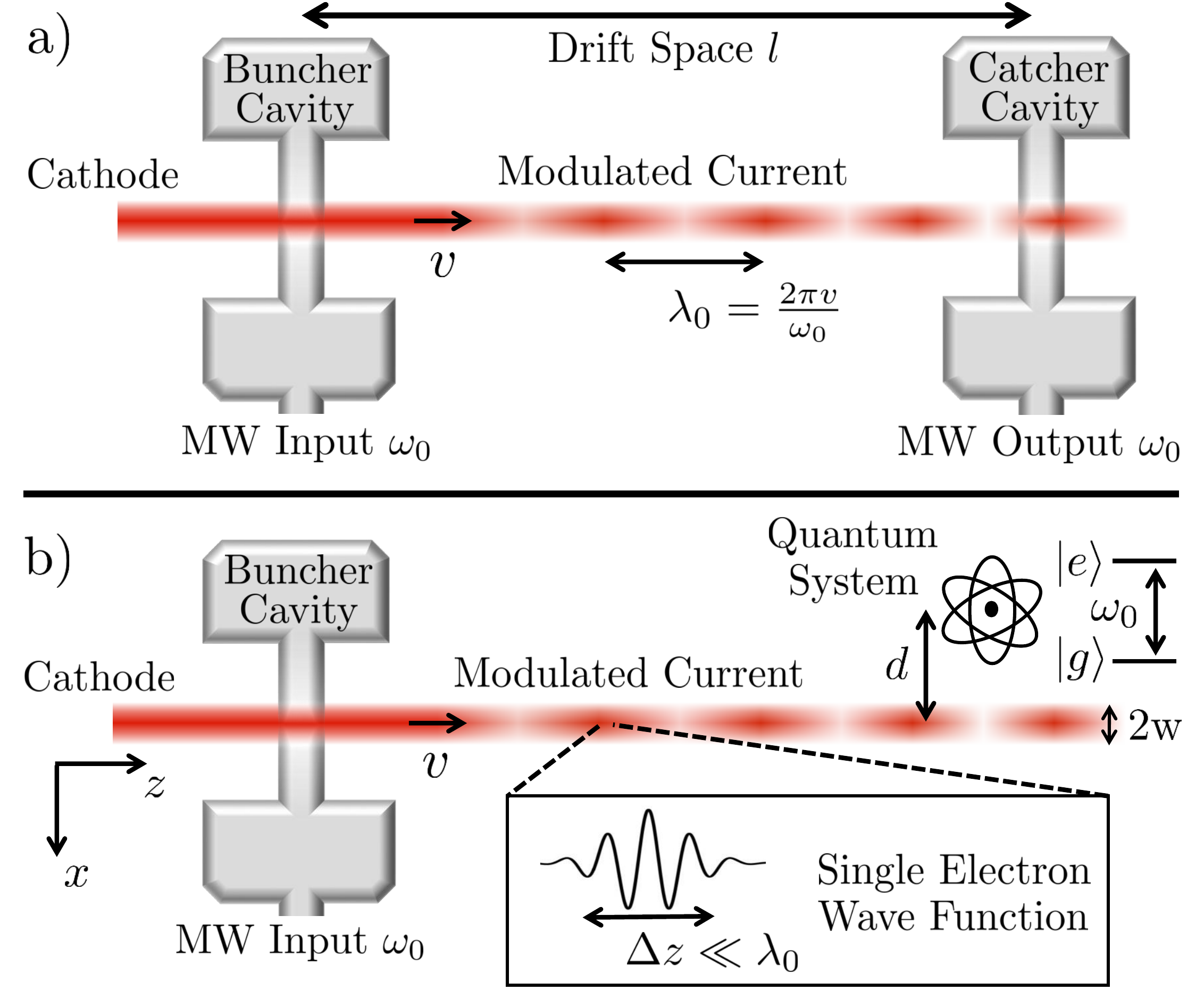}
\caption{\label{fig:klystron} 
a) Schematic view of the conventional klystron: an electron beam is velocity modulated by the electric field of a microwave (MW) cavity, the “buncher cavity", at angular frequency $\omega_0$. 
Through the drift space, the velocity modulation causes a current modulation, which induces amplified microwaves at the “catcher cavity”. The amplified MW radiation is used, for example, to drive atomic transitions coherently and with high fidelity. b) Schematic view of the quantum klystron: the electromagnetic near field of the current-modulated electron beam at the position of the catcher cavity is used directly to drive transitions of quantum systems without the detour of generating electromagnetic radiation. $\text{w}$ is the electron beam waist, $\lambda_0$ is the modulation wavelength, which is much larger than the single electron wave function width $\Delta z$, and $d$ is the distance from the quantum system to the beam center. }
\end{figure}
Here, we demonstrate that the EM near-field of a temporally modulated electron beam provides an alternative route to coherent manipulation of quantum systems. Electromagnetically addressable transitions can be driven by the oscillating EM field surrounding the modulated beam. Due to the small de Broglie wavelength of the electron beam \cite{Reimer2008Transmission}, this mechanism of interaction potentially allows for the addressing of individual quantum systems with nano-scale resolution, 
similarly to the incoherent electron-based spectroscopy methods relying on the same electromagnetic interaction \cite{zhou_2012,krivanek2014vibrational,egerton2011electron, deAbajo2008electron,deAbajo2010optical}. 

The interaction of a temporally modulated electron beam with a quantum system is reminiscent of an RF amplifier type known as the klystron \cite{gilmour2011klystrons}, wherein the electron beam's velocity is modulated by a periodic seed field, resulting in a current modulation downstream of the interaction region. The kinetic energy of the modulated electron beam is then converted into electromagnetic excitation of an RF cavity (Fig. \ref{fig:klystron}a). Here, we propose a quantum counterpart of the klystron, where the kinetic energy of a modulated electron beam is converted into the coherent excitation of a quantum system or an ensemble of such systems, as shown in Fig. \ref{fig:klystron}b. We restrict our considerations to magnetic dipole transitions. Electric dipole transitions and higher multipole transitions can be treated in a similar manner.

While it has been suggested \cite{Favro1971en,robicheaux2000coherent,pindzola2000coherent,Gover2020free} that quantum systems can coherently interact with a stream of electrons with a temporally shaped wave function \cite{Feist2015Quantum, kealhofer2016all, schonenberger2019generation, wang2020coherent}, we theoretically demonstrate that electron-mediated manipulation of quantum systems can be achieved when the longitudinal extent of the individual electron wave packets is much shorter than the wavelength of modulation.
Such ‘classical’ modulation of the electron beam current density is readily achievable in the microwave (MW) frequency range, as it is an integral part of widely used electronic technologies from microwave heaters to radars.

\section{Back-action and the semi-classical regime}

\begin{figure}
\includegraphics[width=\columnwidth,angle=0]{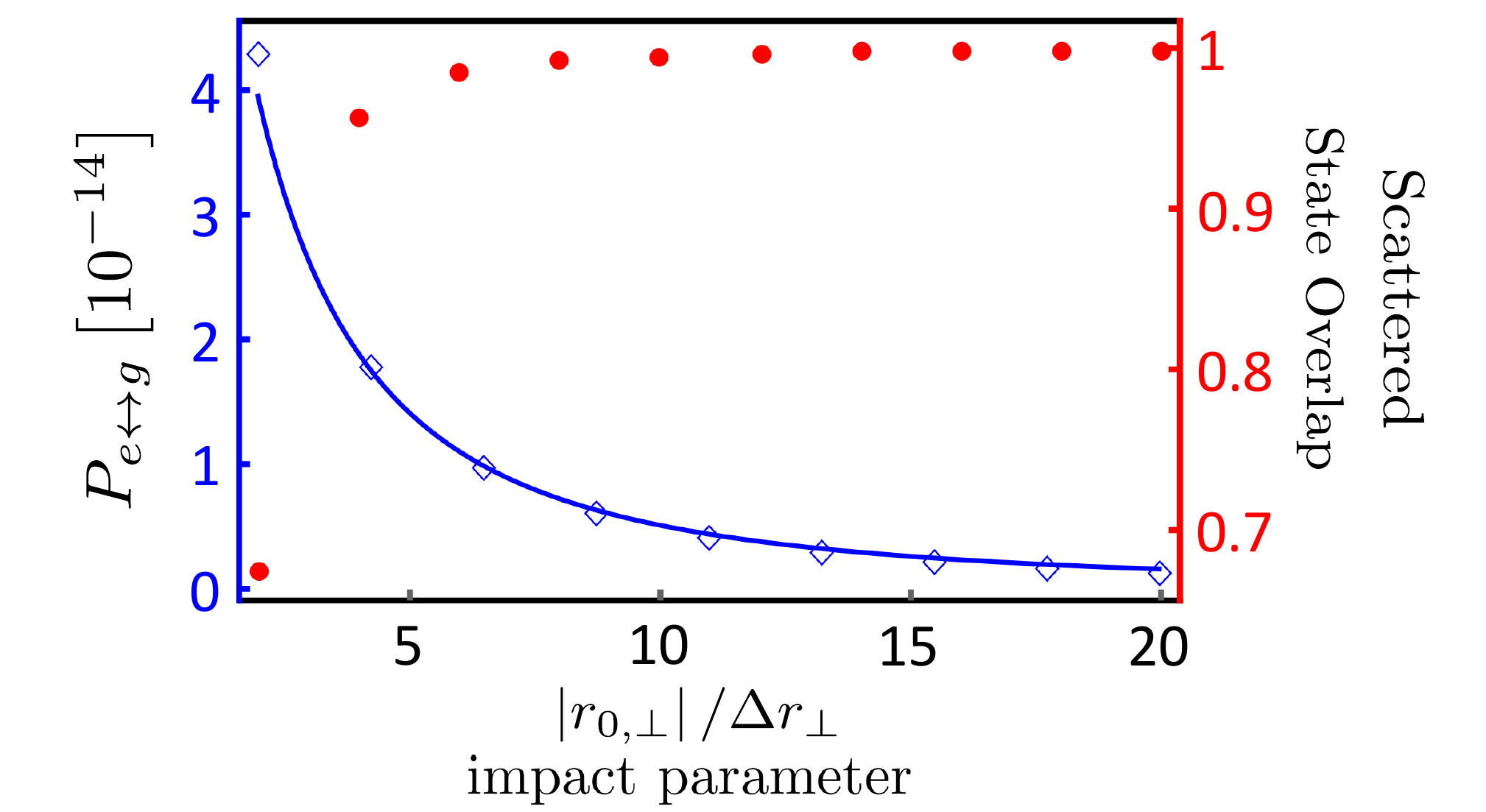}
\caption{\label{fig:overlap_probability} 
Transition induced by a single electron. Left axis: the numerically calculated transition probability (open blue diamonds) and the semi-classical result (blue line) shown in Appendix \ref{sec:transprob}, both as functions of the impact parameter. The numerical result agrees with the analytical result by one percent or less for distances larger than $10\Delta r_\perp$.
Right axis: overlap of the initial and scattered electron states as a function of the distance between the electron and the quantum system (red dots). For distances $\ge 6 \Delta r_\perp$, the overlap can be considered to be one for the purpose of this article. The drop in overlap between these states corresponds to the regime where the wavefunction of the passing electron significantly overlaps with that of the quantum system at the point of closest approach, where we can no longer apply the picture of driving with the near-field of an electron beam. The plots were obtained with $\Delta r_\perp = 5\,$nm, $\Delta z_0 = 100\,$nm, $\omega_0/2\pi = 2.87\,\rm{GHz}$ and an initial kinetic energy of the electron of 2~keV  (details about the model can be found in Appendix \ref{sec:transprob}).}
\end{figure}
A necessary condition for coherent driving of a quantum system is that no information about its state is transferred to the environment. In particular, electron scattering due to quantum back-action causes entanglement between the electron and the quantum system, which disrupts coherent driving. 
Given the overlap $\Lambda$ between an incoming electron state and its scattered state, the probability to find the electron scattered into a state that is orthogonal to the incoming state is $(1-|\Lambda|^2)P_{e\leftrightarrow g}$, where $P_{e\leftrightarrow g}$ (excited state $e$ and ground state $g$, see Fig.~\ref{fig:overlap_probability}) is the probability of the transition of the quantum system induced by a single electron.
The transition rate for coherent transitions, the Rabi frequency, is $\sim|\Lambda|\sqrt{P_{e\leftrightarrow g}}I_{\omega_0}/e$, where $I_{\omega_0}$ is the resonant Fourier component of the modulated beam current (see Eq. \ref{eq:rhochangecont} in the Appendix), where $e$ is the elementary charge. 
Thus even a small overlap is sufficient to preserve coherent driving, in principle, as long as $|\Lambda|\gg \sqrt{P_{e\leftrightarrow g}}$, while an overlap approaching unity is desirable to maximize the driving rate.

We calculate the overlap integral using a QED model of the interaction of a single electron with a two-level quantum system via its magnetic transition dipole moment (relevant to our two examples), see Appendix \ref{sec:transprob}. We consider a situation where the quantum system's dimensions are much smaller than both its distance to the electrons in the beam and the modulation wave length $\lambda_0$. 
We describe the electron field as a Dirac field, and consider an initial spin-unpolarized Gaussian matter-wave packet of transverse width $\Delta r_\perp$ with a propagation axis offset by $|r_{0,\perp}|$ (impact parameter). We numerically evaluate the spin-averaged overlap between an incoming electron state and its  scattered state given that a transition occurred. 

The results of the simulation are shown in Fig.~\ref{fig:overlap_probability}. For instance, for distances $|r_{0,\perp}|\gtrsim 6 \Delta r_\perp$, the overlap is $98\%$ or larger.  In the examples below, we have $|r_{0,\perp}|\gtrsim 6 \Delta r_\perp$, and therefore the overlap factor is close to unity and quantum back-action on the electrons can be neglected. 

\begin{figure*}
\includegraphics[width=\textwidth]{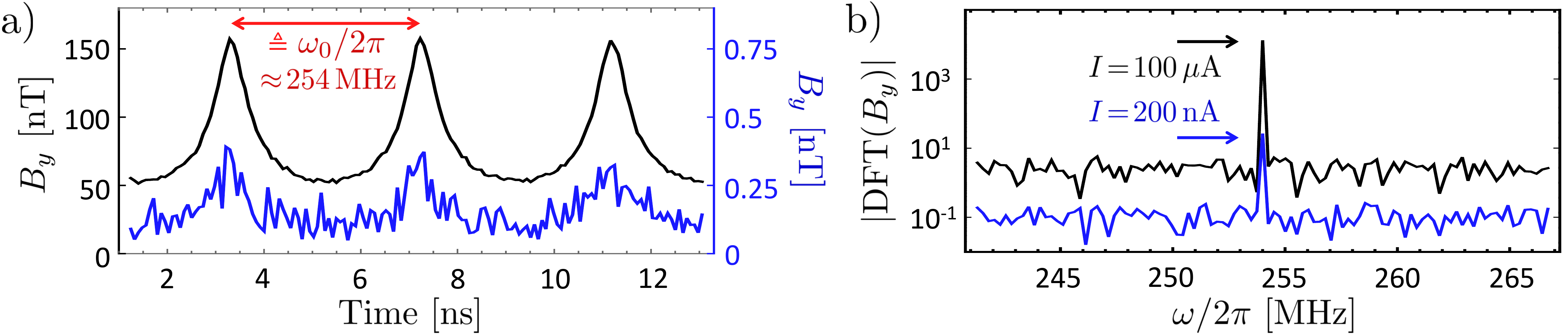}
\caption{\label{fig:discrete3dmovbeam} 
a) Numerical simulation of the magnetic field strength $B_y$ of a current-modulated electron beam with an initial Poissonian distribution of electrons in time and a Gaussian distribution of electrons in the transverse dimensions as could be created in a klystron (see Appendix \ref{sec:elbeamklystron}) with bunching parameter $r_b\approx 0.5$. The waist radius is $\text{w}= 50\,\mu$m and the distance to the center of the beam is $d=250\,\rm{\mu m}$. The plots show the cases of currents of $200\,\rm{n A}$ (blue plot and right axis, corresponding to about $\sim \! 5\,000$ electrons per period) and $100\,\rm{\mu A}$ (black plot purple and left axis, corresponding to $\sim\!2\,500\,000$ electrons per period). The electrons possess a kinetic energy of 18 keV and the base frequency of the modulated electron beam is 254 MHz. It can be seen that the relative strength of shot noise is decreased significantly for the 100 $\mu$A beam in comparison to the weaker 200 nA beam.
b) Fourier limited linewidth: discrete Fourier transform (DFT) of the magnetic field strength $B_y$ with the same parameters as in (a); 200 nA  (blue plot) and 100 $\mu$A  (black plot) evaluated for $10^3$ periods. It can be seen that a decrease in current leads to a decrease in the signal-to-noise ratio but does not affect the linewidth of the modulation.
}
\end{figure*}
We note that the quantum system's transition energy transferred to the electron leads to a momentum shift $\delta p \sim \hbar\omega_0/v = 2\pi\hbar/\lambda_0$,  where $\hbar$ is the reduced Planck constant, $v$ is the velocity of the electrons and $\omega_0$ is the angular transition frequency. %
For the scattered state overlap to be close to one, this momentum shift must be negligible in comparison to the electron wave packet's longitudinal momentum spread $\Delta p_z = \hbar/(2\Delta z_0)$. 
We assume an initial longitudinal wave packet width $\Delta z_0$ of the order of $100\,\rm{nm}$ (which is about the size of the coherence length of typical sources \cite{Reimer2008Transmission,Hasselbach_2009}) and consider MW transition frequencies corresponding to modulation wavelengths (see Fig.~\ref{fig:klystron}) of $\lambda_0\gtrsim1\,\rm{mm}$.
Accordingly, $\delta p\ll \Delta p_z$, which is consistent with the large overlap in Fig.~\ref{fig:overlap_probability}. 

The QED transition probability $P_{e\leftrightarrow g}$ 
is compared with the transition probability induced by the magnetic field of a classical electron (see Appendix \ref{sec:transprob}) in Fig.~\ref{fig:overlap_probability}. The two probabilities converge for distances $\gtrsim 4 \Delta r_\perp$. 
In the following, we consider parameter regimes where this condition is satisfied, which allows us to use the classical description of the electrons' magnetic field. At the same time, electrons from conventional sources can be assumed to be uncorrelated \cite{ferwerda1980coherence}, which allows us to treat the beam as an ensemble of classical Poisson distributed point-like charged particles.

\section{Example applications}

To illustrate the practical feasibility of quantum klystrons, we consider two example applications of this approach.

In the following, we consider an ensemble of electrons with a transverse Gaussian distribution. 
The electrons are longitudinally velocity modulated which, through propagation, leads to bunching further along the beam. This situation corresponds, for example, to that created in a klystron (see Fig.~\ref{fig:klystron}). The base frequency of the current modulation is tuned on resonance with the transition of the quantum system.

In addition to negligible back-action, coherent driving requires dephasing due to noise in the magnetic field of the beam to be limited. The spectral linewidth $\delta \omega$ of klystrons is mainly limited by technical noise \cite{Hemphill1960}; as a conservative estimate, we consider $\delta \omega_0/ \omega_0=10^{-7}$.

\subsection{Driving hyperfine transitions of alkali atoms}

As the first example, we consider driving ground state hyperfine transitions in alkali atoms. Alkali atoms, for example Li, K, and Rb, are especially well suited for a first demonstration of the quantum klystron. This is due to the hydrogen-like level structure with two stable and easily detectable ground state hyperfine levels separated by a transition in the microwave range. 
We consider the transition $F\! =\! 1,\,m_F\!=\!0 \leftrightarrow F\!=\!2,\,m_F\!=\!0$ of optically trapped $^{41}$K atoms as a specific example, where $F$ is the total angular momentum and $m_F$ denotes the Zeeman sublevels. This transition has a frequency of $\sim 254$~MHz and is therefore easily accessible with low-frequency MW electronics.
$^{41}$K atoms can be optically cooled and trapped, and could be controlled on the sub-$\mu$m scale using similar techniques as used for example for $^{40}$K in \cite{Cheuk2015} or for $^{87}$Rb in \cite{Gericke2008High-resolution}.

The results of a numerical simulation of the electron beam's magnetic field for this example are shown in Fig.~\ref{fig:discrete3dmovbeam} (see Appendix \ref{sec:singpartsim}). The effect of shot noise (analyzed in detail in Appendix \ref{sec:singpartfour} and Appendix \ref{sec:singpartauto}) appears as a homogeneous noise floor in the Fourier transform and does not modify the linewidth of coherent oscillations of the magnetic field (see Fig.~\ref{fig:discrete3dmovbeam}b). 

We consider an electron beam waist of $\text{w} = 50\,\rm{\mu m}$, kinetic energy of $18\,\rm{keV}$ and an average current $I_0 = 100\,\mu$A. We consider a bunching parameter $r_b= 0.5$ experimentally attainable at a drift distance of $\sim\!1$m such that $\Delta z\sim 7\,\rm{\mu m}\ll \lambda_0\sim 0.3\,$m in the interaction region. This value of $r_b$ corresponds to a resonant current modulation at the base frequency $\omega_0/2\pi$ of amplitude $I_{\omega_0}= 2I_0 J_1(r_b)\sim 50\,\mu$A (where $J_1$ is the Bessel function of the first kind), and we assume $d=250\,\rm{\mu m}$ between the atom and the beam center. 
In the case under consideration, $\delta \omega/ \omega_0=10^{-7}$ leads to a beam modulation spectral line width of about $25\,\rm{Hz}$. The change of the internal state of the atom will be accompanied by a recoil equivalent in absolute value to the momentum transfer to the electron. We obtain a conservative upper bound for the Lamb Dicke parameter of $\lesssim 4\times 10^{-4}$ for a trap frequency of $\sim\!300\,$kHz (as realized e.g. in \cite{Cheuk2015}) which implies that the recoil is negligible (see Appendix \ref{sec:transprob}).
\begin{figure}[h]
\includegraphics[width=8cm,angle=0]{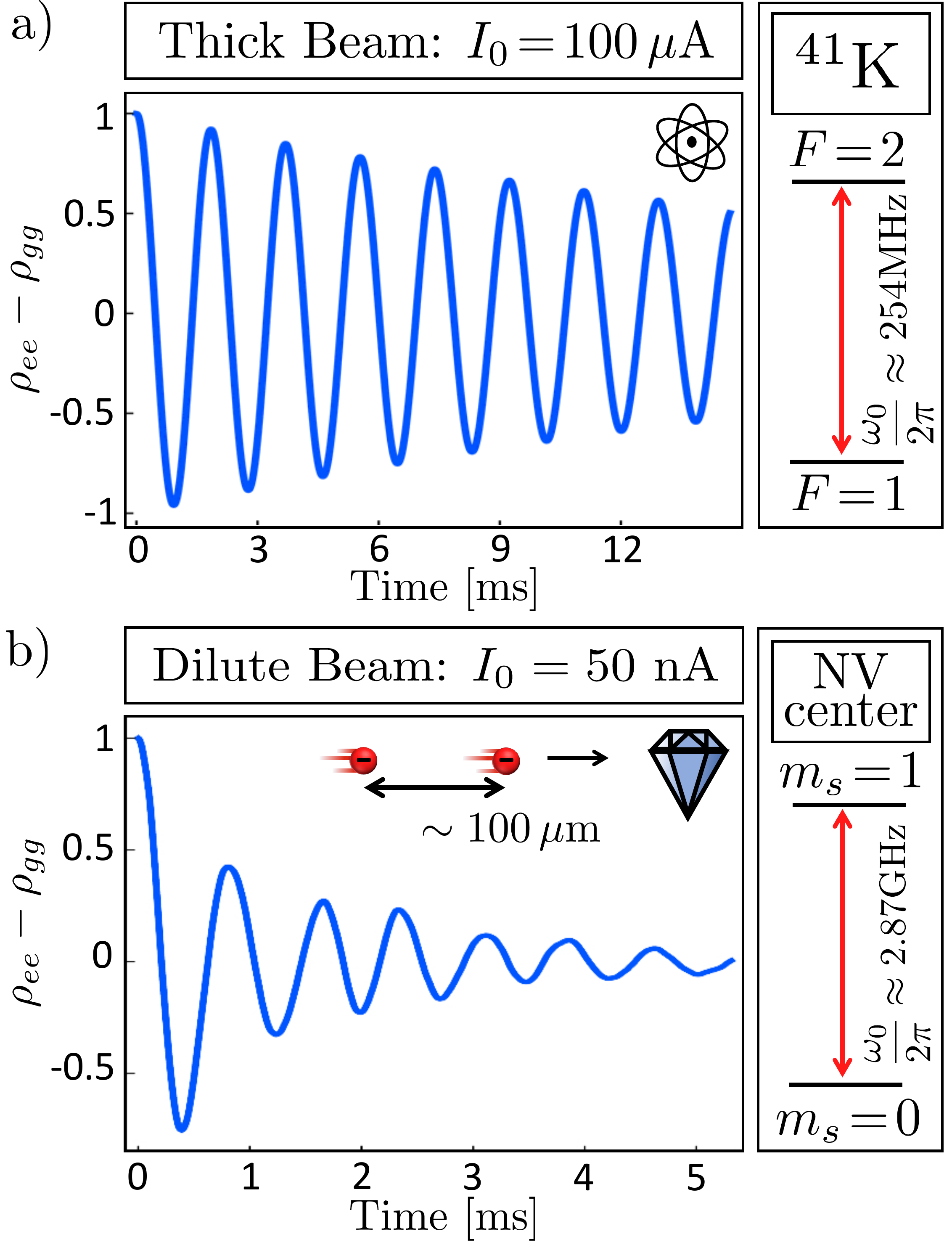}
\caption{\label{fig:blochevolution}  
a) The time-evolution of the inversion $\rho_{ee}-\rho_{gg}$, where $\rho_{ee}$ and $\rho_{gg}$ are the diagonal components of the quantum system's density matrix ($g$ for ground state, $e$ for excited state), for the transition $F\!=\!1,\,m_F\!=\!0 \leftrightarrow F\!=\!2,\,m_F\!=\!0$ for $^{41}$K at a distance $d\!=\!250\,\rm{\mu m}$ from the center of an electron beam of waist $\text{w}\!=\!50\,\rm{\mu m}$, current $100\,\mu$A, bunching parameter $r_b\!=\!0.5$, $E_\rm{kin}=18\,$keV, FWHM linewidth of the electron beam modulation $\delta\omega_0/2\pi =  25\,\rm{Hz}$ and $\Gamma_1 = 2\Gamma_2 \ll \delta\omega_0$.
b) The time-evolution of the inversion for the transition $m_s\!=\!0 \leftrightarrow m_s\!=\!1$ in the $^3A_2$-state of an NV$^-$ center at a distance of $d\!=\!70\,$nm from a beam of waist $\text{w}\!=\!10\,$nm, current $50\,$nA, $E_\rm{kin}=2\,$keV (average distance between electrons $\sim 100\,\mu\rm{m}$) and bunching parameter $r_b\approx 0.5$. We set $1/\Gamma_1\!=\!T_1\!=\!6\,$ms, $1/\Gamma_2=\!T_2\!=\!3\,$ms and the FWHM linewidth of the electron beam modulation $\delta\omega_0/2\pi =   300\,\rm{Hz}$.
} 
\end{figure}

In this example, the fluctuations (e.g. shot noise, modulation phase noise) of the modulated current are small relative to the mean (see Fig.~\ref{fig:discrete3dmovbeam}).  The effect of the mean field can be calculated using the rotating wave approximation. In this case, the evolution equations for the quantum system become the optical Bloch equations with constant coefficient matrix (see Appendix \ref{sec:Bloch}) and Rabi frequency $\Omega \approx g_S\mu_B  \mathcal{B}_{y,\omega_0}/2\hbar$, where $g_S$ is the electron's gyromagnetic ratio, $\mu_B$ is the Bohr magneton and $\mathcal{B}_{y,\omega_0}$ is the amplitude of the Fourier component of the electron beam's magnetic field at the transition frequency  (see \cite{Tiecke2010Properties} and Appendix \ref{sec:matrixelem41K}). Since the distance between the electron beam and the quantum system is larger than 2w, the magnetic field of the electron beam at the position of the quantum system is approximately that of an infinitesimally thin beam, which implies $\Omega \approx r_e I_{\omega_0}/(d\,e) = \sqrt{P_{e\leftrightarrow g}} I_{\omega_0}/e$ (see Appendix \ref{sec:matrixelem41K}), where $r_e=\mu_0 e^2/(4\pi m_e)$ is the classical electron radius, $\mu_0$ is the vacuum permeability and $m_e$ is the electron mass.

The finite spectral linewidth $\delta \omega$ of the driving electromagnetic field, represented by phase fluctuations, results in an increase of the decoherence rate by $b=\delta \omega/2$ (see Appendix \ref{sec:Bloch}). Furthermore, the shot noise of the electron beam is a source of amplitude noise of the driving field, which leads to a dephasing rate $P_{e\leftrightarrow g}I_0/e$, where $I_0$ is the average current. However, this rate is much smaller than the Rabi frequency $\Omega$ provided that $d \gg 2r_e I_0/ I_{\omega_0}$ (see Appendix \ref{sec:blochshot}). 
Since $r_e \sim 10^{-15}\,\rm{m}$, this condition is always fulfilled in practice. 

A plot of the hyperfine state response due to a resonantly modulated electron beam based on numerical evaluation of the optical Bloch equations (see Appendix \ref{sec:driving41K}) can be found in Fig.~\ref{fig:blochevolution}a. Several Rabi oscillations of the hyperfine states are clearly visible, showing that coherent driving with an electron beam is indeed possible. The largest contributor to the decay of coherence is the beam modulation spectral line width; for an analysis of other effects such as incoherent scattering (both elastic and due to other transitions) and beam electron velocity spread, see the Appendix.

\subsection{Addressing NV$^-$ centers in nano-diamond}

In the second example, we consider negatively charged nitrogen vacancy (NV$^-$) centers in nano-diamonds, which could be embedded, for example, in a free-standing nanostructure \cite{Batzer2020}. 
We focus on the transition between the $^3A_2$ ground state magnetic sublevels $m_s=0$ and $m_s= 1$, which are split by $\sim\!2.87\,$GHz. The $m_s = -1$ sublevel is well separated from the $m_s = 1$ sublevel by at least $\sim 4\,$MHz such that the transition $m_s\!=\!0 \leftrightarrow m_s\!=\!1$ can be individually addressed and easily optically detected \cite{Zheng2019zero}. This transition exhibits a coherence time $T_2$ of up to $600\,\rm{ms}$ \cite{Bar-Gill2013}. 

To achieve a sufficiently narrowly focused beam, the electron source could for example be a field emission electron gun with a slightly modulated acceleration voltage. We consider a modulated electron beam current generating a beam waist of $\text{w} = 10\,$nm at $2\,\rm{keV}$, a beam current of $50\, \rm{nA}$ ($\sim\!100$ electrons per modulation period) directed at a distance of $d=70\,\rm{nm}$ next to the NV$^-$ center. These electron beam parameters can be achieved in a standard scanning electron microscope. We assume a bunching parameter $r_b=0.5$ that would be experimentally attainable at a drift distance of $\sim\!3\,\rm{cm}$ such that $\Delta z \sim 400\,\rm{nm}\ll \lambda_0 \sim 9\,\rm{mm}$.
In this situation, the magnetic near field of the electron beam consists of distinct spikes due to the well separated electrons. Therefore, we cannot use the mean magnetic field in the optical Bloch equations. 
Instead, we simulate the effect of the electron beam on the state of the quantum system on the single electron level. Electrons are randomly generated, their kinetic energy is modulated and their propagation over the drift distance $l$ is calculated to obtain the modulated current. For consecutively passing groups of electrons, the optical Bloch equations with time-dependent coefficients are solved iteratively. Details can be found in Appendix \ref{sec:nvcenter}.

A simulation of the expected system evolution is presented in Fig.~\ref{fig:blochevolution}b. Several Rabi oscillations are clearly visible for this example, damped due to the combined effect of the spectral line width of the beam modulation (as a conservative upper bound, we chose $\delta\omega_0 \sim 10^{-7} \omega_0$ as before leading to a damping rate $\delta\omega_0/2 \sim 1\,$kHz; see Appendix \ref{sec:nvcenter}) and the intrinsic decay rates $\Gamma_2=1/T_2\sim 0.3\,$kHz and $\Gamma_1=1/T_1\sim 0.2\,$kHz. Other effects can largely be ignored; the various contributions of these are again discussed in Appendix \ref{sec:nvcenter}. Even though this case is outside of the regime of small fluctuations of the magnetic field, we do not find any notable additional decay due to electron shot noise.

\section{Potential pathway to nano-scale resolution}

The spatial resolution of coherent control utilizing electromagnetic radiation is generally limited by diffraction to centimeters for microwaves, and to hundreds of nanometers for optical frequencies. The need to selectively address individual quantum systems beyond the diffraction limit has been partially met by strategies such as utilizing sub-wavelength antennas \cite{ospelkaus2011microwave, novotny2011antennas}, tunable resonance frequencies \cite{Anders2018,Gardner:1993suboptical} or coupling strengths \cite{Navon2013addressing}, and utilizing multi-photon transitions at a shorter wavelength \cite{dudovich2002single}.  The quantum klystron may provide an alternative pathway towards spectrally selective quantum control with nano-scale spatial resolution as we will argue in the following.

While a temporally modulated beam produces an EM field that scales as
$d^{-1}$ with distance, a stronger localization of the field in the vicinity of the beam can be achieved by employing oscillations of the beam position to generate a driving signal.
Then the oscillating near-field of a moving beam at the first harmonic and the second harmonic (twice the modulation frequency) scale effectively as $d^{-2}$ and $d^{-3}$, respectively. 

Based on this faster decrease of the field, for example, at a distance of $d=15\,$nm to a 1-dimensional array of NV centers, adjacent NV centers with a distance of $\sim 30\,$nm could be individually controlled, in principle (the spatial dependence of the Rabi frequency shows peaks of width $\sim 40\,$nm (FWHM)), plots can be found in Appendix \ref{sec:nvcenter}).
A full Rabi flop may  be performed on an NV center without incurring significant decoherence
by employing low-energetic electron beams (e.g. $200$ eV, see Appendix \ref{sec:nvcenter}). 

Based on the picometer beam widths achievable in scanning electron microscopy \cite{Erni2009Atomic-resolution}, the quantum klystron scheme allows for the creation of spatially structured oscillating electromagnetic near-fields in two dimensions on the atomic scale (by choosing an electron impact parameter $d$ of several \r{A}ngstr\"om) which, in principle, could be employed for coherent control of systems that are sufficiently robust against electric field noise such as nuclear spins \cite{Anders2018}.

\section{Conclusions}
 
In summary, our results show that the electromagnetic near-field of a classically modulated electron beam can be used to coherently drive quantum systems and potentially provides a pathway to nano-scale resolution. 

The quantum klystron could be combined with a scanning electron microscope \cite{Verhoeven2018High} to perform coherent spectroscopic investigations by directing a modulated electron beam next to a sample to selectively drive a quantum transition of interest. This can be realized in a setup similar to aloof Electron Energy Loss Spectroscopy (EELS) \cite{
krivanek2014vibrational,EGERTON201595}, where the electron beam is directed at a distance of tens of nanometer from the sample to reduce radiation damage. 

In the MW frequency range, the excitation levels at every beam position could be read out by an optical channel \cite{Doherty2013}, or microwave sensors \cite{Anders2018}. This is particularly interesting as the MW range is inaccessible to the incoherent conventional methods, where the spectral resolution is determined by the energy spread of a monochromated electron beam \cite{egerton2011electron,krivanek2014vibrational}. For electric dipole transitions in the far infrared frequency range, the read-out of the sample's excitation could also be performed with an additional monochromated electron beam applying the aloof EELS method.
The coherent control provided by the quantum klystron would potentially enable increased spectral resolution by employing the method of Ramsey spectroscopy \cite{Ramsey1950}, which is widely used in spectroscopic applications from NMR \cite{Anders2018} to time-keeping \cite{ludlow2015optical}.

An appealing feature of the general approach to electron spectroscopy described above is that the coherence-preserving scattering investigated in this article leads to a coherent addition of excitation levels from individual electrons \footnote{This is the origin of Rabi oscillation in our examples (see Appendix \ref{sec:transprob} for details).}. This results in a quadratic dependence of the excitation level on the number of electrons close to $\rho_{ee}-\rho_{gg}=\pm 1$  (demonstrated e.g. in Fig.~\ref{fig:blochevolution} by the quadratic decrease/increase of the transition probability with time). It also leads to an increase of the  transition  rate  per  electron  by orders of magnitude in comparison to incoherent scattering \footnote{The transition rate due to coherence-preserving scattering reaches its maximum at $\rho_{ee}-\rho_{gg}=0$ and where it is larger by $1/\sqrt{P_{e\leftrightarrow g}}$ than the largest change obtained with incoherent scattering (see Appendix \ref{sec:transprob}).}. This scheme could be, for example, utilized to investigate specimens with reduced electron dose.

Finally, we note that the electron-based control of quantum systems can be extended to electric and magnetic multipole transitions. Such transitions are driven by the corresponding spatial derivatives of the electric and magnetic fields. Due to the strong dependence on the beam's near field on $d$, at nanometer-scale distances from the beam, the multipole transition rates are enhanced by orders of magnitude compared to addressing the same transitions by free-space EM fields. This property can be used to study e.g. quadrupole vibrational transitions in homonuclear diatomic molecules \cite{tao2018toward}, or to directly address the quadrupole transitions serving as the basis of optical clocks \cite{ludlow2015optical}.

\section*{acknowledgements}
We thank Holger M\"uller, J\"org Schmiedmayer, Peter Schattschneider, Thomas Schachinger, Hannes Matuschek, Philipp Thomas, Matthias Sonnleitner, Gregor Pieplow, Tim Schr\"oder, Michael St\"oger-Pollach, Kurt Busch, Kiri Mochrie, Ralf Menzel, Francesco Intravaia, Igor Mazets, Matthias Kolb, Thomas Weigner, Michael Scheucher, Thomas Juffmann, Stephanie Manz, Arne Wickenbrock and Thomas Kiel for helpful remarks and discussions. DR and PH acknowledge the hospitality of the Erwin Schr\"odinger Institute in the framework of their ``Research in Teams" project. DR thanks the Humboldt Foundation and the Marie Skłodowska-Curie Action IF program ("Phononic Quantum Sensors for Gravity" grant number 832250 — PhoQuS-G) for support. PH thanks the Austrian Science Fund (FWF): J3680, Y1121. We acknowledge financial support by the ESQ (Erwin Schrödinger Center for Quantum Science \& Technology) Discovery programme 2019 "Quantum Klystron (QUAK)", hosted by the Austrian Academy of Sciences (\"OAW).

\bibliography{modulated_beam}

\onecolumngrid

\appendix

\section{Transition probabilities, back-action and decoherence}\label{sec:transprob}

If the change of state of the electron due to back-action is in principle detectable, decoherence occurs in the reduced state of the quantum system. Here, we analyze this effect. 
\newline

\subsection{Transition probability due to a classical electron}
The transition probability of the excited quantum system due to the magnetic field of a passing point-like classical electron can be calculated with the interaction Hamiltonian $H_\rm{int}=-\hat{\mu}\cdot \vec{B}$. We obtain the out-state to first order
\begin{equation}
	|\rm{out}\rangle_\rm{qs} \approx \left(\mathbb{I} + \frac{i}{\hbar}\int_{-\infty}^\infty dt \, \hat{\mu}\cdot \vec{B} \right) |e\rangle\,. 
\end{equation}
For an electron moving with a velocity $v$ parallel to the $z$-axis with a displacement of $\vec r_\perp=(x,y)$, whose trajectory pierces the $z=0$ plane at time $t_j$, the magnetic field is (equation 11.152 of \cite{jackson}, translated and rotated)
\begin{eqnarray}\label{eq:magsingrel}
	\vec B_j(0,t) &=& \left[\begin{array}{c} y \\  -x \\ 0 \end{array}\right] \frac{\mu_0 e \gamma v}{4\pi(r_\perp^2 + \gamma^2 v^2 (t-t_j)^2)^{3/2}}
\end{eqnarray}
where $r_\perp = |\vec r_\perp|$ is the minimal distance between the single electron and the quantum system (impact parameter) and $\gamma = (1 - v^2/c^2)^{-1/2}$ is the Lorentz factor. For the transition probability, we find
\begin{eqnarray}\label{eq:transprobmethods}
\nonumber	P_{e\rightarrow g} &=& \frac{1}{\hbar^2}\left| \langle g| \int_{-\infty}^\infty dt \, \hat{\mu}\cdot \vec{B}|e\rangle\right|^2 \\
	&=& \left(\frac{\mu_0 e |y \mu^x_{ge} - x \mu^y_{ge}| \omega}{2\pi \hbar \,r_\perp \gamma v} K_1\left(\frac{\omega \,r_\perp }{\gamma v}\right) \right)^2 \,,
\end{eqnarray}
where $\mu^x_{ge}$ and $\mu^y_{ge}$ are the components of the transition dipole moment. For the plot in Fig.2  of the main text, we consider $x=0$, $y=r_\perp$ and the magnetic dipole moment oriented in the $x$-direction. The same results are obtained for the inverse process $g\rightarrow e$, implying that $P_{g\rightarrow e} = P_{e\rightarrow g}$ for the purposes of this article.

\subsection{QED model for back-action}

We consider the transition from the excited to the ground state first and obtain the opposite case by the replacement $\omega_0\rightarrow -\omega_0$.
We assume that the quantum system's dimensions are much smaller than its distance to the center of the electron beam $d$ and the modulation wave length $\lambda_0=2\pi v/\omega_0$, where $v$ is the average velocity of the electrons and $\omega_0$ is the radian frequency of both the modulation and the transition of the quantum system. Thus, we can consider the quantum system as point-like.
Furthermore, we consider an initial Gaussian matter-wave packet of longitudinal size much smaller than $\lambda_0$, such that the beam modulation is not on the level of the single electron wave function but corresponds to correlations between electrons.   
The transversal width of the Gaussian matter-wave packet is bound from above by the focal width of the beam as $\Delta r_\perp < \text{w}/2$.
\newline
We describe the electron field as a Dirac field normalized with the charge
\begin{equation}
	Q(\psi,\psi') = \int d^3r \,\psi^\dagger \psi'  =: \langle \psi ,\psi'\rangle  \,,
\end{equation}
such that $Q(\psi_{p,s},\psi_{p',s'}) = (2\pi\hbar)^3 \delta^{(3)}(\vec{p} - \vec{p}^{\,\prime}) \delta_{ss'}$.
Furthermore, we define the momentum eigenstates as
\begin{eqnarray}
	\psi_{p,s}(\vec{r}) =  \left(\begin{array}{c}  \sqrt{\frac{\hbar\omega_p + mc^2}{2\hbar\omega_p}} \chi_s \\ \frac{\vec\sigma\cdot \vec{p} c}{\sqrt{2\hbar\omega_p (\hbar\omega_p + mc^2)}} \chi_s \end{array}\right) e^{i \vec{p}\cdot \vec{r}/\hbar}\,,
\end{eqnarray}
where $\omega_p = c\sqrt{|\vec{p}|^2 + m^2 c^2}/\hbar$, $\vec\sigma = (\sigma_1,\sigma_2,\sigma_3)$ is the vector of Pauli matrices and $\chi_s$ are the two-spinors $\chi_+ = (1,0)$ and $\chi_- = (0,1)$. We restrict our considerations to the particle solutions of positive energy.
In the following, we will also use the momentum eigenstates in the bra-ket notation $|\vec p,s\rangle$ that are defined such that $ \psi_{p,s}(\vec{r}) = \langle \vec{r} | \vec{p},s\rangle$ and $\langle \vec{p},s|\vec{p}^\prime,s'\rangle =  (2\pi\hbar)^3 \delta^{(3)}(\vec{p} - \vec{p}^{\,\prime}) \delta_{ss'}$.

We assume that the electromagnetic field stays in the vacuum throughout the process (no spontaneous emission) and that the quantum system is initially in the excited state $|e\rangle$.
For the electrons, we consider an initial state that is un-polarized
\begin{equation}
	\rho_{\rm{el},\rm{in}} =   \int \frac{d^3p}{(2\pi\hbar)^3}\frac{d^3p'}{(2\pi\hbar)^3} \, \phi_{\rm{in}}(\vec p)^* \phi_{\rm{in}}(\vec p^{\,\prime})  \frac{1}{2}  \sum_s |\vec p,s\rangle \langle\vec p^{\,\prime},s|
\end{equation}
where $\phi_\rm{in}(\vec p)$ is the polarization-independent single electron wave function in the momentum representation. For explicit calculations, we will use the $z$-axis as the spin quantization direction later. However, we note that the initial state is independent of the choice of spin basis. 
\newline
We consider an initial wave function $\phi_\rm{in}(\vec p)$ in the interaction region at $\vec{r}=0$ that factorizes into a transversal Gaussian wave packet with width $\Delta p_\perp$ and a longitudinal Gaussian wave packet with width $\Delta p_z$, that is
\begin{equation}
	\phi_{\rm{in}}(\vec p) = \phi_{\rm{in},z}( p_z) \phi_{\rm{in},\perp}(\vec p_\perp)e^{-i\omega_p (l_\rm{tot}/v - t_0)}  \,,
\end{equation}
where  $l_\rm{tot}/v$ is the time for the propagation of the wave packet from the source and we decomposed $\vec{p} = \vec{p}_z + \vec{p}_\perp$, where $\vec{p}_z$ and $\vec{p}_\perp$ are parallel and perpendicular to the $z$-axis, respectively.
For the longitudinal direction, we consider the Gaussian wave packet
\begin{equation}\label{eq:instatelong}
	\phi_{\rm{in},z}(p_z) = \left(\frac{(2\pi)^{1/2}\hbar}{\Delta p_z}\right)^{1/2} e^{-\frac{(p_z - p_{z,0})^2}{4\Delta p_z^2}}  e^{ip_z l_\rm{tot}/\hbar} \,
\end{equation}
where the phase $e^{ip_z l_\rm{tot}/\hbar}$ incorporates the propagation from the source at $z=-l_\rm{tot}$ to the interaction region. We assume that the initial transversal state of the electron in real space is a Gaussian wave packet displaced by $\vec{r}_{0,\perp}$, that is
\begin{equation}
	\psi_{\rm{in},\perp}^e(\vec{r}_\perp,t) = \tilde\psi_{\rm{in},\perp}^e(\vec{r}_\perp - \vec{r}_{0,\perp},t)
\end{equation}
and $\tilde\psi_{\rm{in},\perp}^e$ reaches its minimal extension at $t_0$ at the position of the quantum system at $z=0$. In momentum space, the displacement leads to a factor $e^{-i \vec p_\perp \cdot \vec{r}_{0,\perp}/\hbar}$. The transversal momentum spread is small enough to consider the transversal dispersion non-relativistically. In particular, the energy can be approximated as
\begin{equation}
	\omega_p = \frac{c}{\hbar} \sqrt{|\vec{p}|^2 + m^2 c^2} \approx  \frac{c}{\hbar} \sqrt{p_z^2 + m^2 c^2} + \frac{c |\vec{p}_\perp|^2}{2\hbar \sqrt{p_z^2 + m^2 c^2}} 
\end{equation}
which implies the following form of the transversal wave function in momentum space
\begin{equation}\label{eq:instatetrans}
	\phi_{\rm{in},\perp}(\vec p_\perp) = \frac{\sqrt{2\pi}\hbar}{\Delta p_\perp} e^{-i \vec p_\perp \cdot \vec{r}_{0,\perp}/\hbar} e^{-\frac{|\vec p_\perp|^2}{4\Delta p_\perp^2}} e^{i \frac{c |\vec{p}_\perp|^2}{2\hbar \sqrt{p_z^2 + m^2 c^2}} \frac{l_\rm{tot}}{v} } \,,
\end{equation}
where the last factor has been added to cancel the transversal dispersion terms induced by the time evolution at $t=t_0$.
The full in-state can be written as
\begin{equation}
	\rho_{\rm{in}} =  \frac{1}{2}  \sum_s \left(\vphantom{\sum}|e\rangle\otimes |\rm{in}_s\rangle_\rm{el}\right)\left(\vphantom{\sum}\langle e|\otimes  {}_\rm{el}\langle\rm{in}_s|\right)
\end{equation}
where the state vectors $|\rm{in}_s\rangle_\rm{el}$ are defined as
\begin{equation}
	|\rm{in}_s\rangle_\rm{el}  = \int \frac{d^3p}{(2\pi\hbar)^3} \, \phi_{\rm{in}}(\vec p) |\vec p,s\rangle\,.
\end{equation}

To lowest order, after the electron and quantum system interact, the full state will be
\begin{equation}
	\rho_{\rm{out}} = \frac{1}{2}  \sum_s |\rm{out}_s\rangle \langle \rm{out}_s|\,,
\end{equation}
where \footnote{The normalization of the out-state is imposed by hand to second order in $\sqrt{P_{e\rightarrow g}}$ leading to the factor in front of the first term.}
\begin{eqnarray}
	|\rm{out}_s\rangle &=& \sqrt{1 - P_{e\rightarrow g}(s)}|e\rangle\otimes |\rm{in}_s\rangle_\rm{el} + \sqrt{P_{e\rightarrow g}(s)}\,|g\rangle \otimes |\rm{scatt}_s\rangle_\rm{el}\,.
\end{eqnarray}
$P_{e\rightarrow g}(s)$ is the probability for the transition from the excited to the ground state for a fixed initial electron spin $s$ defined as
\begin{eqnarray}
    P_{e\rightarrow g}(s) &=& \sum_{s'} \int \frac{d^3p'}{(2\pi\hbar)^3} \frac{d^3p}{(2\pi\hbar)^3}\frac{d^3p''}{(2\pi\hbar)^3} \phi_{\rm{in}}^*(\vec p)\mathcal{S}^*(\vec p,s;\vec p^{\,\prime},s') \phi_{\rm{in}}(\vec p^{\,\prime\prime}) \mathcal{S}(\vec p^{\,\prime\prime},s;\vec p^{\,\prime},s') \,,
\end{eqnarray}
where  $\mathcal{S}(\vec{p},s;\vec{p}^{\,\prime},s') = \langle g, \vec p^{\,\prime},s' ,\rm{vac}|\hat{S}^{(2)}|e, \vec p,s ,\rm{vac}\rangle$ is the scattering matrix element for the transition from momentum $\vec p$ and spin $s$ to $\vec p^{\,\prime}$ and $s'$ in second order perturbation theory. Furthermore, we define the normalized scattered state of the electron
\begin{eqnarray}
	|\rm{scatt}_s\rangle_\rm{el} &=&  \sum_{s'} \int \frac{d^3p'}{(2\pi\hbar)^3}\,\phi_{\rm{scatt},s}(\vec p^{\,\prime},s') |\vec p^{\,\prime},s'\rangle \,,
\end{eqnarray}
where
\begin{equation}
	\phi_{\rm{scatt},s}(\vec p^{\,\prime},s') = (P_{e\rightarrow g}(s))^{-1/2} \int \frac{d^3p}{(2\pi\hbar)^3}\, \phi_{\rm{in}}(\vec p) \mathcal{S}(\vec{p},s,\vec{p}^{\,\prime},s') \,.
\end{equation} 
The driving process can only be coherent if the reduced density matrix of the quantum system (with the partial trace taken over the electron Hilbert space) is close to that of a pure state. We find
\begin{equation}
    \varrho_\rm{qs} = \rm{Tr}_\rm{el}[\rho_{\rm{out}}]=\frac{1}{2}  \sum_s \sum_{s'}  \int \frac{d^3p}{(2\pi\hbar)^3} \langle\vec p,s'|   \rm{out}_s\rangle \langle \rm{out}_s |\vec p,s'\rangle
\end{equation}
and
\begin{eqnarray}
	\langle\vec p,s'|   \rm{out}_s\rangle &=& \delta_{ss'} \phi_{\rm{in}}(\vec p) |e\rangle + \phi_{\rm{scatt},s}(\vec p,s') |g\rangle\,.
\end{eqnarray} 
Therefore,
\begin{eqnarray}\label{eq:reddens}
    \nonumber \varrho_\rm{qs} &=&\frac{1}{2}\sum_s \Big( (1 - P_{e\rightarrow g}(s))|e\rangle\langle e| + \\
  \nonumber  && \hphantom{=} \sqrt{1 - P_{e\rightarrow g}(s)}\sqrt{P_{e\rightarrow g}(s)}\int\frac{d^3p}{(2\pi\hbar)^3} \phi_{\rm{in}}(\vec p)^*\phi_{\rm{scatt},s}(\vec p,s) \,|e\rangle\langle g| + c.c.  + P_{e\rightarrow g}(s) |g\rangle \langle g| \Big) \\
    &\approx & \left(\begin{array}{cc} 1 - \frac{1}{2}\sum_s P_{e\rightarrow g}(s) &   \frac{1}{2}\sum_s \sqrt{P_{e\rightarrow g}(s)} \langle \rm{in}_s | \rm{scatt}_s\rangle \\  \frac{1}{2}\sum_s \sqrt{P_{e\rightarrow g}(s)} \langle \rm{in}_s | \rm{scatt}_s\rangle^* & \frac{1}{2}\sum_s P_{e\rightarrow g}(s)
    \end{array}\right)\,.
\end{eqnarray}
to second order in $\sqrt{P_{e\rightarrow g}(s)}$.
We will find later that $2|P_{e\rightarrow g}(+1/2)-P_{e\rightarrow g}(-1/2)|/|P_{e\rightarrow g}(+1/2)+P_{e\rightarrow g}(-1/2)|$ is at most of the order of $10^{-9}$ in the case that we consider here. So in the following, we assume that $P_{e\rightarrow g}(+1/2)=P_{e\rightarrow g}(-1/2) = P_{e\rightarrow g}$. Furthermore, we find that $P_{e\rightarrow g} = P_{g\rightarrow e}$. Since $\sqrt{P_{e\rightarrow g}}\gg P_{e\rightarrow g}$, the change of the reduced density matrix in equation (\ref{eq:reddens}) is dominated by the off-diagonal terms. Therefore, for coherent driving to be possible in principle, we need 
\begin{equation}\label{eq:overlap_condition}
	\left|\frac{1}{2}\sum_s \langle \rm{in}_s | \rm{scatt}_s\rangle \right| \gg \sqrt{P_{e\rightarrow g}} \,.
\end{equation}

\subsection{The scattering matrix element}

For the S-matrix, we have (see Chapter VIII of \cite{LandauLifshitzQED} or §104 of \cite{LandauLifshitzShorterQM} for details)
\begin{equation}
	\hat S = \mathcal{T}\exp\left(-\frac{i}{\hbar c}\int d^4x\, \hat{J}^\mu \hat{A}_\mu\right)\,,
\end{equation}
where $\mathcal{T}$ denotes time ordering, $\hat{J}^\mu$ contains all of the currents of charged particles and $\hat{A}_\mu$ is the electromagnetic 4-potential operator. The lowest order interaction term for our process occurs at the second order, for which we find
\begin{equation}
	\hat S^{(2)} = -\frac{1}{2(\hbar c)^2}\int d^4x \int d^4x'\, \mathcal{T} \left(\hat{J}^\mu(x) \hat{J}^\nu(x')\right) \mathcal{T}\left(\hat{A}_\mu(x) \hat{A}_\nu(x')\right)\,.
\end{equation}
For the process under consideration, the electromagnetic field stays in the vacuum state and we use
\begin{equation}
    \langle \rm{vac}| \mathcal{T}\left(\hat{A}_\mu(x) \hat{A}_\nu(x')\right) |\rm{vac}\rangle = i D^\rm{F}_{\mu\nu}(x-x')\,,
\end{equation}
where $D^\rm{F}_{\mu\nu}(x-x')$ is the Feynman propagator of the electromagnetic field. We are considering a regime where the quantum system only changes its internal state and the free electrons remain free, that is, we are neglecting any violent effects such as ionization. Then, the charged current operator can be split into the current operator of the free electron and that of the quantum system as
\begin{equation}
    \hat{J}^\mu(x) = [\hat{J}^\rm{el}(x)]^\mu + [\hat{J}^\rm{qs}(x)]^\mu
\end{equation}
and we obtain
\begin{equation}
	\mathcal{S}(\vec p,s;\vec p^{\,\prime},s') = \langle g, \vec p^{\,\prime},s',\rm{vac}|\hat S^{(2)}|e,\vec p,s,\rm{vac}\rangle =	-\frac{i}{(\hbar c)^2}\int d^4x\,d^4x'\, [J_{p,s \rightarrow p',s'}^\rm{el}(x)]^\mu D^\rm{F}_{\mu\nu}(x-x')[J^\rm{qs}_{e \rightarrow g}(x')]^\nu\,,
\end{equation}
where $J_{p,s\rightarrow p',s'}^\rm{el}(x)$ is the transition current of the free electron and $J^\rm{qs}_{e\rightarrow g}(x')$ is the transition current of the quantum system. We consider the Feynman propagator in a specific gauge where only the spatial components do not vanish:
\begin{equation}
	D^\rm{F}_{ij}(x-x') = - \mu_0\hbar^3 c \lim_{\epsilon \rightarrow 0^+} \int \frac{d^4q}{(2\pi\hbar)^4} e^{i \vec{q} \cdot (\vec{r}-\vec{r'})/\hbar} e^{-i q_0 c(t - t')/\hbar} \frac{1}{q_0^2 - |\vec q|^2 + i\epsilon}\left(\delta_{ij} - \frac{q_iq_j}{q_0^2}\right)\,
\end{equation}
(the real space version of the expressions given in \cite{dzyaloshinskii1959vanderwaals} equation (A.7b) and \cite{LandauLifshitzQED} §76).
Therefore, we can restrict our considerations to the spatial components of the transition currents.
We describe the free electron as a charged spin-1/2 field with the current (using $\vec\alpha = \gamma^0(\gamma^1,\gamma^2,\gamma^3)$ and $\gamma^\mu$ the Dirac matrices)
\begin{eqnarray}\label{eq:elcurrent}
	\nonumber \vec J_{p,s \rightarrow p',s'}^\rm{el} = - ec\,\psi_{p',s'}^\dagger \,\vec{\alpha}\, \psi_{p,s}   & = & -\frac{ec^2}{2\hbar\sqrt{\omega_p\omega_{p'}}} e^{-i(\vec{p}^{\,\prime} - \vec{p})\cdot \vec{r}/\hbar} e^{i(\omega_{p'} - \omega_p)t} \left(\vphantom{\chi_{s'}^\dagger} \left(\kappa_{p',p} \,\vec{p} + \kappa_{p,p'} \,\vec{p}^{\,\prime}\right) \delta_{ss'}\right.   \\
	&&  \hphantom{=} + i \left.\left( \kappa_{p',p} \,\vec{p} - \kappa_{p,p'} \,\vec{p}^{\,\prime}\right)\times \chi_{s'}^\dagger \vec{\sigma} \chi_s  \right)
\end{eqnarray}
where 
\begin{eqnarray}
	\kappa_{p',p} &=& \sqrt{\frac{\hbar\omega_{p'} + mc^2}{\hbar\omega_{p} + mc^2}}	= (\kappa_{p,p'})^{-1}\,.
\end{eqnarray}
We remark that the quantization volume is omitted throughout the calculations, so $\vec J_{p \rightarrow p'}^\rm{el}$ has the dimensions of a current instead of a current density. 
\newline
To describe the schemes proposed in this work, we restrict our considerations to magnetic dipole transitions of the quantum system (we also consider the effect of electric dipole transitions in Appendix \ref{sec:electrictransitions}). Furthermore, we assume that the quantum system is much heavier than the electrons (e.g. $m_e/m_{\,^{41}K}\sim 10^{-5}$ for our first example and much smaller for the NV-center in diamond) such that we can ignore the momentum recoil on the quantum system for this calculation. In this case, the spatial wave function of the quantum system is always unchanged to a good approximation. Then, the transition current of the quantum system can be approximated as that of a point-like magnetic dipole (see page 79 of \cite{itzykson2012quantum})
\begin{equation}
	\vec J^\rm{qs}_{e \rightarrow g}(x) = - e^{-i \omega_0 t}  \vec{\mu} \times \nabla \delta^{(3)}(\vec{r}) \,,
\end{equation}
where $\vec{\mu}$ is the magnetic transition dipole moment.
Partial integration and execution of the Fourier transforms leads to
\begin{eqnarray}
    \nonumber \mathcal{S}(\vec p,s;\vec p^{\,\prime},s') &=& -\frac{e\mu_0 c^2}{2\hbar\sqrt{\omega_p\omega_{p'}}} \lim_{\epsilon \rightarrow 0^+}    \frac{1}{(\hbar\omega_0/c)^2 - |\vec{p}^{\,\prime} - \vec{p}|^2 + i\epsilon} \Big( \epsilon^{jkl}  (\kappa_{p',p}\,p_j + \kappa_{p,p'}\,p^{\prime}_j)   \mu_k (p'_l - p_l) \,\delta_{ss'}  \\
	&& \hphantom{=} + i \left(\delta^{mk}\delta^{nl} - \delta^{ml}\delta^{nk}\right) ( \kappa_{p',p} \,p_m - \kappa_{p,p'} \,p'_m) \mu_k (p'_l - p_l) \chi_{s'}^\dagger \sigma_n \chi_s \Big) \, 2\pi \delta(\omega_{p'} - \omega_p - \omega_0)\,.
\end{eqnarray}
Since $\omega_p/\omega_0 = c\sqrt{|\vec{p}|^2 + m^2c^2}/\hbar\omega_0 \gtrsim 10^{10} \gg 1$, energy conservation implies 
\begin{equation}
	\kappa_{p',p} = \sqrt{\frac{\hbar\omega_{p'} + mc^2}{\hbar\omega_{p'} - \hbar\omega_0 + mc^2}} \approx 1 + \frac{\hbar\omega_0}{2(\hbar\omega_{p'} + mc^2)} \approx 1\,,
\end{equation}
and $\kappa_{p,p'} \approx 1$, which leads to
\begin{eqnarray}\label{eq:Smatrixelementapp}
 \nonumber \mathcal{S}(\vec p,s;\vec p^{\,\prime},s')	&\approx &  -\frac{e\mu_0 c^2}{2\hbar\sqrt{\omega_p\omega_{p'}}}  \Big( \epsilon^{jkl}  (p^{\prime}_j + p_j )   \mu_k (p'_l - p_l) \,\delta_{ss'} - i \left(\delta^{mk}\delta^{nl} - \delta^{ml}\delta^{nk}\right) ( p'_m - p_m) \mu_k (p'_l - p_l) \chi_{s'}^\dagger \sigma_n \chi_s \Big) \\
 &&   \frac{1}{(\hbar\omega_0/c)^2 - |\vec{p}^{\,\prime} - \vec{p}|^2 }   \,2\pi \delta(\omega_{p'} - \omega_p - \omega_0)   	 \,,
\end{eqnarray}
where the limit has been taken trivially since $\hbar\omega_0/c \neq |\vec{p}^{\,\prime} - \vec{p}|$ due to the energy conservation condition $\omega_{p'} - \omega_p - \omega_0$.

\subsection{The scattered state}

For the scattered state, we find
\begin{eqnarray}
	\nonumber \bar{\phi}_{\rm{scatt},s}(\vec{p}^{\,\prime},s') &: =&  \int \frac{d^3p}{(2\pi\hbar)^3}\, \phi_\rm{in}(\vec p) \mathcal{S}(\vec p,s;\vec p^{\,\prime},s') \\
	&=&  \frac{e\mu_0 c^2}{2\hbar}   \sum_s \int \frac{d^3p}{(2\pi\hbar)^3}\, \phi_\rm{in}(\vec p) \frac{1}{\sqrt{\omega_p \omega_{p'}}} \frac{1}{(\hbar \omega_0/c)^2-|\vec{p}^{\,\prime} - \vec{p}|^2 } 2\pi \delta(\omega_{p'} - \omega_p - \omega_0)  \\
	\nonumber && \Big( 2 \epsilon^{jkl}  p^{\prime}_j \mu_k p_l \,\delta_{ss'} + i \Big( (p'_m - p_m) \mu^m (p'_l - p_l) \sigma_{ss'}^l  - |\vec{p}^{\,\prime} - \vec{p}|^2 \mu_l \sigma_{ss'}^l \Big)  \Big)  \,.
\end{eqnarray}
where $\sigma_{ss'}^l = \chi_{s'}^\dagger \sigma^l \chi_s$.
Note that we have defined $\bar{\phi}_{\rm{scatt},s}(\vec{p},s)$ as the unnormalised scattered wave function for notational convenience, i.e. $\bar{\phi}_{\rm{scatt},s}(\vec{p},s)=\sqrt{P_{e\rightarrow g}}\phi_{\rm{scatt},s}(\vec{p},s)$. We define
\begin{eqnarray}
	p_{z,\rm{sol}}(\vec{p}_\perp,\vec{p}^{\,\prime}) &=& \left( \frac{\hbar^2}{c^2}(\omega_{p'} - \omega_0)^2 - m^2c^2 - |\vec{p}_\perp|^2 \right)^{1/2}\,.
\end{eqnarray}
Energy conservation implies $p_z = p_{z,\rm{sol}}(\vec{p}_\perp,\vec{p}^{\,\prime})$ and we obtain
\begin{equation}
	\delta(\omega_{p'} - \omega_p - \omega_0) = \left|\frac{\partial\omega_p}{\partial p_z}\right|^{-1} \delta(p_z - p_{z,\rm{sol}}(\vec{q}_\perp,\vec{p}^{\,\prime})) = \frac{\hbar^2}{c^2} \frac{\omega_{p'}-\omega_0}{p_{z,\rm{sol}}(\vec{p}_\perp,\vec{p}^{\,\prime})}  \delta(p_z - p_{z,\rm{sol}}(\vec{q}_\perp,\vec{p}^{\,\prime}))
\end{equation}
and for the scattered state
\begin{eqnarray}
	\nonumber \bar{\phi}_{\rm{scatt},s}(\vec{p}^{\,\prime},s') &=&  \frac{e\mu_0}{2}  \sqrt{\frac{\omega_{p'} - \omega_0}{ \omega_{p'}}}\, e^{-i(\omega_{p'}-\omega_0) (l_\rm{tot}/v - t_0)}  \int \frac{d^2p_\perp}{(2\pi\hbar)^2}\, \phi_{\rm{in},z}(p_{z,\rm{sol}}(\vec{p}_\perp,\vec{p}^{\,\prime})) \,\phi_{\rm{in},\perp}(\vec p_\perp) \frac{1}{a(\vec{p}_\perp,\vec{p}^{\,\prime})^2+|\vec{p}_\perp^{\,\prime}-\vec{p}_\perp|^2} \\
	\nonumber && \Bigg( 2\epsilon^{zkl}\mu_k\, \left( p_{\perp,l}'  -  p_{\perp,l} \frac{ p_z'    }{p_{z,\rm{sol}}(\vec{p}_\perp,\vec{p}^{\,\prime})}  \right) \,\delta_{ss'}  + i \frac{ (p_z' - p_{z,\rm{sol}}(\vec{p}_\perp,\vec{p}^{\,\prime}))^2 + |\vec{p}^{\,\prime}_\perp - \vec{p}_\perp|^2  }{p_{z,\rm{sol}}(\vec{p}_\perp,\vec{p}^{\,\prime})} \mu_l \sigma_{ss'}^l   \\
	\nonumber &&  - i \frac{ 1 }{p_{z,\rm{sol}}(\vec{p}_\perp,\vec{p}^{\,\prime})}  \Big((p_z' - p_{z,\rm{sol}}(\vec{p}_\perp,\vec{p}^{\,\prime})) \mu^z  +  (p'_{\perp,m} - p_{\perp,m}) \mu^m \Big)\\
	&& \Big((p_z' - p_{z,\rm{sol}}(\vec{p}_\perp,\vec{p}^{\,\prime})) \sigma_{ss'}^z  +  (p'_{\perp,l} - p_{\perp,l}) \sigma_{ss'}^l\Big)     \Bigg) \,,
\end{eqnarray}
where
\begin{equation}
	a(\vec{p}_\perp,\vec{p}^{\,\prime})^2 = (p_z'- p_{z,\rm{sol}}(\vec{p}_\perp,\vec{p}^{\,\prime}))^2-(\hbar \omega_0/c)^2\,.
\end{equation}
We assume that the magnetic dipole moment $\vec{\mu}$ of the transition under consideration is pointing into the $x$-direction. If we take into account that we have chosen the spin-quantization direction along the $z$-axis, for the spin preserving case we find
\begin{eqnarray}
	\nonumber \bar{\phi}_{\rm{scatt},s}(\vec{p}^{\,\prime},s) &=&   \frac{e\mu_0}{2} |\vec\mu| \sqrt{\frac{\omega_{p'} - \omega_0}{ \omega_{p'}}}\, e^{-i(\omega_{p'}-\omega_0) (l_\rm{tot}/v - t_0)} \int \frac{d^2p_\perp}{(2\pi\hbar)^2}\,   \frac{\phi_{\rm{in},z}(p_{z,\rm{sol}}(\vec{p}_\perp,\vec{p}^{\,\prime})) \,\phi_{\rm{in},\perp}(\vec p_\perp) }{a(\vec{p}_\perp,\vec{p}^{\,\prime})^2+|\vec{p}_\perp^{\,\prime}-\vec{p}_\perp|^2} \\
	&& \Bigg( 2 \left( p_{\perp,y}'  -  p_{\perp,y} \frac{ p_z'    }{p_{z,\rm{sol}}(\vec{p}_\perp,\vec{p}^{\,\prime})}  \right) +  (-1)^{s-1/2}   i (p'_{\perp,x} - p_{\perp,x})   \left( 1 - \frac{ p_z'  }{p_{z,\rm{sol}}(\vec{p}_\perp,\vec{p}^{\,\prime})} \right)     \Bigg) \,,
\end{eqnarray}
and for the spin-flip transition,
\begin{eqnarray}
      \bar{\phi}_{\rm{scatt},s}(\vec{p}^{\,\prime},-s) &=& \frac{e\mu_0}{2} |\vec\mu| \sqrt{\frac{\omega_{p'} - \omega_0}{ \omega_{p'}}}\, e^{-i(\omega_{p'}-\omega_0) (l_\rm{tot}/v - t_0)}  \int \frac{d^2p_\perp}{(2\pi\hbar)^2}\,   \frac{\phi_{\rm{in},z}(p_{z,\rm{sol}}(\vec{p}_\perp,\vec{p}^{\,\prime})) \,\phi_{\rm{in},\perp}(\vec p_\perp)}{a(\vec{p}_\perp,\vec{p}^{\,\prime})^2+|\vec{p}_\perp^{\,\prime}-\vec{p}_\perp|^2} \\
 \nonumber	&& i \frac{1}{p_{z,\rm{sol}}(\vec{p}_\perp,\vec{p}^{\,\prime})}\Bigg( (p_z' - p_{z,\rm{sol}}(\vec{p}_\perp,\vec{p}^{\,\prime}))^2 +  (p'_{\perp,y} - p_{\perp,y}) \left( (p'_{\perp,y} - p_{\perp,y})  - (-1)^{s-1/2}  i (p'_{\perp,x} - p_{\perp,x}) \right) \Bigg) \,.
\end{eqnarray}
In terms of the un-normalized scattered wave functions, the transition probability is
\begin{eqnarray}
	P_{e\rightarrow g}(s) &=& \sum_{s'} \int \frac{d^3p'}{(2\pi\hbar)^3} |\bar{\phi}_{\rm{scatt},s}(\vec{p}^{\,\prime},s')|^2 \,.
\end{eqnarray}
Taking $P_{e\rightarrow g}(+1/2)=P_{e\rightarrow g}(-1/2) = P_{e\rightarrow g}$ for granted, the spin averaged overlap becomes
\begin{eqnarray}
	\nonumber  \frac{1}{2}\sum_s \langle \rm{in}_s | \rm{scatt}_s\rangle  &=& \int\frac{d^3p}{(2\pi\hbar)^3} \phi_{\rm{in}}(\vec p)^*\frac{1}{2}\sum_s\phi_{\rm{scatt},s}(\vec p,s) \\
	\nonumber &=&  \frac{1}{\sqrt{P_{e\leftrightarrow g}}}\int\frac{d^3p}{(2\pi\hbar)^3} \phi_{\rm{in}}(\vec p)^*\frac{1}{2}\sum_s \bar\phi_{\rm{scatt},s}(\vec p,s)  \\
	\nonumber &=& \frac{e\mu_0|\vec\mu|}{\sqrt{P_{e\leftrightarrow g}}}  e^{i\omega_0 (l_\rm{tot}/v - t_0)} \int\frac{d^3p'}{(2\pi\hbar)^3}   \phi_{\rm{in},z}(p_z')^* \,\phi_{\rm{in},\perp}(\vec p_\perp^{\,\prime})^* \\
	&&  \sqrt{\frac{\omega_{p'} - \omega_0}{ \omega_{p'}}}\,  \int \frac{d^2p_\perp}{(2\pi\hbar)^2}\, \phi_{\rm{in},z}(p_{z,\rm{sol}}(\vec{p}_\perp,\vec{p}^{\,\prime})) \,\phi_{\rm{in},\perp}(\vec p_\perp)  \frac{\left( p_{\perp,y}'  -  p_{\perp,y} \frac{ p_z'    }{p_{z,\rm{sol}}(\vec{p}_\perp,\vec{p}^{\,\prime})}  \right)}{a(\vec{p}_\perp,\vec{p}^{\,\prime})^2+|\vec{p}_\perp^{\,\prime}-\vec{p}_\perp|^2}  \,.
\end{eqnarray}
where the spin term has canceled out.

\subsection{Numerical treatment}

We consider a Gaussian envelope for the input wavefunction $\phi_\rm{in}$ given in equations (\ref{eq:instatetrans}) and (\ref{eq:instatelong}). We rewrite the above equations in terms of dimensionless quantities $\vec{\pi}^{\,\prime}_\perp = \vec{p}^{\,\prime}_\perp/(2\Delta p_\perp)$, $\vec{\pi}_\perp = \vec{p}_\perp/(2\Delta p_\perp)$, $\pi_z' = p_z'/(2\Delta p_z)$, $\pi_{z,0}=p_{z,0}/(2\Delta p_z)$ and $\vec{\rho}_{0,\perp} = \vec{r}_{0,\perp} 2\Delta p_\perp/\hbar = \vec{r}_{0,\perp}/\Delta r_\perp$ so
\begin{eqnarray}
	 \nonumber   \bar{\phi}_{\rm{scatt},s}(\vec{\pi}^{\,\prime},s) &=& \mathcal{F}  \, \sqrt{\frac{\Omega_{\pi'} - \Omega_0}{ \Omega_{\pi'}}}  \,  e^{-i(\Omega_{p'}-\Omega_0) (\tau - \tau_0)} \int d^2\pi_\perp\,  e^{-(\pi_{z,\rm{sol}}(\vec{\pi}_\perp,\vec{\pi}^{\,\prime}) - \pi_{z,0})^2}  e^{i\pi_{z,\rm{sol}}(\vec{\pi}_\perp,\vec{\pi}^{\,\prime}) \tilde{l} }     \\
	 \nonumber && e^{-i \vec{\pi}_\perp \cdot \vec{\rho}_{0,\perp}} e^{-\vec{\pi}_\perp^2}  e^{i \xi^2 \frac{ |\vec{\pi}_\perp|^2}{2\sqrt{\pi_{z,\rm{sol}}(\vec{\pi}_\perp,\vec{\pi}^{\,\prime})^2 + M^2}} \tau }   \frac{1}{\bar{a}(\vec{\pi}_\perp,\vec{\pi}^{\,\prime})^2 + |\vec{\pi}_\perp^{\,\prime} - \vec{\pi}_\perp|^2}  \\
	 && \Bigg( 2\left(\pi_{\perp,y}' -  \pi_{\perp,y} \frac{ \pi_z'}{\pi_{z,\rm{sol}}(\vec{\pi}_\perp,\vec{\pi}^{\,\prime})}\right)  + (-1)^{s-1/2} i ( \pi'_{\perp,x} - \pi_{\perp,x} )  \left(  1 - \frac{\pi_z'}{\pi_{z,\rm{sol}}(\vec{\pi}_\perp,\vec{\pi}^{\,\prime})} \right)  \Bigg)   \,,
\end{eqnarray}
and
\begin{eqnarray}
	 \nonumber  \bar{\phi}_{\rm{scatt},s}(\vec{\pi}^{\,\prime},-s) &=& \mathcal{F} \, \sqrt{\frac{\Omega_{\pi'} - \Omega_0}{ \Omega_{\pi'}}}  \, e^{-i(\Omega_{p'}-\Omega_0) (\tau - \tau_0)}  \int d^2\pi_\perp\,  e^{-(\pi_{z,\rm{sol}}(\vec{\pi}_\perp,\vec{\pi}^{\,\prime}) - \pi_{z,0})^2}  e^{i\pi_{z,\rm{sol}}(\vec{\pi}_\perp,\vec{\pi}^{\,\prime}) \tilde{l} }  \\
	\nonumber  && e^{-i \vec{\pi}_\perp \cdot \vec{\rho}_{0,\perp}} e^{-\vec{\pi}_\perp^2}  e^{i \xi^2 \frac{ |\vec{\pi}_\perp|^2}{2\sqrt{\pi_{z,\rm{sol}}(\vec{\pi}_\perp,\vec{\pi}^{\,\prime})^2 + M^2}} \tau }  
	 \frac{1}{\bar{a}(\vec{\pi}_\perp,\vec{\pi}^{\,\prime})^2 + |\vec{\pi}_\perp^{\,\prime} - \vec{\pi}_\perp|^2}    \\
	  && \frac{i}{\xi \pi_{z,\rm{sol}}(\vec{\pi}_\perp,\vec{\pi}^{\,\prime})} \Bigg(  (\pi_z' - \pi_{z,\rm{sol}}(\vec{\pi}_\perp,\vec{\pi}^{\,\prime}))^2 + \xi^2 (\pi'_{\perp,y} - \pi_{\perp,y}) \\
	 \nonumber && \left( (\pi'_{\perp,y} - \pi_{\perp,y}) - (-1)^{s-1/2} i(\pi'_{\perp,x} - \pi_{\perp,x}) \right)    \Bigg)   \,,
\end{eqnarray}
where
\begin{eqnarray}
	\bar{a}(\vec{\pi}_\perp,\vec{\pi}^{\,\prime})^2 &=& \frac{1}{\xi^2} \left( (\pi_z' - \tilde\pi_{z,\rm{sol}}(\vec{\pi}_\perp,\vec{\pi}^{\,\prime}))^2 - \Omega_0^2 \right) \,,\\
	 \pi_{z,\rm{sol}}(\vec{\pi}_\perp,\vec{\pi}^{\,\prime}) &=& \left( (\Omega_{\pi'} - \Omega_0)^2 - M^2 - \xi^2 \vec{\pi}_\perp^2  \right)^{1/2} \,,\\
	 \Omega_{\pi'} &=& \sqrt{\pi_z^{\,\prime 2} + \xi^2 \vec{\pi}_\perp^{\,\prime 2} + M^2}\,,
\end{eqnarray}
$\Omega_0 = \frac{\hbar \omega_0}{2 c\Delta p_z }$, $M=\frac{mc}{2\Delta p_z}$, $\xi = \Delta p_\perp/\Delta p_z$,  $\tilde{l} = 2\Delta p_z l_\rm{tot}/\hbar = l_\rm{tot}/\Delta z_0 $, $\tau = 2\Delta p_z c l_\rm{tot} /(\hbar v)$, $\tau_0 = 2\Delta p_z c t_0 /\hbar$ 
and 
\begin{eqnarray}
	\mathcal{F} &=& \frac{ e\mu_0 |\vec{\mu}| }{(2\pi\hbar)^2}    \left(\frac{(2\pi)^{1/2}\hbar}{\Delta p_z}\right)^{1/2} \sqrt{2\pi}\hbar = \frac{ e\mu_0 |\vec{\mu}| }{(2\pi)^{5/4} (\hbar \Delta p_z)^{1/2}}   	\,.
\end{eqnarray}
As above, the transversal dispersion can be treated perturbatively. In particular, we have
\begin{eqnarray}
	\pi_{z,\rm{sol}}(\vec{\pi}_\perp,\vec{\pi}^{\,\prime}) &\approx& \left( (\Omega_{\pi'} - \Omega_0)^2 - M^2 \right)^{1/2} - \frac{1}{2} \xi^2 \vec{\pi}_\perp^2 \left( (\Omega_{\pi'} - \Omega_0)^2 - M^2 \right)^{-1/2}
\end{eqnarray} 
to first order in $\xi^2 \vec{\pi}_\perp^2/\pi_{z,0}^2=\vec{p}_\perp^2/p_0^2$
and
\begin{eqnarray}
	\frac{\tau}{\tilde{l}}  = \frac{c}{v} = \frac{\gamma M}{\pi_{z,0}} = \frac{(M^2 + \pi_{z,0}^2)^{1/2}}{\pi_{z,0}} &\approx&   \frac{\Omega_{\pi}}{(\Omega_{\pi}^2 - M^2)^{1/2}} = \frac{\Omega_{\pi'} - \Omega_0}{((\Omega_{\pi'} - \Omega_0)^2 - M^2)^{1/2}}\quad\rm{and}\\
	 \frac{ \xi^2 \vec{\pi}_\perp^2}{\sqrt{\pi_{z,\rm{sol}}(\vec{\pi}_\perp,\vec{\pi}^{\,\prime})^2 + M^2}} &=& \frac{\xi^2 \vec{\pi}_\perp^2}{\Omega_{\pi'} - \Omega_0}
\end{eqnarray} 
to zeroth order and first order in $\xi^2 \vec{\pi}_\perp^2/\pi_{z,0}^2$, respectively. Then to first order in $\xi^2 \vec{\pi}_\perp^2/\pi_{z,0}^2$, we find for the sum of the dispersion phases
\begin{eqnarray}
	\pi_{z,\rm{sol}}(\vec{\pi}_\perp,\vec{\pi}^{\,\prime}) \tilde{l} + \xi^2 \frac{ \vec{\pi}_\perp^2}{ 2\sqrt{\pi_{z,\rm{sol}}(\vec{\pi}_\perp,\vec{\pi}^{\,\prime})^2 + M^2}} \tau  \approx \left( (\Omega_{\pi'} - \Omega_0)^2 - M^2 \right)^{1/2} \tilde{l}\,,
\end{eqnarray}
which is independent of $\vec{\pi}_\perp^2$ and can be pulled out of the integral. Then, the transversal dispersion of the wave packet over the drift distance is canceled by the phases we introduced when we defined the initial state just for this purpose since we consider the electron beam to be focused in the interaction region. The final dispersion phase cancels in the probability integral. It only appears in the overlap integral. We define the dispersion free scattered state functions
\begin{eqnarray}
	 \nonumber   \tilde{\phi}_{\rm{scatt},s}(\vec{\pi}^{\,\prime},s) &=& \mathcal{F}  \, \sqrt{\frac{\Omega_{\pi'} - \Omega_0}{ \Omega_{\pi'}}}  \,  \int d^2\pi_\perp\,  e^{-(\pi_{z,\rm{sol}}(\vec{\pi}_\perp,\vec{\pi}^{\,\prime}) - \pi_{z,0})^2} e^{-i \vec{\pi}_\perp \cdot \vec{\rho}_{0,\perp}} e^{-\vec{\pi}_\perp^2} \frac{1}{\bar{a}(\vec{\pi}_\perp,\vec{\pi}^{\,\prime})^2 + |\vec{\pi}_\perp^{\,\prime} - \vec{\pi}_\perp|^2}  \\
	 && \Bigg( 2\left(\pi_{\perp,y}' -  \pi_{\perp,y} \frac{ \pi_z'}{\pi_{z,\rm{sol}}(\vec{\pi}_\perp,\vec{\pi}^{\,\prime})}\right)  + (-1)^{s-1/2} i ( \pi'_{\perp,x} - \pi_{\perp,x} )  \left(  1 - \frac{\pi_z'}{\pi_{z,\rm{sol}}(\vec{\pi}_\perp,\vec{\pi}^{\,\prime})} \right)  \Bigg)   \,,
\end{eqnarray}
and
\begin{eqnarray}
	 \nonumber  \tilde{\phi}_{\rm{scatt},s}(\vec{\pi}^{\,\prime},-s) &=& \mathcal{F} \, \sqrt{\frac{\Omega_{\pi'} - \Omega_0}{ \Omega_{\pi'}}}  \,  \int d^2\pi_\perp\,  e^{-(\pi_{z,\rm{sol}}(\vec{\pi}_\perp,\vec{\pi}^{\,\prime}) - \pi_{z,0})^2} e^{-i \vec{\pi}_\perp \cdot \vec{\rho}_{0,\perp}} e^{-\vec{\pi}_\perp^2}	 \frac{1}{\bar{a}(\vec{\pi}_\perp,\vec{\pi}^{\,\prime})^2 + |\vec{\pi}_\perp^{\,\prime} - \vec{\pi}_\perp|^2}    \\
	  && \frac{i}{\xi \pi_{z,\rm{sol}}(\vec{\pi}_\perp,\vec{\pi}^{\,\prime})} \Bigg(  (\pi_z' - \pi_{z,\rm{sol}}(\vec{\pi}_\perp,\vec{\pi}^{\,\prime}))^2 + \xi^2 (\pi'_{\perp,y} - \pi_{\perp,y}) \\
	 \nonumber && \left( (\pi'_{\perp,y} - \pi_{\perp,y}) - (-1)^{s-1/2} i(\pi'_{\perp,x} - \pi_{\perp,x}) \right)    \Bigg)   \,.
\end{eqnarray}
Then, the probability is obtained as
\begin{eqnarray}
	P_{e\rightarrow g}(s) &=& \bar{\mathcal{F}}^2  \sum_{s'} \int d^3\pi' | \tilde{\phi}_{\rm{scatt},s}(\vec{\pi}^{\,\prime},s')/\mathcal{F}|^2 \,,
\end{eqnarray}
where
\begin{eqnarray}
	\bar{\mathcal{F}} = \frac{2\Delta p_\perp (2\Delta p_z)^{1/2} }{(2\pi\hbar)^{3/2}} \frac{ e\mu_0 |\vec{\mu}| }{(2\pi)^{5/4} (\hbar \Delta p_z)^{1/2}} = \frac{e\mu_0 |\vec{\mu}|}{ \hbar \Delta r_\perp  \pi^{1/2} (2\pi)^{9/4}}\,.
\end{eqnarray}
Using the semi-classical transition probability found in equation (\ref{eq:transprobmethods}) above implies
\begin{equation}
	\frac{\bar{\mathcal{F}}^2}{P_{e\rightarrow g}(s)} = \left(\frac{r_{0,\perp}}{\Delta r_\perp}\right)^2 \frac{1}{4 \pi^3 (2\pi)^{1/2}}\,.
\end{equation}
Performing the same approximations as above for the initial state, we find
\begin{eqnarray}
	 \nonumber   \phi_{\rm{in}}(\vec{\pi}^{\,\prime}) &=&  \frac{\sqrt{2\pi}\hbar}{\Delta p_\perp}\left(\frac{(2\pi)^{1/2}\hbar}{\Delta p_z}\right)^{1/2}   e^{-(\pi_z' - \pi_{z,0})^2}  e^{i\pi_z' \tilde{l} }   e^{-i \vec{\pi}_\perp \cdot \vec{\rho}_{0,\perp}} e^{-\vec{\pi}_\perp^2} e^{i \xi^2 \frac{ |\vec{\pi}_\perp|^2}{\sqrt{\pi_z^{\prime\,2} + M^2}} \tau }   e^{-i\Omega_{p'} (\tau - \tau_0)} \\
	 &\approx & \frac{\sqrt{2\pi}\hbar}{\Delta p_\perp}\left(\frac{(2\pi)^{1/2}\hbar}{\Delta p_z}\right)^{1/2}   e^{-(\pi_z' - \pi_{z,0})^2}   e^{-i \vec{\pi}_\perp \cdot \vec{\rho}_{0,\perp}} e^{-\vec{\pi}_\perp^2} e^{i\left( \Omega_{\pi'}^2 - M^2 \right)^{1/2} \tilde{l}} \,e^{-i\Omega_{p'} (\tau - \tau_0)}\,.
\end{eqnarray}
Then, the spin-averaged overlap is
\begin{eqnarray}
	\nonumber  \frac{1}{2}\sum_s \langle \rm{in}_s | \rm{scatt}_s\rangle & =  &   e^{i\Omega_0 (\tau - \tau_0)}  \frac{1}{2}  \sum_{s'} \int d^3\pi'\, \phi_{\rm{in}}(\vec{\pi}^{\,\prime})^* \bar\phi_{\rm{scatt},s''}(\vec{\pi}^{\,\prime},s') \,.
\end{eqnarray}
Under the integral, the remaining relative phase between in-state and scattered state can be approximated as
\begin{equation}
	\left(\left( (\Omega_{\pi'} - \Omega_0)^2 - M^2 \right)^{1/2} - \left( \Omega_{\pi'}^2 - M^2 \right)^{1/2}\right) \tilde{l} \approx  - \Omega_0 \frac{\Omega_{\pi'}}{\left( \Omega_{\pi'}^2 - M^2 \right)^{1/2}} \tilde{l} =  - \frac{\Omega_{\pi'}}{\left( \Omega_{\pi'}^2 - M^2 \right)^{1/2}} 2\pi \frac{l_\rm{tot}}{\lambda_0} \,.
\end{equation}
Then, we define the un-dispersed wave packet
\begin{eqnarray}
	 \nonumber   \tilde\phi_{\rm{in}}(\vec{\pi}^{\,\prime},s) &=&  \frac{\sqrt{2\pi}\hbar}{\Delta p_\perp}\left(\frac{(2\pi)^{1/2}\hbar}{\Delta p_z}\right)^{1/2}   e^{-(\pi_z' - \pi_{z,0})^2}   e^{-i \vec{\pi}_\perp \cdot \vec{\rho}_{0,\perp}} e^{-\vec{\pi}_\perp^2} \,,
\end{eqnarray}
and calculate the spin-averaged overlap as
\begin{eqnarray}
	\nonumber \frac{1}{2}\sum_s \langle \rm{in}_s | \rm{scatt}_s\rangle &=&  \left(  \int d^3\pi |\tilde\phi_{\rm{in}}(\vec{\pi})|^2 \right)^{-1/2}  \left(  \sum_{s'} \int d^3\pi' | \tilde{\phi}_{\rm{scatt},s}(\vec{\pi}^{\,\prime},s')|^2 \right)^{-1/2}  \\
	&&  e^{i\Omega_0 (\tau - \tau_0)}   \frac{1}{2}  \sum_{s''} \int d^3\pi''\, \tilde\phi_{\rm{in}}(\vec{\pi}^{\,\prime\prime})^* \tilde\phi_{\rm{scatt},s''}(\vec{\pi}^{\,\prime\prime},s'')  e^{- i\Omega_0 \frac{\Omega_{\pi'}}{\left( \Omega_{\pi'}^2 - M^2 \right)^{1/2}} \tilde{l}}  \,.
\end{eqnarray}
For the numerical evaluation, we set $\tau_0=0$.

Due to the Gaussian shape of the initial state, the momentum distribution can be restricted to the intervals given by $-n \lesssim \pi_z - \pi_{z,0} \lesssim n$ and $-n \lesssim |\vec{\pi}_\perp | \lesssim n$. We chose $n=5$, which implies that contributions smaller than $e^{-25}$ are neglected. 
\begin{figure}[hbt!]
\hspace*{-3mm}
\includegraphics[width=7.8cm,angle=0]{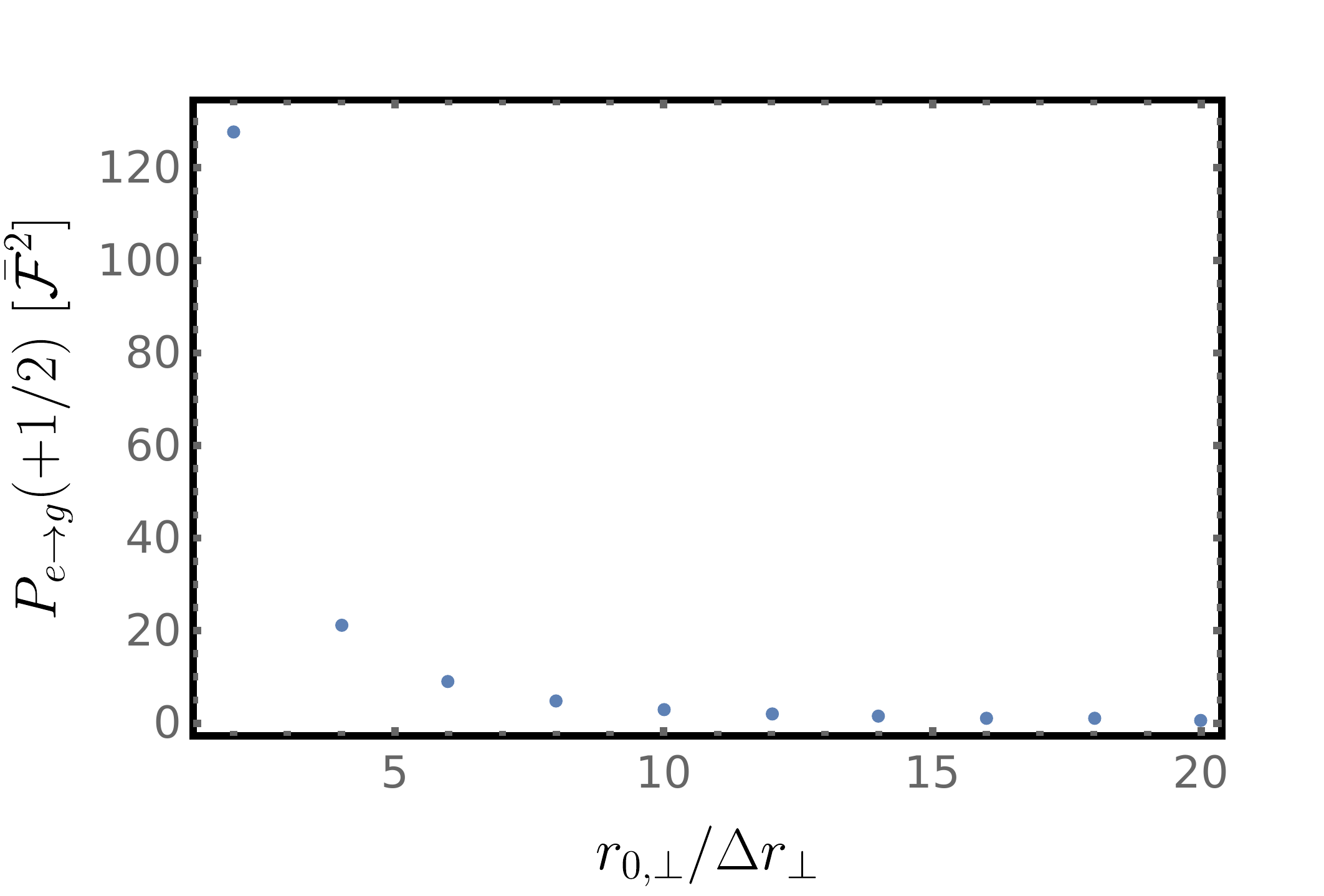}
\includegraphics[width=9cm,angle=0]{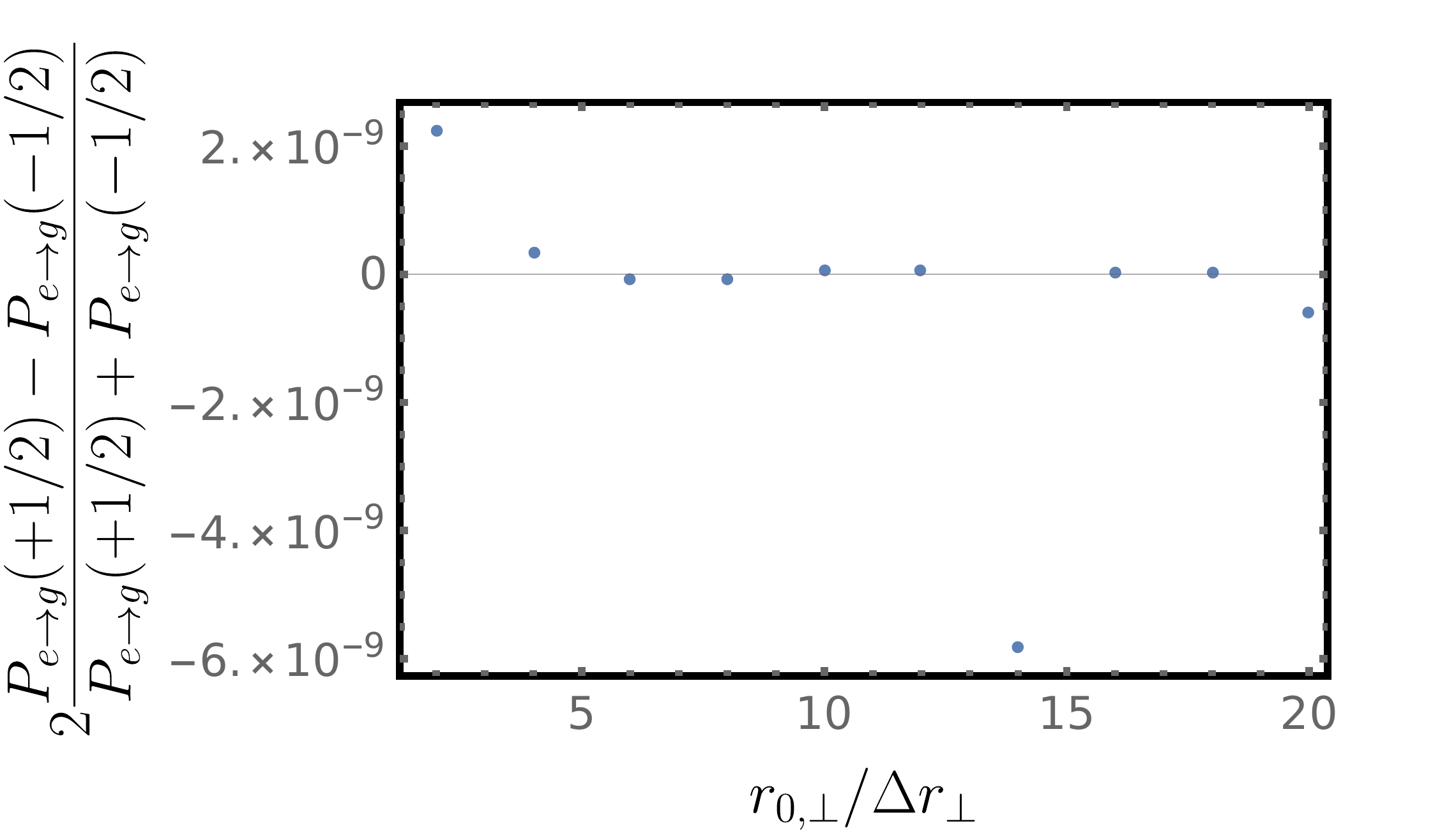}
\includegraphics[width=7.8cm,angle=0]{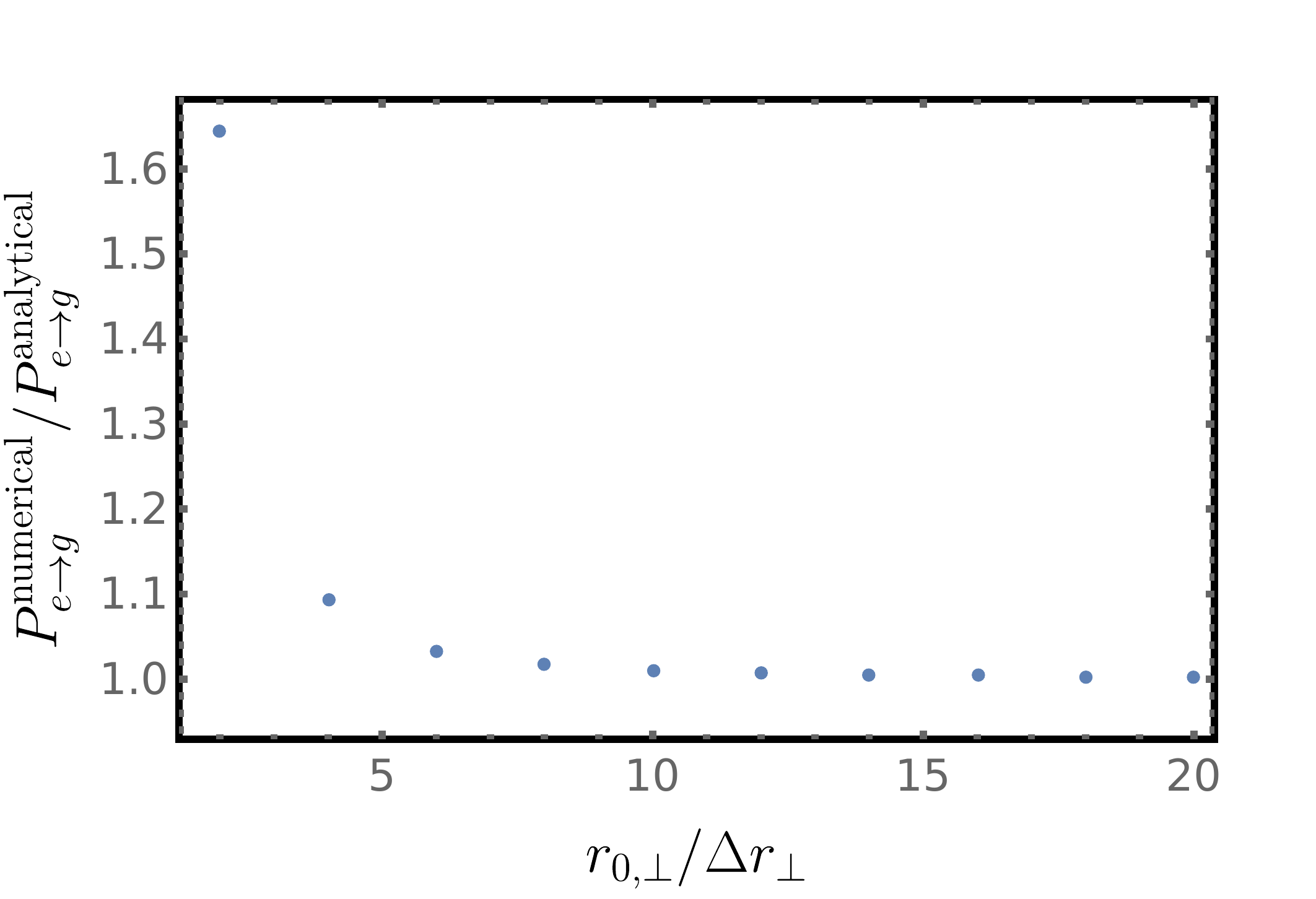}
\includegraphics[width=8cm,angle=0]{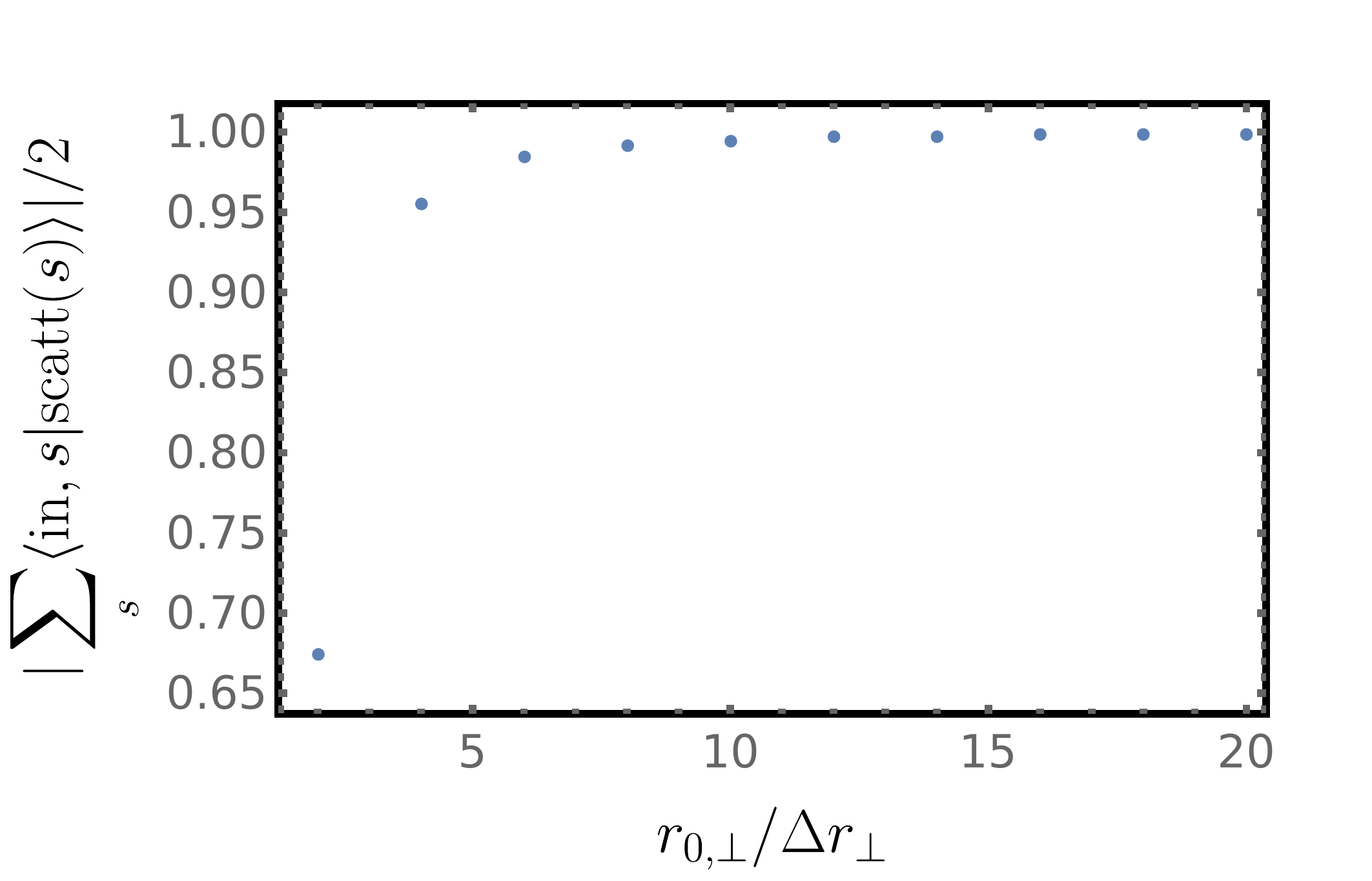}
\caption{\label{fig:probability} Upper left plot: the transition probability for initial spin $+1/2$ plotted for different distances between the quantum system and the beam line. Upper right plot: the relative difference between the transition probabilities for the initial spin states. This plot shows fluctuations that are due to numerical noise. We can conclude that the relative difference is at most of the order of $10^{-9}$ and potentially even lower. Therefore, probabilities can be considered as equal for the purpose of this article. Lower left plot: the quotient of the numerically calculated transition probability and the semi-classical result above in equation (\ref{eq:transprobmethods}). The numerical result agrees with the analytical result by one percent or less for distances larger than $10\Delta r_\perp$. Lower right plot: the overlap of the initial electron state and the final electron state. For distances $\ge 6 \Delta r_\perp$, the overlap can be considered to be one for the purpose of this article. The plots were obtained for $\Delta r_\perp = 5\,$nm, $\Delta z_0  = 100\,$nm, $\omega_0 = 2\pi\times 2.87\times 10^9\,\rm{rad/s}$, an initial kinetic energy of the electron of 2~keV and a total propagation distance from the electron source $l_\rm{tot} = 1\,$m. For $\vec{r}_{0,\perp}$ we considered $(0,r_{0,\perp})$, that is the quantum system and the center of the electron wave packet lie in the $y$-$z$-plane. }
\end{figure}
The corresponding initial range of significant electron energies is
\begin{equation}
	-n\frac{\pi_{z,0}}{\Omega_{\pi_{z,0}}} \lesssim \Omega_{\pi} -  \Omega_{\pi_{z,0}} \lesssim \frac{2n\pi_{z,0} + n^2 (1+\xi^2)}{2\Omega_{\pi_{z,0}}}\,.
\end{equation}
where $\Omega_{\pi_{z,0}}=\sqrt{\pi_{z,0}^{2} + M^2}$ and we used the fact that $\pi_{z,0}\gg \xi n$. Due to energy conservation, it follows that the integration of the final wave function can be restricted to 
\begin{eqnarray}
	|\pi_z^{\,\prime}| & \lesssim & \pi_{z,0} + \frac{\Omega_{\pi_{z,0}}\Omega_0}{\pi_{z,0}}  + n + \frac{n^2(1+\xi^2)}{2\pi_{z,0}} \\
	|\vec{\pi}_\perp^{\,\prime}| & \lesssim & \xi^{-1}\left|\pi_{z,0}^2 + 2\Omega_{\pi_{z,0}}\Omega_0 - \pi_z^{\,\prime 2} + 2n\pi_{z,0} + n^2 (1+\xi^2) \right|^{1/2}\,,
\end{eqnarray}
and if $\pi_{z,0}^2 + 2\Omega_{\pi_{z,0}}\Omega_0 - \pi_z^{\,\prime 2} - 2n\pi_{z,0} >0$,
\begin{eqnarray}
  |\vec{\pi}_\perp^{\,\prime}| &\gtrsim & \xi^{-1}\left|\pi_{z,0}^2 + 2\Omega_{\pi_{z,0}}\Omega_0 - \pi_z^{\,\prime 2} - 2n\pi_{z,0}\right|^{1/2}  \,,
\end{eqnarray}
where we have taken into account that $\Omega_0\ll \Omega_{\pi_{z,0}}$. Plots of the results are given in Fig. \ref{fig:probability} for the parameters used for the second
explicit example presented in this article; an NV center in diamond driven by an electron beam on the nano-scale.  Similar plots are obtained for the parameters considered for the first example.
We unsurprisingly find no visible effect of the initial spin on the transition probability. Furthermore, for distances of about $10\Delta r_\perp$, the numerical result for the transition probability approaches the result of the semi-classical calculation found in equation (\ref{eq:transprobmethods}), a result which can be assumed to continue for even larger distances. The condition on the overlap of in-state and out-state in equation (\ref{eq:overlap_condition}) 
is fulfilled for all distances in the plot. This behavior can also be expected to continue for larger distances. Our numerical result shows that, for $\tau_0=0$, the overlap is purely imaginary (the real part cannot be distinguished from numerical noise).
\newline
For the transition from the ground to the excited state, we must perform the replacement $\omega_0\rightarrow -\omega_0$ in all equations above. Performing the numerical analysis as before leads to the conclusion that 
\begin{eqnarray}
    \left|\frac{1}{2}\sum_s \langle \rm{in}_s | \rm{scatt}_s\rangle_{e\rightarrow g}\right| \approx \left|\frac{1}{2}\sum_s \langle \rm{in}_s | \rm{scatt}_s\rangle_{g\rightarrow e} \right|\,,
\end{eqnarray}
where we defined $| \rm{scatt}_s\rangle_{e\rightarrow g}$ as the scattered state of the electron after inducing a transition from the ground state to the excited state and analogously $| \rm{scatt}_s\rangle_{g\rightarrow e}$.
Furthermore, we obtain that the overlap is imaginary for $\tau_0=0$ and positive as for the transition $e\rightarrow g$. This implies
\begin{eqnarray}\label{eq:overlapgeandeg}
    \frac{1}{2}\sum_s \langle \rm{in}_s | \rm{scatt}_s\rangle_{e\rightarrow g} \approx -\left(\frac{1}{2}\sum_s \langle \rm{in}_s | \rm{scatt}_s\rangle_{g\rightarrow e} \right)^{*}\,,
\end{eqnarray}

\subsection{General initial state} 

In this section, we investigate the effect on the quantum system starting from a general separable pure state
\begin{equation}
\left|\rm{in}\right>=\left(\alpha\left|e\right>+\beta\left|g\right>\right)\otimes\left|\rm{in}\right>_\rm{el}\,.
\end{equation}
From our results above, we conclude that the spin of the electron can be neglected and that $P_{e\rightarrow g}=P_{e\leftarrow g}$ and we will write $P$ for both in the following. After the scattering, the state becomes
\begin{eqnarray}
\left|\rm{out}\right> &=& \alpha\left[\sqrt{1-P} \left|e\right>\otimes\left|\rm{in}\right>_\rm{el} + \sqrt{P}\left|g\right>\otimes\left|\rm{scatt}\right>_{\rm{el},e\rightarrow g} \right] + \beta \left[\sqrt{1-P} \left|g\right>\otimes\left|\rm{in}\right>_\rm{el}  + \sqrt{P}\left|e\right>\otimes\left|\rm{scatt}\right>_{\rm{el},g\rightarrow e}\right]\,.
\end{eqnarray}
Taking into account equation (\ref{eq:overlapgeandeg}) with spin-independent overlaps and defining $ i\Lambda_1 = {}_\rm{el}\langle \rm{in} | \rm{scatt} \rangle_{\rm{el},e\rightarrow g} = - \,{}_\rm{el}\langle \rm{in} | \rm{scatt} \rangle_{\rm{el},g\rightarrow e}^*$ and
$ \Lambda_2 = {}_{\rm{el},e\rightarrow g}\langle \rm{scatt} | \rm{scatt} \rangle_{\rm{el},g\rightarrow e}$,
we find for the reduced density matrix
\begin{eqnarray}
   \varrho_\rm{qs} &=&  (1-P) \left(\begin{array}{cc}
    |\alpha|^2    &  \alpha\beta^* \\
    \alpha^*\beta & |\beta|^2 
    \end{array}\right)   + i\sqrt{P} \left(\begin{array}{cc}
     - \Lambda_1\alpha\beta^* + \Lambda_1^*\alpha^*\beta   & -\Lambda_1^*|\alpha|^2 + \Lambda_1^*|\beta|^2  \\
    \Lambda_1|\alpha|^2 - \Lambda_1|\beta|^2  &  \Lambda_1\alpha\beta^* - \Lambda_1^*\alpha^*\beta
    \end{array}\right)  + P \left(\begin{array}{cc}
    |\beta|^2    &  \Lambda_2 \alpha^*\beta \\
    \Lambda_2^* \alpha\beta^* &  |\alpha|^2
    \end{array}\right) \,.
\end{eqnarray}
The change of the reduced density matrix due to the scattering event can be written in terms of the vector of components of the density matrix (in the co-rotating frame as we started in the interaction picture)  $(\tilde\rho_{eg},\tilde\rho_{ge},\tilde\rho_{ee},\tilde\rho_{gg})=(\alpha\beta^*,\alpha^*\beta,|\alpha|^2,|\beta|^2)$ as 
\begin{eqnarray}\label{eq:rhochange}
	  \Delta \left[\begin{array}{c}
		\tilde\rho_{eg} \\ \tilde\rho_{ge} \\ \tilde\rho_{ee} \\ \tilde\rho_{gg}
	\end{array}\right] &=& 
	\left[ \begin{array}{cccc} 
			-P  & P\Lambda_2 & -i\Lambda_1^* \sqrt{P} & i\Lambda_1^* \sqrt{P} \\
			P\Lambda_2^* & -P  &  i\Lambda_1 \sqrt{P} & 	-i\Lambda_1 \sqrt{P} \\
			  - i\Lambda_1\sqrt{P} & i\Lambda_1^*\sqrt{P} & -P  & P \\
			  i\Lambda_1\sqrt{P} & - i\Lambda_1^*\sqrt{P} &  P  & -P  	
	\end{array}  \right] 
\left[\begin{array}{c}
		\tilde\rho_{eg} \\ \tilde\rho_{ge} \\ \tilde\rho_{ee} \\ \tilde\rho_{gg}
	\end{array}\right] \,.
\end{eqnarray}
There are two extremal situations; when the electron state is either unchanged to a good approximation ($|\Lambda_1|=|\Lambda_2|=1$, corresponding to the case we obtain here) or so strongly affected so as to be in an almost orthogonal state ($|\Lambda_1|=|\Lambda_2|=0$), so that the reduced state of the quantum system remains approximately in a pure state (coherence preserving scattering) or becomes a mixed state (decohering scattering), respectively. 

If the electron state changes substantially, that is $\Lambda_1 \ll \sqrt{P}$, we obtain the case of incoherent scattering
\begin{eqnarray}\label{eq:rhochangeincoherent}
	  \Delta \left[\begin{array}{c}
		\rho_{eg} \\ \rho_{ge} \\ \rho_{ee} \\ \rho_{gg}
	\end{array}\right] &=& 
	\left[ \begin{array}{cccc} 
			-P  & 0 & 0 & 0 \\
		0 & -P  &  0  & 0 \\
			  0 & 0 & -P  & P \\
			  0 & 0 &  P  & -P  	
	\end{array}  \right] 
\left[\begin{array}{c}
		\rho_{eg} \\ \rho_{ge} \\ \rho_{ee} \\ \rho_{gg}
	\end{array}\right] \,,
\end{eqnarray}
where the quantum system is driven by a stochastic force into a steady state of vanishing inversion $\rho_{ee}-\rho_{gg}=0$ and vanishing off-diagonal components which is a maximally mixed state.

When the state change of the electron is negligible up to a phase, we find
\begin{eqnarray}\label{eq:rhochangecoherent}
	  \Delta \left[\begin{array}{c}
		\tilde\rho_{eg} \\ \tilde\rho_{ge} \\ \tilde\rho_{ee} \\ \tilde\rho_{gg}
	\end{array}\right] &=& 
	\left[ \begin{array}{cccc} 
			-P  & P e^{-2i\phi} & -ie^{-i\phi} \sqrt{P} & ie^{-i\phi} \sqrt{P} \\
			P e^{2i\phi} & -P  &  ie^{i\phi} \sqrt{P} & 	-ie^{i\phi} \sqrt{P} \\
			  - ie^{i\phi} \sqrt{P} & ie^{-i\phi}\sqrt{P} & -P  & P \\
			  ie^{i\phi}\sqrt{P} & - ie^{-i\phi}\sqrt{P} &  P  & -P  	
	\end{array}  \right] 
\left[\begin{array}{c}
		\tilde\rho_{eg} \\ \tilde\rho_{ge} \\ \tilde\rho_{ee} \\ \tilde\rho_{gg}
	\end{array}\right] \,.
\end{eqnarray}
where we have defined ${}_\rm{el}\langle \rm{in} | \rm{scatt} \rangle_{\rm{el},e\rightarrow g} = - \,{}_\rm{el}\langle \rm{in} | \rm{scatt} \rangle_{\rm{el},g\rightarrow e}^* = ie^{i\phi}$, and therefore, $\Lambda_2 = {}_{\rm{el},e\rightarrow g}\langle \rm{scatt} | \rm{scatt} \rangle_{\rm{el},g\rightarrow e} = e^{-2i\phi}$. Note that the phase $\phi$ encodes the arrival time of the center of the wave packet, that is $\phi=\omega_0 t$. 
Considering consecutive interactions with single electrons with arrival times that are equivalent modulo $2\pi/\omega_0$, we obtain oscillating dynamics, Rabi oscillations, of the quantum system's state inversion $\rho_{ee}-\rho_{gg}$.

The appearance of Rabi oscillations can be seen also analytically by an approximate continuum treatment: considering consecutive interactions with single electrons of a current $I(t)$ with fixed temporal distance between electrons $\Delta t$ and taking the limit $\Delta\rho_\rm{qs}/\Delta t\rightarrow d\rho_\rm{qs}/dt$ while keeping $\sqrt{P}/\Delta t$ finite, equation (\ref{eq:rhochange}) leads to
\begin{eqnarray}\label{eq:rhochangecont}
	  \frac{d}{d t} \left[\begin{array}{c}
		\tilde\rho_{eg} \\ \tilde\rho_{ge} \\ \tilde\rho_{ee} \\ \tilde\rho_{gg}
	\end{array}\right] &=& i\frac{|\Lambda_1|\sqrt{P} I(t)}{e}
	\left[ \begin{array}{cccc} 
			0  & 0 & -e^{-i\omega_0 t}  & e^{-i\omega_0 t}  \\
			0 & 0  &  e^{i\omega_0 t}& 	-e^{i\omega_0 t}  \\
			  - e^{i\omega_0 t} & e^{-i\omega_0 t} & 0  & 0 \\
			  e^{i\omega_0 t} & - e^{-i\omega_0 t} &  0  & 0  	
	\end{array}  \right] 
\left[\begin{array}{c}
		\tilde\rho_{eg} \\ \tilde\rho_{ge} \\ \tilde\rho_{ee} \\ \tilde\rho_{gg}
	\end{array}\right] \,.
\end{eqnarray}
where we assumed $\Lambda_1 = |\Lambda_1|e^{i\omega_0 t}$.
This is equivalent to the optical Bloch equations without damping (see Sec.\ref{sec:Bloch}). The rotating wave approximation reveals a Rabi frequency that depends on the overlap as $\Omega = \sqrt{P}|{}_\rm{el}\langle \rm{in} | \rm{scatt} \rangle_{\rm{el},e\rightarrow g}|I_{\omega_0}/e$, where $I_{\omega_0}$ is the resonant Fourier component of the electron current. 

From the vanishing diagonal blocks in the continuum limit (\ref{eq:rhochangecont}), we also conclude that the terms in the diagonal blocks in equation (\ref{eq:rhochange}) can be associated with the discreteness of the beam electrons. 

Using the parameterization of the complex parameters of the initial state
\begin{equation}
\alpha=\cos\varphi\,,\,\beta=e^{i\rho}\sin\varphi\,,
\end{equation} 
the change of inversion for the two extremal cases is 
\begin{equation}
\Delta\left(\tilde\rho_{ee}-\tilde\rho_{gg}\right) = \begin{cases}
-2P\cos\left(2\varphi\right)-2\sqrt{P}\sin\left(2\varphi\right)\sin\rho & \left|\rm{out}\right>_\rm{el}\rightarrow\left|\rm{in}\right>_\rm{el}\\
-2P\cos\left(2\varphi\right) & \left|\rm{out}\right>_\rm{el}\nrightarrow\left|\rm{in}\right>_\rm{el}
\end{cases}
\end{equation}
which shows that the inversion change due to coherence preserving scattering reaches its maximum of $2\sqrt{P}$ at $\varphi=\pi/4$, where $\tilde\rho_{ee}-\tilde\rho_{gg}=0$, while the inversion change due to decohering scattering has maximums of $2P$ when $\varphi$ reaches multiples of $\pi/2$, where $\tilde\rho_{ee}-\tilde\rho_{gg}=\pm 1$. In particular, the maximal effect of coherence-preserving scattering on the quantum system's state inversion is larger by a factor $1/\sqrt{P}$ than the maximal effect of decohering scattering, which is usually several order of magnitude (11 orders of magnitude in our first example).

\subsection{Recoil of the quantum system}

The change of the internal state of the quantum system will be accompanied by a recoil equivalent in absolute value to the momentum transfer to the electron. We can give a bound on the momentum transfer to the electron based on the numerical overlap between the electron in-state and the scattered state. For the overlap to be close to one, the transversal momentum change has to be much smaller than the transversal momentum spread of the initial wave packet. From our simulations, we find an overlap of more than $0.99$ even for a wave packet width of $\Delta r_\perp = 50\,\rm{\mu m}$ leading to a bound on the transversal momentum transfer to the quantum system $\delta p_\perp \lesssim 2 \times 10^{-30}\,\rm{kg\,m/s}$.  For trapped atoms with a trap frequency $\sim\!300\,$kHz, this is much smaller than the momentum difference between the motional ground and excited state of the potassium atoms. Explicitly, the Lamb Dicke parameter is $\delta p_\perp/\sqrt{2m\hbar\omega_\rm{trap}}\lesssim 4\times 10^{-4}$. Hence, for a setup as described in \cite{Cheuk2015}, we can conclude that the atoms will stay in the motional ground state during the interaction time.

\section{Other inelastic scattering processes}
\label{sec:electrictransitions}

We can estimate the total electron scattering probability using the total scattering cross section (comprising ionization, elastic and inelastic scattering \cite{schattschneider2012fundamentals}). For potassium atoms exposed to an 18 keV electron beam, we find $\sigma_{tot}\approx 1.5 \cdot 10^{-17} \rm{cm^2} $, extrapolated from \cite{Inokuti1971Inelastic}. The current density $j$ of a Gaussian beam at $5\text{w}$ would theoretically lead to a negligible scattering rate $\sigma_{tot}j/e$, where $e$ is the elementary charge. Therefore, all short range interactions can be neglected.
\newline
To reduce unwanted effects due to electrons scattering on the diamond structure of an NV center \cite{Tanuma2011} to a negligible level, the electron beam intensity at the position of the NV center should be reduced by a factor of $10^{-6}$ compared to its maximum. At this intensity, the number of electrons scattering within a radius of 1 nm around the NV center per period of the Rabi oscillation is less than one.
This can be achieved for a Gaussian beam at a distance of five waists or more, which is fulfilled for the parameters considered here.
%

\subsection{Electric dipole transitions}

In addition to the transition we consider, there are electric dipole transitions that can lead to an effective loss of the quantum system. Since these processes are incoherent, the probability for these transitions can be estimated using the transition probability for a single scattering event.
Nearly all electrons in the beam are further away from the quantum system than the transversal width of their wave packets. Therefore, we model the interaction of the electron with the quantum system as that of a point-like charged particle, as above. For electric dipole transitions, we start from the interaction Hamiltonian $H_\rm{int}=-\hat{d}\cdot \vec{E}$ to obtain the atomic out-state to first order;
\begin{equation}
	|\rm{out}\rangle_\rm{qs} \approx \left(\mathbb{I} + \frac{i}{\hbar}\int_{-\infty}^\infty dt \, \hat{d}(t)\cdot \vec{E}(t) \right) |e\rangle\, 
\end{equation}
in the interaction picture. For the transition probability, we find
\begin{eqnarray}
	P^d_{g\rightarrow o} &=& \frac{1}{\hbar^2}\left| \langle g| \int_{-\infty}^\infty dt \, \hat{d}(t)\cdot \vec{E}(t)|e\rangle\right|^2 = \frac{1}{\hbar^2}\left|  \int_{-\infty}^\infty dt \, e^{-i\omega t}\vec{d}_{go}\cdot \vec{E}(t) \right|^2\,
\end{eqnarray}
for a transition from the ground state $|g\rangle$ to an "other" state $|o\rangle$. The electric field of the electron moving parallel to the $z$-axis in the $y$-$z$-plane is
\begin{eqnarray}
	E_y(0,t) &=&  \frac{ e \gamma  r_\perp}{4\pi \epsilon_0 (r_\perp^2 + \gamma^2 v^2 (t-t_j)^2)^{3/2}} \\
	E_z(0,t) &=&  \frac{ e \gamma v (t-t_j) }{4\pi \epsilon_0 (r_\perp^2 + \gamma^2 v^2 (t-t_j)^2)^{3/2}}\,
\end{eqnarray}
where $\gamma$ is the Lorentz factor. Then,
\begin{equation}
    \int_{-\infty}^\infty dt \, e^{-i\omega t}\vec{d}_{go}\cdot \vec{E}(t) = \frac{e \omega}{2\pi \epsilon_0 \gamma v^2} e^{-i\omega t_j} \left( d^y_{go}  K_1\left(\frac{\omega r_\perp}{\gamma v}\right) - i  \frac{d^z_{go}}{\gamma} K_0\left(\frac{\omega r_\perp}{\gamma v}\right)  \right)   \,
\end{equation}
is the Fourier transform of the electric field due to a single point-like electron. For the transition probability, we find
\begin{eqnarray}\label{eq:probabilityelectric}
	\nonumber P^d_{g\rightarrow o} &=& \left(\frac{e\omega}{2\pi \hbar \, \epsilon_0 \gamma v^2} \right)^2 \left( \left(d^y_{go}  K_1\left(\frac{\omega r_\perp}{\gamma v}\right) \right)^2 + \left( \frac{d^z_{go}}{\gamma}  K_0\left(\frac{\omega r_\perp}{\gamma v}\right)\right)^2      \right)  \\
	&=& \left(\frac{\lambda_e}{2\pi r_\perp} \frac{2c}{v} \right)^2 \left( \left(\frac{d^y_{go}}{e a_0} \frac{\omega r_\perp}{\gamma v} K_1\left(\frac{\omega r_\perp}{\gamma v}\right) \right)^2   + \left(\frac{d^z_{go}}{\gamma e a_0} \frac{\omega r_\perp}{\gamma v} K_0\left(\frac{\omega r_\perp}{\gamma v}\right) \right)^2 \right)  \,
\end{eqnarray}
where $\lambda_e$ is the electron Compton wavelength and $a_0$ is the Bohr radius. 
For the parameters used in example 1 to generate Fig.~\ref{fig:blochevolution}a, we obtain $r_\perp/(\gamma v) \sim 10^{-12}$. Therefore, electric dipole transitions of potassium atoms are exponentially suppressed and can be neglected. The transition energy of the electric dipole transition from the NV$^-$ center $^3A_2$ ground state to the next excited state (the zero phonon line) has the transition energy $1.945\,$eV. This leads to exponential suppression and $P^{d}_{g->o}\sim 10^{-13}$ for the parameters used to generate the results presented in Fig. 4b. These parameters correspond to $\sim10^8$ electrons passing the quantum system per Rabi cycle leading to a total electric transition probability per Rabi cycle on the order of $P^{d}_{g->o}\sim 10^{-5}$, which can be neglected.

\subsection{Full QED calculation for electric dipole transitions}

The derivation of the scattering matrix elements and the scattered state in full QED works along the same lines as for 
magnetic dipole transitions above. We assume that the transition of the quantum system is due to a state change of an electron of the quantum system.
Then, the transition current can be expressed as
\begin{equation}
	\vec J^\rm{qs}_{e \rightarrow g}(x) = -\frac{i e\hbar}{2m} e^{-i\omega_j t} \left( \psi_e \nabla \psi_g^* - \psi_g^* \nabla \psi_e \right) \,.
\end{equation}
Partial integration and execution of the Fourier transforms leads to
\begin{eqnarray}
      \nonumber  \mathcal{S}(\vec p,\vec p^{\,\prime})	&=& - \frac{ e\hbar}{2m} \frac{e\mu_0 c^2}{2\sqrt{\omega_p\omega_{p'}}} \Big((\kappa_{p',p}\,p^i + \kappa_{p,p'}\,p^{\prime i})\chi_{s'}^\dagger \chi_s + i \epsilon^{imn} ( \kappa_{p',p} \,p_m - \kappa_{p,p'} \,p'_m) \chi_{s'}^\dagger \sigma_n \chi_s \Big)  \\
	\nonumber && 2\pi \delta(\omega_{p'} - \omega_p - \omega_{eg})\, \lim_{\epsilon \rightarrow 0^+}   \frac{1}{(\hbar\omega_{eg}/c)^2 - |\vec{p}^{\,\prime} - \vec{p}|^2 + i\epsilon}\left(\delta_{ij} - \frac{(p'_i - p_i)(p'_j - p_j)}{(\hbar\omega_{eg}/c)^2}\right)  \\
	&& \int d^3r'\, e^{- i (\vec{p}^{\,\prime} - \vec{p}) \vec{r}^{\,\prime}/\hbar} \left( \psi_e \nabla \psi_g^* - \psi_g^* \nabla \psi_e \right)^j(\vec{r}^{\,\prime}) \,.
\end{eqnarray}
We assume that the quantum system is strongly localized in comparison to the distance between the electron and the quantum system and we neglect recoil on the quantum system. Furthermore, we assume that the transition of the quantum systems is between different energy states of an electron of the quantum system and that the Hamiltonian defining the electronic level structure of the quantum system contains the momentum operator $-i\hbar \nabla$ only in the kinetic term $-\hbar^2\nabla^2/2m$. For example, the Hamiltonian modeling the electronic structure of NV-centers is of this form \cite{Doherty_2011}.
Then, we find
\begin{eqnarray}
	\nonumber (E_e-E_g)\int d^3r'\,   \psi_g^*\, \vec{r} \, \psi_e &=& \int d^3r'\, \psi_g^* [\hat{\vec{r}},\hat{H}] \psi_e \\
	&=& \frac{\hbar^2}{m} \int d^3r'\, \psi_g^* \nabla \psi_e 
\end{eqnarray}
such that
\begin{eqnarray}
	\int d^3r'\,  \psi_g^* \nabla \psi_e = \frac{ m \omega_{eg}}{\hbar} \int d^3r'\,  \psi_g^*\, \vec{r} \, \psi_e = \frac{ m \omega_{eg}}{\hbar e}  \vec{\mathcal{D}}_{eg} \,,
\end{eqnarray}
We employ the dipole approximation as
\begin{eqnarray}
	\int d^3r'\, e^{- i (\vec{p}^{\,\prime} - \vec{p}) \cdot \vec{r}^{\,\prime}/\hbar} \left( \psi_e \nabla \psi_g^* - \psi_g^* \nabla \psi_e \right) &\approx&  \int d^3r'\,  \left( \psi_e \nabla \psi_g^* - \psi_g^* \nabla \psi_e \right) = - \frac{ 2 m \omega_{eg} }{e\hbar } \vec{\mathcal{D}}_{eg}\,.
\end{eqnarray}
We obtain the scattering matrix element
\begin{eqnarray}
  \nonumber  \mathcal{S}(\vec p,\vec p^{\,\prime})	&=& - \frac{e\mu_0 c^2}{2\sqrt{\omega_p\omega_{p'}}} \Big((\kappa_{p',p}\,p^i + \kappa_{p,p'}\,p^{\prime i})\chi_{s'}^\dagger \chi_s + i \epsilon^{imn} ( \kappa_{p',p} \,p_m - \kappa_{p,p'} \,p'_m) \chi_{s'}^\dagger \sigma_n \chi_s \Big) 2\pi \delta(\omega_{p'} - \omega_p - \omega_{eg})\, \\
    &&  \lim_{\epsilon \rightarrow 0^+}   \frac{1}{(\hbar\omega_{eg}/c)^2 - |\vec{p}^{\,\prime} - \vec{p}|^2 + i\epsilon}   \left(-\omega_{eg}  \mathcal{D}_{eg,i}  + \frac{c^2}{\hbar^2\omega_{eg}} (\vec{p}^{\,\prime} - \vec{p})_i\, (\vec{p}^{\,\prime} - \vec{p})\cdot \vec{\mathcal{D}}_{eg}  \right) \,.
\end{eqnarray}
Since $\omega_p/\omega_{eg} = c\sqrt{|\vec{p}|^2 + m^2c^2}/\hbar\omega_{eg} \gtrsim 10^{10} \gg 1$, energy conservation implies 
\begin{equation}
	\kappa_{p',p} = \sqrt{\frac{\hbar\omega_{p'} + mc^2}{\hbar\omega_{p'} - \hbar\omega_{eg} + mc^2}} \approx 1 + \frac{\hbar\omega_{eg}}{2(\hbar\omega_{p'} + mc^2)} \approx 1\,
\end{equation}
and likewise for $\kappa_{p,p'}$, which leads to
\begin{eqnarray}\label{eq:Smatrixelementappel}
    \nonumber \mathcal{S}(\vec p,\vec p^{\,\prime})	&\approx & -\frac{e\mu_0 c^2}{2 \omega_{p'}}  \Big( \left( - \omega_{eg}(\vec{p}^{\,\prime} + \vec{p} ) +  2\omega_{p'}\, (\vec{p}^{\,\prime} - \vec{p})\right) \cdot  \vec{\mathcal{D}}_{eg} \, \chi_{s'}^\dagger \chi_s    \\
  &&  + i\omega_{eg} \epsilon^{imn} \mathcal{D}_{eg,i}  (p'_m - p_m) \chi_{s'}^\dagger \sigma_n \chi_s \Big)  \,2\pi \delta(\omega_{p'} - \omega_p - \omega_{eg})   \lim_{\epsilon \rightarrow 0^+}    \frac{1}{(\hbar\omega_{eg}/c)^2 - |\vec{p}^{\,\prime} - \vec{p}|^2 + i\epsilon} \,.
\end{eqnarray}
We used the energy conservation condition $\omega_{p'} - \omega_p - \omega_{eg}$, that $\omega_{p} \gg \omega_{eg}$ and $mc^2 \gg \hbar \omega_{eg}$, and
\begin{equation}
	|\vec{p}^{\,\prime}|^2 - |\vec{p}|^2 \approx 2\omega_{p'}\omega_{eg}\frac{\hbar^2}{c^2}\,.
\end{equation}
With the $z$-direction as the quantization direction, we obtain the un-normalized scattered state without a spin flip
\begin{eqnarray}
	\nonumber \bar{\phi}_{\rm{scatt},s}(\vec{p}^{\,\prime},s) &\approx &   -\frac{e\mu_0\hbar}{2} \,  \int \frac{d^2p_\perp}{(2\pi\hbar)^2}\, \phi_\rm{in}(p_{z,\rm{sol}}(\vec{p}_\perp,\vec{p}^{\,\prime}), \vec p_\perp) \frac{1}{a(\vec{p}_\perp,\vec{p}^{\,\prime})^2+|\vec{p}_\perp^{\,\prime}-\vec{p}_\perp|^2} \frac{ 1 }{p_{z,\rm{sol}}(\vec{p}_\perp,\vec{p}^{\,\prime})} \\
	\nonumber && \Bigg( \Big( \left( - \omega_{eg}(p_z' + p_{z,\rm{sol}}(\vec{p}_\perp,\vec{p}^{\,\prime}) ) +  2\omega_{p'}\, (p_z' - p_{z,\rm{sol}}(\vec{p}_\perp,\vec{p}^{\,\prime}))\right) \mathcal{D}_{eg}^z  \\
	&& +  \left( - \omega_{eg}(\vec{p}^{\,\prime}_\perp + \vec{p}_\perp ) +  2\omega_{p'}\, (\vec{p}^{\,\prime}_\perp - \vec{p}_\perp)\right) \cdot  \vec{\mathcal{D}}_{eg,\perp} \Big)   + (-1)^{s-1/2} i\omega_{eg}  \epsilon^{imz} \mathcal{D}_{eg,i}  ( \vec{p}^{\,\prime}_\perp - \vec{p}_\perp)_m  \Bigg) \,,
\end{eqnarray}
and the scattered state with a spin flip
\begin{eqnarray}
	\nonumber \bar{\phi}_{\rm{scatt},s}(\vec{p}^{\,\prime},-s) &\approx &   i\frac{e\mu_0\hbar\omega_{eg}}{2} \,  \int \frac{d^2p_\perp}{(2\pi\hbar)^2}\, \phi_\rm{in}(p_{z,\rm{sol}}(\vec{p}_\perp,\vec{p}^{\,\prime}), \vec p_\perp)  \frac{1}{a(\vec{p}_\perp,\vec{p}^{\,\prime})^2+|\vec{p}_\perp^{\,\prime}-\vec{p}_\perp|^2} \frac{ 1 }{p_{z,\rm{sol}}(\vec{p}_\perp,\vec{p}^{\,\prime})} \\
	\nonumber &&   \Big( \epsilon^{zin} \mathcal{D}_{eg,i}  ( p_z' - p_{z,\rm{sol}}(\vec{p}_\perp,\vec{p}^{\,\prime}) )    - \epsilon^{imn} \mathcal{D}_{eg,i}  ( \vec{p}^{\,\prime}_\perp - \vec{p}_\perp)_m \Big)\left(\delta^x_n + i(-1)^{s-1/2} \delta^y_n\right) \\
	& = &   i\frac{e\mu_0\hbar\omega_{eg}}{2} \,  \int \frac{d^2p_\perp}{(2\pi\hbar)^2}\, \phi_\rm{in}(p_{z,\rm{sol}}(\vec{p}_\perp,\vec{p}^{\,\prime}), \vec p_\perp) \frac{1}{a(\vec{p}_\perp,\vec{p}^{\,\prime})^2+|\vec{p}_\perp^{\,\prime}-\vec{p}_\perp|^2} \frac{ 1 }{p_{z,\rm{sol}}(\vec{p}_\perp,\vec{p}^{\,\prime})} \\
	\nonumber &&   \Big( -(\mathcal{D}_{eg,y} -  i(-1)^{s-1/2} \mathcal{D}_{eg,x})  ( p_z' - p_{z,\rm{sol}}(\vec{p}_\perp,\vec{p}^{\,\prime}) )   +  \mathcal{D}_{eg,z}  ( (p'_y - p_y) -  i(-1)^{s-1/2} (p'_x - p_x) ) \Big) \,.
\end{eqnarray}

\begin{figure}[hbt!]
\hspace*{-3mm}
\includegraphics[width=7.8cm,angle=0]{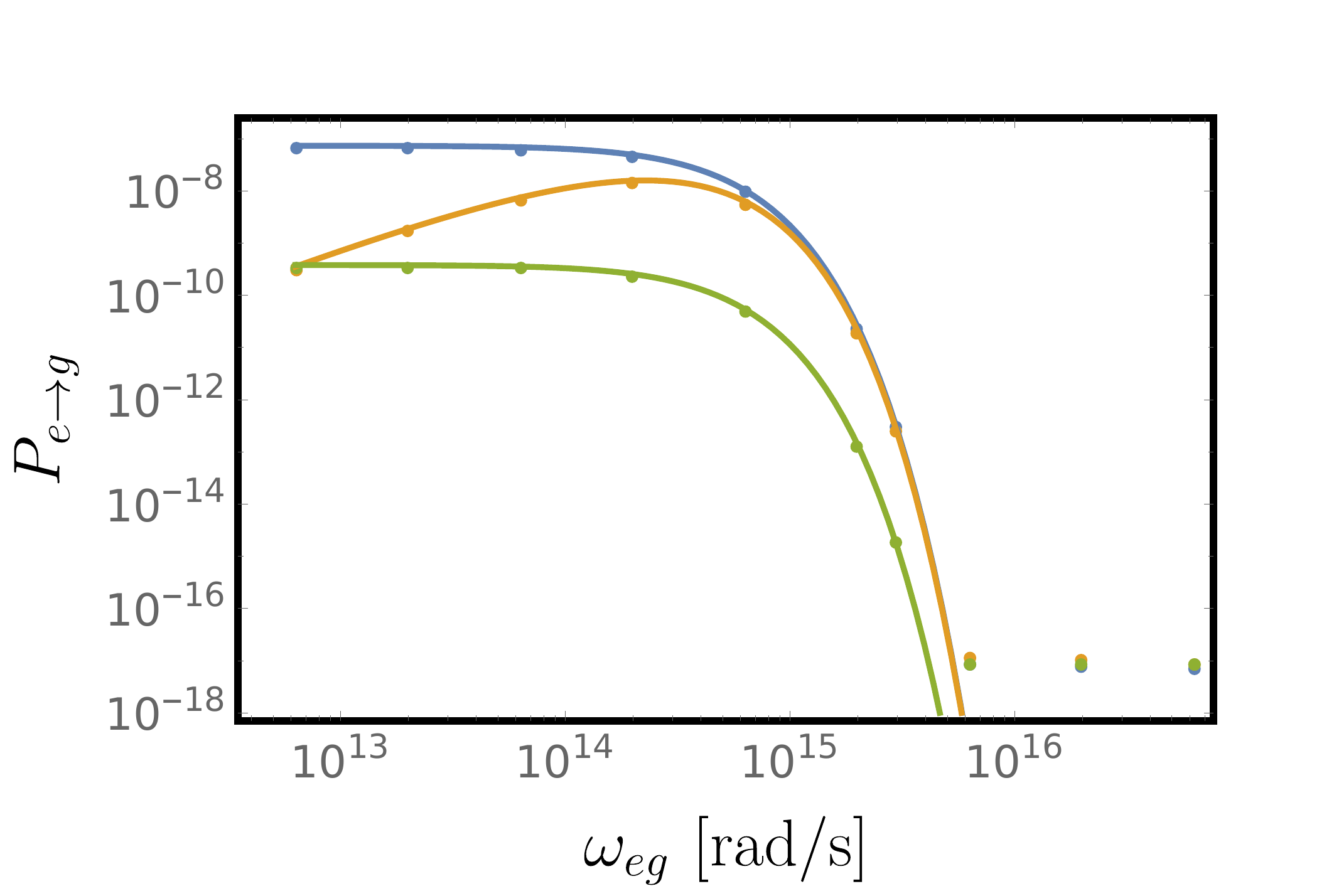}
\caption{\label{fig:probability_el} The transition probability for an electric dipole transition for an electric dipole moment oriented in the $x$-direction (green), $y$-direction (blue) and $z$-direction (orange). The dots are generated from the full QED result neglecting the spin terms (i.e. equation (\ref{eq:el_dip_spincons}) without the last term). The solid lines are generated from the semi-classical result in equation (\ref{eq:probabilityelectric}).
The plots were obtained for $\Delta r_\perp = 5\,$nm, $r_{\perp,0}=70\,$nm, $\Delta z_0 = 100\,$nm, $|\vec{\mathcal{D}}_{eg}|=2.27 ea_0$ (corresponding to the electric dipole moment of the $^3A_2$ to $^3E$ transition (zero phonon line or ZPL) of a NV$^-$ center at $1.945\,$eV \cite{Alkauskas:2014first}) and an initial kinetic energy of the electron of 2~keV. For $\vec{r}_{0,\perp}$ we considered $(0,r_{0,\perp})$, that is the quantum system and the center of the electron wave packet lie in the $y$-$z$-plane. We find that the two results agree up to the plateau beyond PHz frequencies which is due to the numerical noise floor of our algorithm.
We note the strong suppression of the scattering probability for angular frequencies above $10^{15}\,\rm{rad/s}$. }
\end{figure}

\subsection{Numerical treatment}

We again use the Gaussian envelopes given in equations (\ref{eq:instatetrans}) and (\ref{eq:instatelong}). We rewrite the above equations in terms of the dimensionless quantities $\vec{\pi}^{\,\prime}_\perp = \vec{p}^{\,\prime}_\perp/(2\Delta p_\perp)$, $\vec{\pi}_\perp = \vec{p}_\perp/(2\Delta p_\perp)$, $\pi_z' = p_z'/(2\Delta p_z)$ and $\vec{\rho}_{0,\perp} = \vec{r}_{0,\perp} 2\Delta p_\perp/\hbar = \vec{r}_{0,\perp}/\Delta r_\perp$. As noted in the numerical treatment of magnetic dipole transitions, the dispersion phase does not enter into the probabilities to first order in $|\vec p_\perp|^2/p_{z,0}^2$. Since we are only interested in the probabilities here, we only consider the dispersion free scattered wave functions
\begin{eqnarray}\label{eq:el_dip_spincons}
	 \nonumber \tilde{\phi}_{ \rm{scatt},s}(\vec{\pi}^{\,\prime},s) &=& -\mathcal{F}_\mathcal{D} \,  \int d^2\pi_\perp\,  e^{-(\pi_{z,\rm{sol}}(\vec{\pi}_\perp,\vec{\pi}^{\,\prime}) - \pi_{z,0})^2}   e^{-i \vec{\pi}_\perp \cdot \vec{\rho}_{0,\perp}} e^{-\vec{\pi}_\perp^2} 
	 \frac{1}{\bar{a}(\vec{\pi}_\perp,\vec{\pi}^{\,\prime})^2 + |\vec{\pi}_\perp^{\,\prime} - \vec{\pi}_\perp|^2}  \\
	 \nonumber && \frac{ 1 }{\pi_{z,\rm{sol}}(\vec{\pi}_\perp,\vec{\pi}^{\,\prime})} \Bigg( \Big(  \xi^{-1}\left( - \Omega_{eg}(\pi_z' + \pi_{z,\rm{sol}}(\vec{\pi}_\perp,\vec{\pi}^{\,\prime}) ) +  2\Omega_{\pi'}\, (\pi_z' - \pi_{z,\rm{sol}}(\vec{\pi}_\perp,\vec{\pi}^{\,\prime}))\right) D_{eg}^z  \\
	&& +  \left( - \Omega_{eg}(\vec{\pi}^{\,\prime}_\perp + \vec{\pi}_\perp ) +  2\Omega_{\pi'}\, (\vec{\pi}^{\,\prime}_\perp - \vec{\pi}_\perp)\right) \cdot  \vec{D}_{eg,\perp} \Big)   +  (-1)^{s-1/2} i\Omega_{eg}  \epsilon^{jmz} D_{eg,j}  ( \vec{\pi}^{\,\prime}_\perp - \vec{\pi}_\perp)_m  \Bigg) \,
\end{eqnarray}
and
\begin{eqnarray}
	 \nonumber \tilde{\phi}_{ \rm{scatt},s}(\vec{\pi}^{\,\prime},-s) &=& \mathcal{F}_\mathcal{D} \,  \int d^2\pi_\perp\,  e^{-(\pi_{z,\rm{sol}}(\vec{\pi}_\perp,\vec{\pi}^{\,\prime}) - \pi_{z,0})^2}   e^{-i \vec{\pi}_\perp \cdot \vec{\rho}_{0,\perp}} e^{-\vec{\pi}_\perp^2} 
	 \frac{1}{\bar{a}(\vec{\pi}_\perp,\vec{\pi}^{\,\prime})^2 + |\vec{\pi}_\perp^{\,\prime} - \vec{\pi}_\perp|^2}  \\
	 \nonumber  && i \frac{\Omega_{eg}}{\pi_{z,\rm{sol}}(\vec{\pi}_\perp,\vec{\pi}^{\,\prime})} \Big( -\xi^{-1}(\mathcal{D}_{eg,y} -  i(-1)^{s-1/2} \mathcal{D}_{eg,x})  ( \pi_z' - \pi_{z,\rm{sol}}(\vec{\pi}_\perp,\vec{\pi}^{\,\prime}) )   \\
	 && +  \mathcal{D}_{eg,z}  ( (\pi'_y - \pi_y) -  i(-1)^{s-1/2} (\pi'_x - \pi_x) ) \Big) \,,    
\end{eqnarray}
where 
\begin{eqnarray}
	\mathcal{F}_\mathcal{D} &=& \frac{e\mu_0 c  |\vec{\mathcal{D}}|}{(2\pi\hbar)^2}\left(\frac{(2\pi)^{1/2}\hbar}{\Delta p_z}\right)^{1/2}\sqrt{2\pi}\hbar
\end{eqnarray}
and $\Omega_{eg} = \frac{\hbar \omega_{eg}}{2 c\Delta p_z }$ and $\vec{D}_{eg}= \vec{\mathcal{D}}_{eg}/|\vec{\mathcal{D}}|$.
The probability is obtained as
\begin{eqnarray}
	P_{e\rightarrow g}(s) &=& \bar{\mathcal{F}}_\mathcal{D}^2  \sum_{s'} \int d^3\pi' | \tilde{\phi}_{ \rm{scatt},s}(\vec{\pi}^{\,\prime},s')/\mathcal{F}_\mathcal{D}|^2 \,,
\end{eqnarray}
where
\begin{eqnarray}
	\bar{\mathcal{F}}_\mathcal{D} = \frac{2\Delta p_\perp (2\Delta p_z)^{1/2} }{(2\pi\hbar)^{3/2}} \frac{ e\mu_0 c  |\vec{\mathcal{D}}| }{(2\pi)^{5/4} (\hbar \Delta p_z)^{1/2}} = \frac{e\mu_0 c  |\vec{\mathcal{D}}|}{ \hbar \Delta r_\perp  \pi^{1/2} (2\pi)^{9/4}}\,.
\end{eqnarray}
The numerical results can be found in Fig.~\ref{fig:probability_el}.

\section{Electron beam of a klystron}
\label{sec:elbeamklystron}

In the following, we will give some details on the current-modulated electron beam in a klystron based on the article \cite{Webster1939Cathode-ray} by Webster.
The electron beam modulation of a klystron is achieved by modulating the kinetic energy, or, effectively, the velocity of electrons in the buncher.
This can be done, for example, with a microwave field in a resonator (the buncher cavity) that leads to an electric field in the beam propagation direction \cite{farago1974the}.
Under the assumption of monochromatic oscillations of the kinetic energy modulation in the buncher and approximating the buncher as infinitesimally short, for the velocity of an electron passing the buncher at time $t_1$, we can approximate
\begin{equation}
	v = v_0 + v_1 \sin(\omega_0 t_1).
\end{equation}
When the amplitude of the kinetic energy modulation $\delta E_\rm{kin}$ is small in comparison to the average kinetic energy, we have $v_1 \approx \delta E_\rm{kin}/(\gamma^3 m_e v_0)$. The arrival time of the electron at the target (e.g. an atom) is
\begin{equation}
	t_2 = t_1 + \frac{l}{v_0 + v_1 \sin(\omega_0 t_1)} \approx  t_1 + \frac{l}{v_0} - \frac{lv_1}{v_0^2} \sin(\omega_0 t_1)\,.
\end{equation}
where $l$ is the distance to the buncher. Charge conservation can be written as $I(t_1,z_1) dt_1 = I(t_2,z_2) dt_2$ and leads to 
\begin{equation}
	I_2 = \frac{I_1}{1-r_b\cos(\omega_0 t_1)} = \frac{I_0}{1 - r_b\cos(\omega_0 t_1)}
\end{equation}
where $I_0$ is the un-modulated stationary current and 
\begin{equation}
	r_b = l\omega_0 v_1/v_0^2
\end{equation}
is the bunching parameter. 
\begin{figure}[h]
\includegraphics[width=7cm,angle=0]{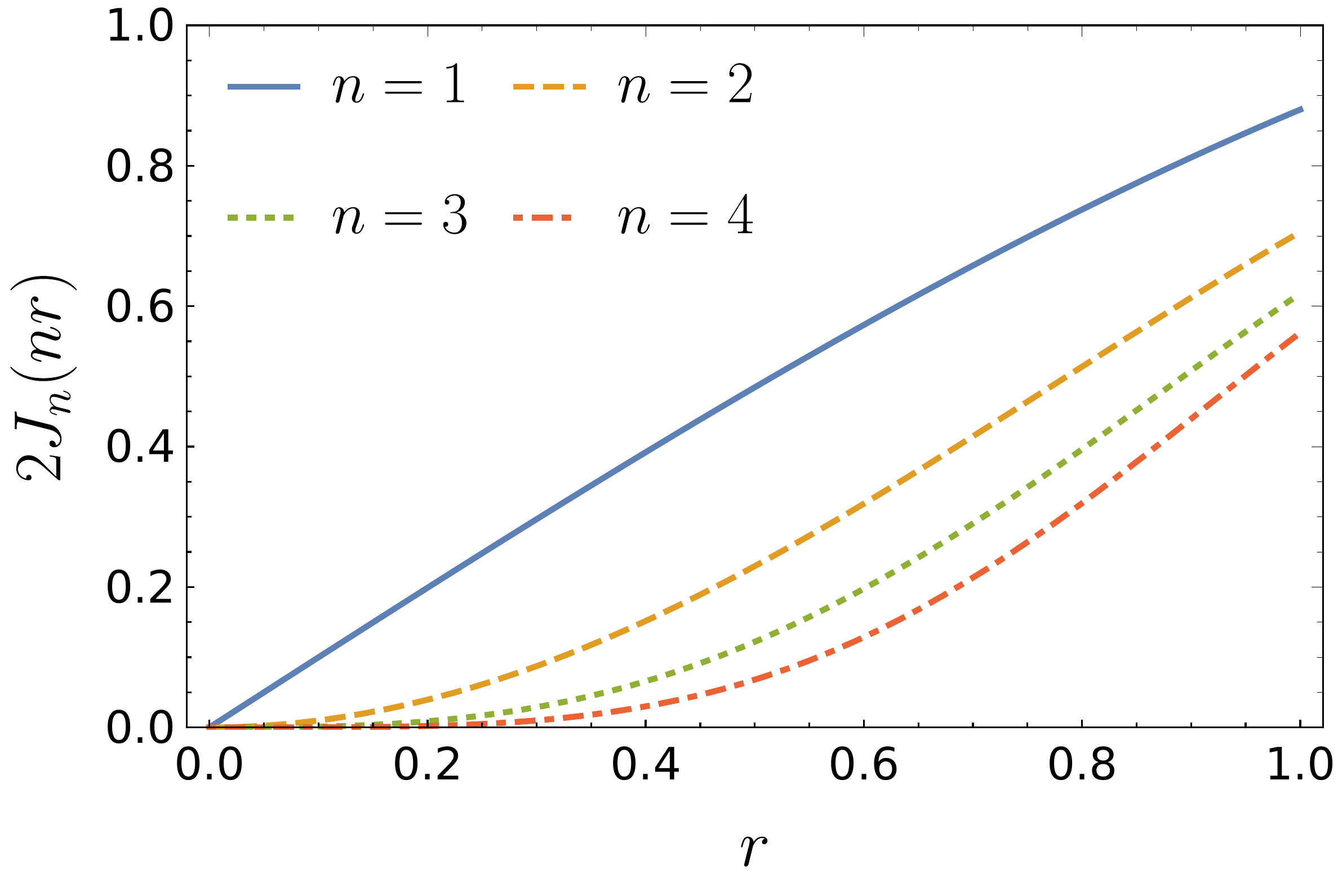}
\includegraphics[width=7cm,angle=0]{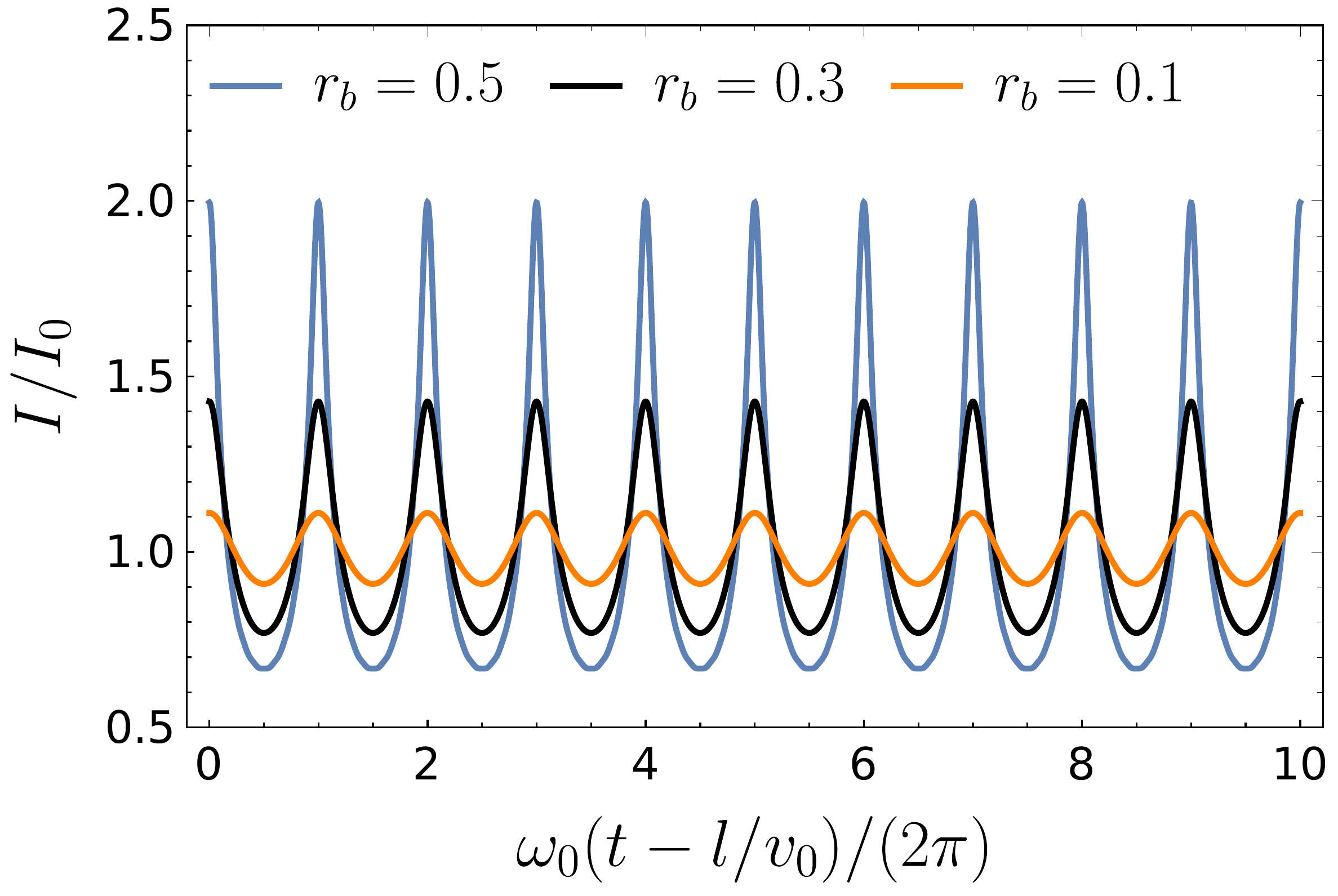}
\caption{\label{fig:fourierbunching} Left plot: Fourier coefficients of the modulated current for frequencies $n\omega_0$ as a function of the bunching parameter $r_b$.
Right plot: Time dependent current for different bunching parameters.
}
\end{figure}
Writing the distance 
to the buncher as a coordinate $l = z - z_0$, where $z_0$ is the position of the buncher, we obtain
\begin{equation}\label{eq:klycurr}
	I(z,t) = \frac{I_0}{1-r_b(z)\cos\theta(z,t)}\,
\end{equation}
where $\theta(t)$ is the solution of the equation
\begin{equation}
	\theta(z,t) - r_b(z) \sin\theta(z,t) = \omega_0\left( t - \frac{(z-z_0)}{v_0}\right) \,.
\end{equation}
It has been shown that the current in equation (\ref{eq:klycurr}) can be expressed as a Fourier series with the coefficients \cite{Webster1939Cathode-ray}
\begin{equation}\label{eq:klycurrfour}
	I(z,t) = \left[I_0 +  \sum_{n = 1}^{\infty} I_{n\omega_0}(z) \cos\left(n\omega_0\left( t - \frac{(z-z_0)}{v_0}\right)\right) \right]\,,
\end{equation}
where $I_{n\omega_0}(z)=2I_0 J_n(n\,r_b(z))$ and $J_n$ are the Bessel functions of the first kind. The Fourier coefficients as a function of the bunching
parameter are plotted in Fig.~\ref{fig:fourierbunching}. The beam current as a function of the distance from the buncher cavity is plotted in Fig. \ref{fig:bunchingspace}.
\begin{figure}[h]
\includegraphics[width=7cm,angle=0]{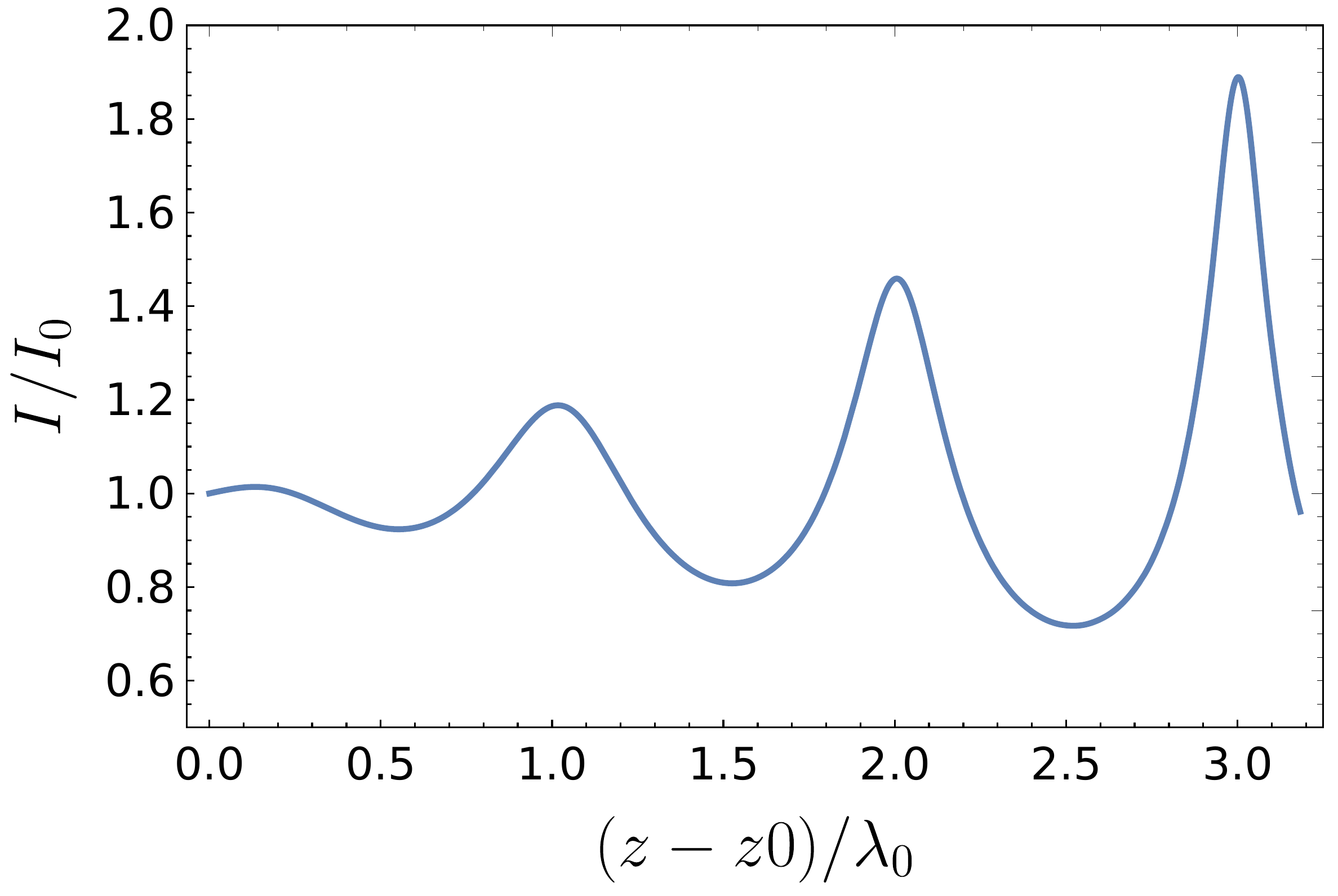}
\caption{\label{fig:bunchingspace}Spatial change of the electron beam modulation for $v_1/v_0 = 1/40$ at $t=0$ plotted as a function of the distance from the buncher cavity in units of the modulation wavelength $\lambda_0 = 2\pi v_0/\omega_0$. The spatial range plotted corresponds to the interval $r_b=0$ to $r_b=0.5$.
}
\end{figure}

\subsection{The effect of the electron velocity distribution}

The finite width of the electron velocity distribution affects the beam modulation because the bunching of the beam depends on the initial velocity. We consider a kinetic energy distribution with a width $\Delta E_\rm{kin}$ of about $1$ eV. 
Approximately, we have $\Delta E_\rm{kin} = \gamma^3 m v_0 \Delta v_0$. This leads to a distribution in the bunching parameter $\Delta r_b = 2 l\omega_0 v_1 \Delta E_\rm{kin}/(\gamma^3 m v_0^4)$ and $\Delta r_b/r_b = 2 \Delta E_\rm{kin}/(\gamma^3 m v_0^2)$. This ratio decreases for increasing kinetic energies. For $E_\rm{kin} = 18\, \rm{keV}$, we have $v_0/c \sim 1/4$ and we can find $\Delta r_b/r_b = \Delta E_\rm{kin}/E_\rm{kin} \sim 10^{-4} $. We find a minor correction to the bunching parameters which implies a minor correction to the amplitudes of the Fourier components corresponding to the distinct lines in the modulation spectrum. It is also important to note that this effect will not broaden the spectral linewidth of the modulated near field affecting the quantum system.

\section{Single particle beam simulation}\label{sec:singpartsim}

To analyze systematic effects due to shot noise, we model the electron beam as a collection of single electrons generated in a homogeneous Poisson process. Electrons are generated after waiting times 
that are exponentially distributed \cite{tijms2003first} with the mean given by the inverse of the rate $\sigma$ of electron creation at the cathode. We simulate the beam of a Klystron by modulating the kinetic energy of the particles and propagating them over a drift distance $l$ to obtain the current modulation. The modulation of the kinetic energy is sinusoidal: $E_\rm{kin}(t) = E_{\rm{kin},0} + \delta E_\rm{kin} \sin(\omega_0 t)$. 

We wrote a numerical algorithm
using Python that generates a set of electron positions representing the beam. For an electron moving at $\vec r_\perp=(x,y)$ parallel to the $z$-axis and arriving at $z=0$ at time $t_j$, the only non-vanishing component of the magnetic field strength at the origin becomes (take equation 11.152 from \cite{jackson} and shift and rotate)
\begin{eqnarray}\label{eq:magsingrelsup}
	\vec B_j(0,t) &=& \frac{\mu_0 e \gamma v}{4\pi} \left[\begin{array}{c} y \\  -x \\ 0 \end{array}\right] \frac{1}{(r_\perp^2 + \gamma^2 v^2 (t-t_j)^2)^{3/2}}\,.
\end{eqnarray}
where $\gamma =(1-v^2/c^2)^{-1/2}$ is the Lorentz factor and $r_\perp = |\vec{r}_\perp|$. These contributions to the total magnetic field strength are summed for all electron positions.  We simulated the beam over a length of $L\sim 100 d$ centered at $z=0$ to limit the numerical effort. This approximation is well justified as electrons in the beam affect the quantum system significantly only in an interaction region on the length scale of several $d/\gamma$ centered at $z=0$. For larger $z$, the effect on the quantum system decays as $|z|^{-3}$. The electrons have an initial kinetic energy of 18 keV.
\begin{figure}[h]
\includegraphics[width=5.2cm,angle=0]{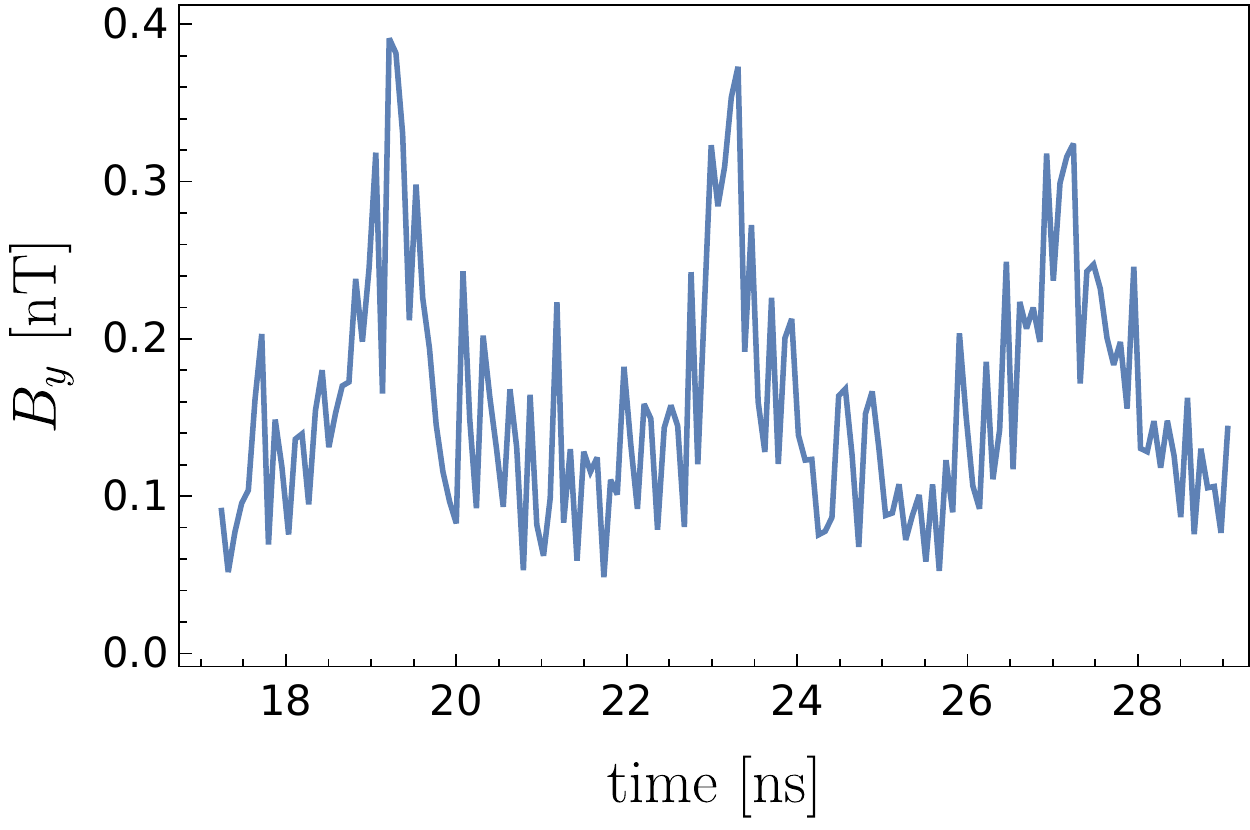}
\includegraphics[width=5.4cm,angle=0]{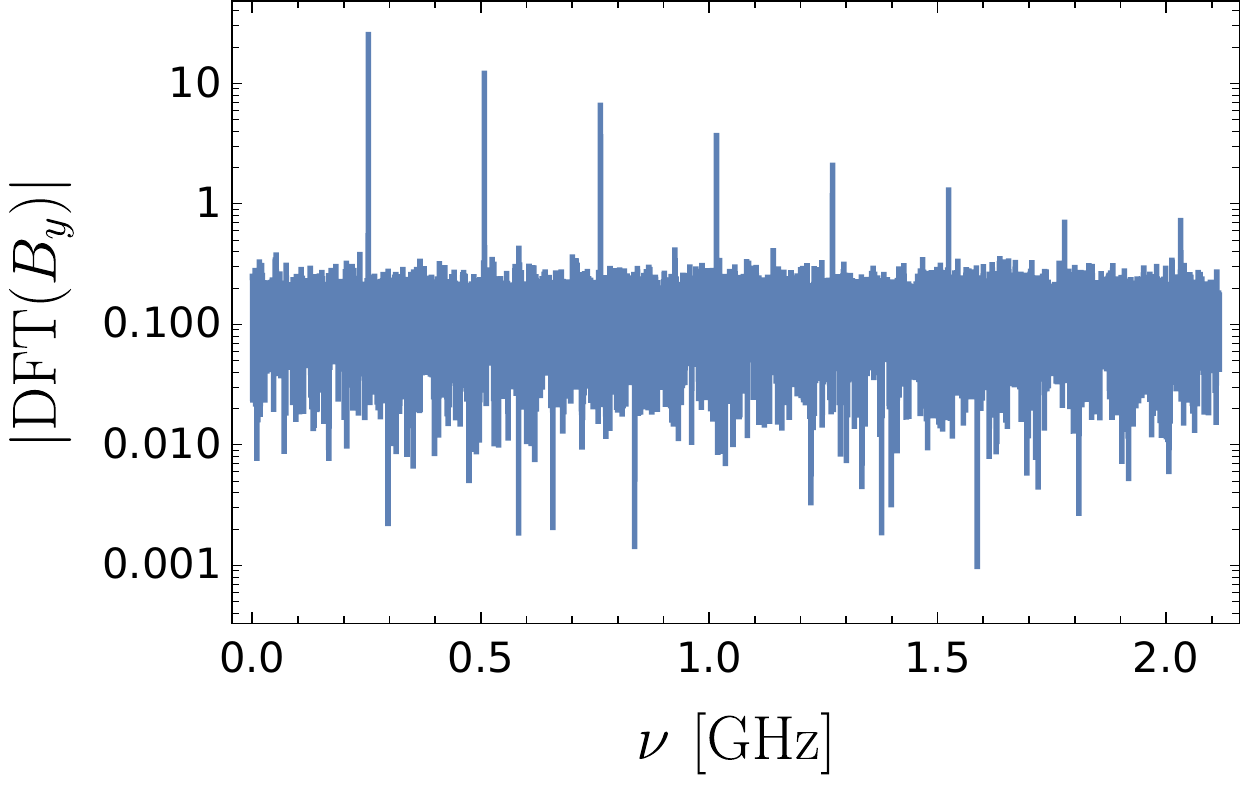}
\includegraphics[width=5.4cm,angle=0]{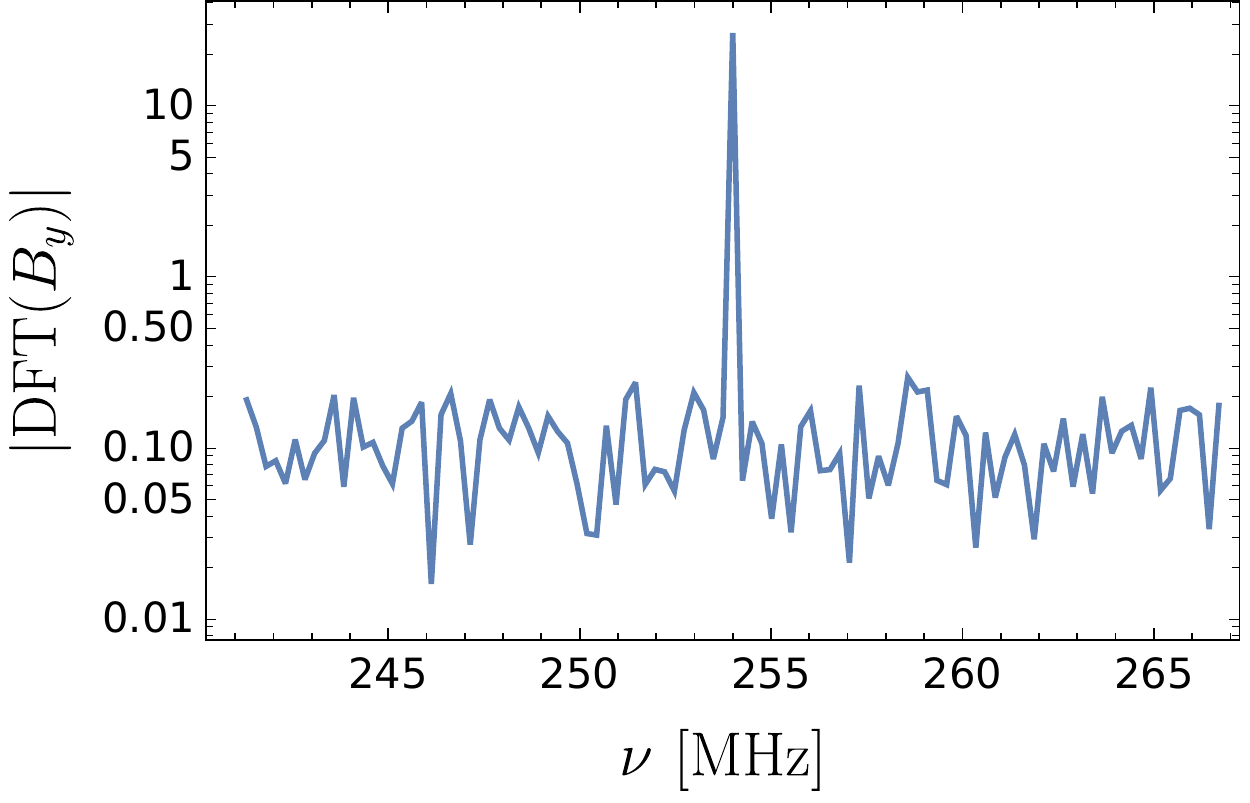}
\includegraphics[width=5.2cm,angle=0]{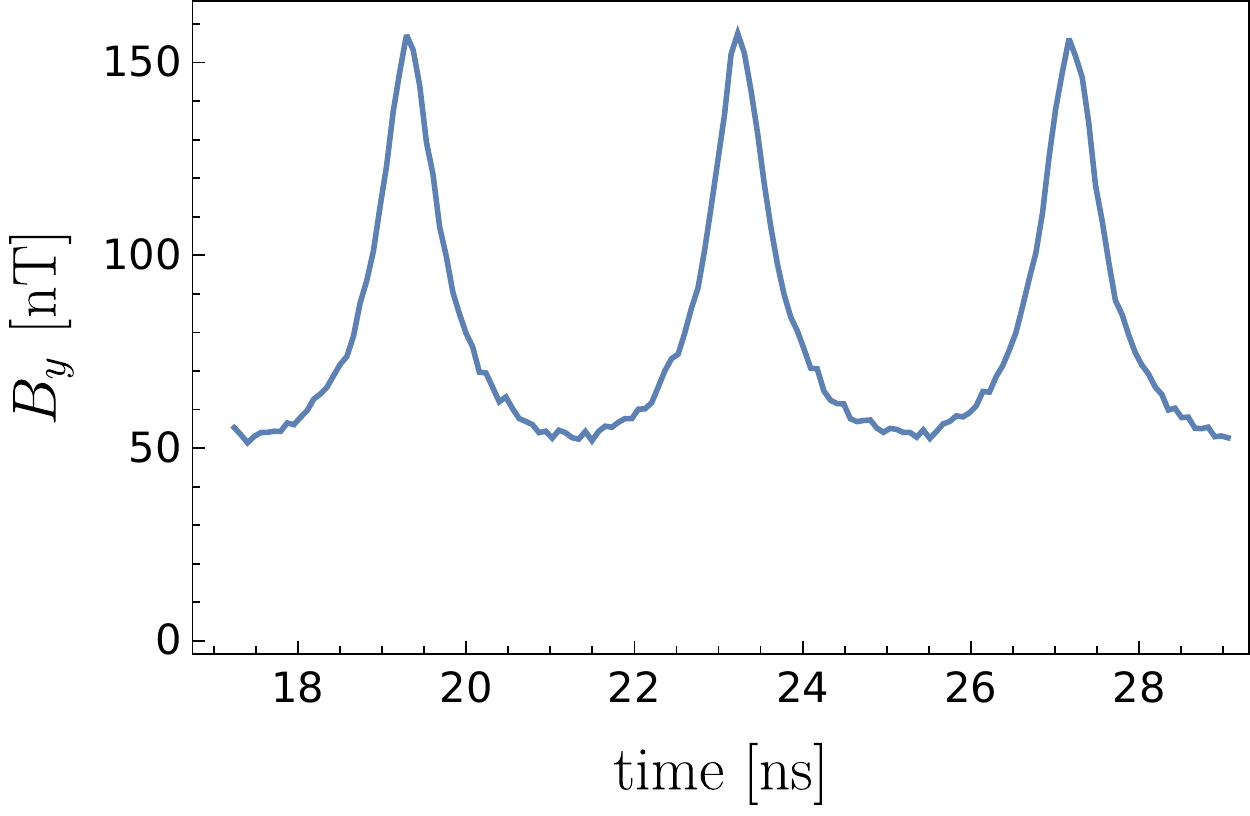}
\includegraphics[width=5.4cm,angle=0]{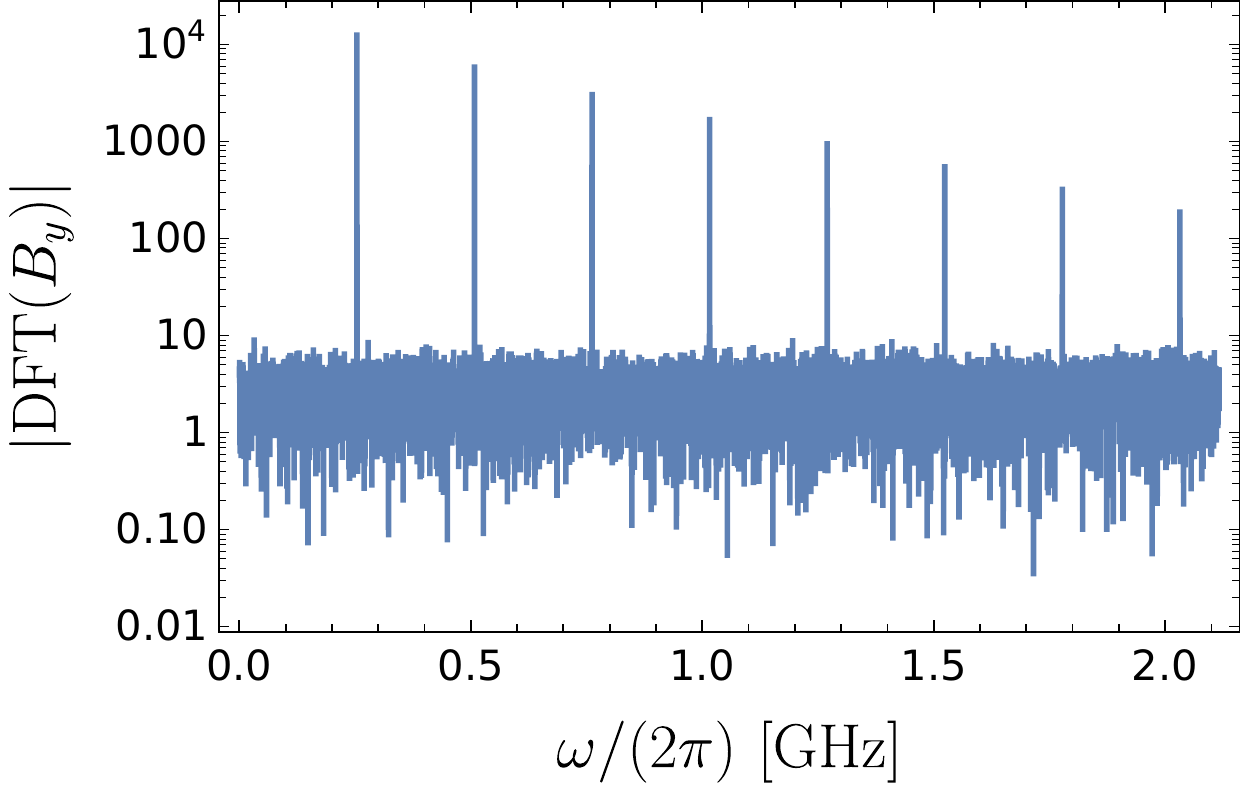}
\includegraphics[width=5.4cm,angle=0]{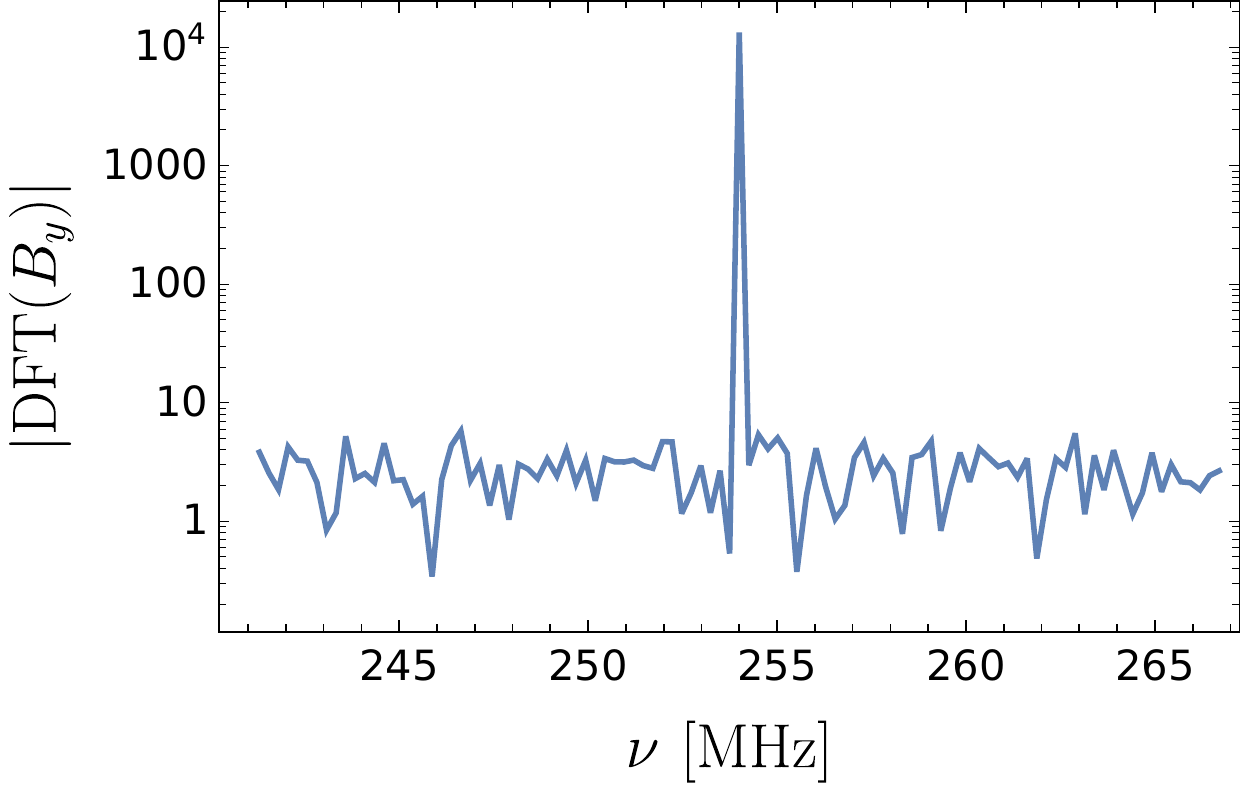}
\caption{\label{fig:discrete3dmovbeamsupklystron}  Numerical simulation of the magnetic field strength $B^y$ due to an electron beam at a distance $d = 5\text{w} = 250\,\rm{\mu m}$ to its center, where $\text{w}=50\,\rm{\mu m}$ is the beam waist radius (left plots), and the corresponding DFT (middle and right plots) for a beam with $200\,\rm{n A}$ (upper plots) and $100\,\rm{\mu A}$ (lower plots). The beam current is modulated with a base frequency of $\nu=\omega/2\pi = 254\,\rm{MHz}$ by varying the electron velocity (as e.g. in a Klystron). The higher harmonics can be seen in the middle plot. The bunching parameter $r_b$ is approximately $0.5$ ($l =1$m and $\delta E_\rm{kin}/E_\rm{kin}=1/20$), which implies that the Fourier component corresponding to the modulation at base frequency has an amplitude of $50\%$ of the average current.
}
\end{figure}
The initial transversal position of electrons is modeled as a normal distribution, where the variance $\sigma = \text{w}/2$ is given by the waist radius $\text{w}$ of the beam.
To restrict the beam to a finite radius $\Delta r_{\perp,0}$ and to avoid numerical singularities, the transversal distribution is generated from a uniform distribution by mapping values of $y$ between $0$ and $1$ with the function $r_\perp(y) =  2^{-1/2} \text{w} \sqrt{\ln((1-y(1-\exp(-2 \Delta r_{\perp,0}^2/\text{w}^2))^{-1})}$ to a radius and generating the azimuthal distribution with another uniform distribution. 

Furthermore, we note that the magnetic field of each electron effectively acts on a length scale $|v(t-t_j)| \sim r_\perp/\gamma$ as can be seen from equation (\ref{eq:magsingrelsup}). Therefore, the beam divergence, specified by the divergence angle $\theta$, can be neglected if the change of beam width is on the length scale of the interaction region $\theta r_\perp/\gamma \ll \text{w}$. Later, we will consider $d = 5\text{w}$, and the above condition is fulfilled for almost all electrons in the beam if $\theta \ll \gamma / 5$. For a strongly focused Gaussian electron beam, $\theta$ is given by the wave properties of electrons as $\theta = \lambda_\rm{dB}/(\pi \text{w})$, where the de Broglie wavelength is $\lambda_\rm{dB}= 2\pi \hbar /(\gamma m_e v)$. We obtain a condition for the waist $\text{w} \gg 10\hbar/(m_e\gamma^2 v)$. The right hand side decreases monotonously with increasing $v$, and, therefore, with increasing kinetic energy. For a velocity $c/4$ (corresponding to $18\,\rm{keV}$), the right hand side becomes approximately $40\hbar/(m_e c) \sim  10^{-11}\,\rm{m}$, and the condition on $\text{w}$ will be always fulfilled in the context of our proposal. For a wide beam with a given transversal kinetic energy spread of about $\Delta E_\rm{kin}$ or less, we obtain the corresponding transversal velocity spread of $\Delta v \approx \sqrt{2 \Delta E_\rm{kin}/m_e}$ and the divergence angle $\theta =  \Delta v/v$. For $v = c/4$ and $\Delta E_\rm{kin}\sim 1\,\rm{eV}$, we find $\theta \sim 10^{-2} \ll 1/5$ which falls within standard parameters for electron microscopes \cite{Reimer2008Transmission,egerton2011electron}, and the above condition is fulfilled.  
The results are shown in Fig.~\ref{fig:discrete3dmovbeamsupklystron}. To generate the data for the plots, the simulation was run for $10^3$ periods of the modulation.

\section{The single electron Fourier transform and the magnetic field spectrum}\label{sec:singpartfour}

The Fourier transform of the magnetic field due to the electron beam can also be calculated directly from the Fourier transform of a single electron's magnetic field. The minimal distance (impact parameter) between the electrons and the origin is $r_\perp := (x^2 + y^2)^{1/2}$.
At time $t$ and position $\vec{r}=0$, the magnetic field caused by an electron moving with velocity $v_j$ and arriving at $z=0$ at time $t_j$ is
\begin{eqnarray}\label{eq:magsingrot}
	B^y_j(\vec{r} = 0, t) & = & \frac{e\mu_0 \gamma_j v_j}{4\pi}    \frac{r_{\perp} }{(r_{\perp}^2 +  \gamma_j^2 v_j^2(t_j - t)^2)^{3/2}}\,.
\end{eqnarray}
The temporal Fourier transform of the magnetic field of a single electron is
\begin{eqnarray}\label{eq:magsingrotft}
	\nonumber \mathcal{F}_t\left[ B^y_j(\vec{r} = 0, t)\right]& = & \frac{e\mu_0  \gamma_j  v_j}{4\pi} \frac{1}{\sqrt{2\pi}} \int_{-\infty}^\infty dt\, e^{i\omega t} \frac{r_{\perp} }{(r_{\perp}^2 +  \gamma_j^2 v_j^2(t_j - t)^2)^{3/2}} \\
	&=& \frac{ e\mu_0 \omega }{4\pi \gamma_j v_j } \sqrt{\frac{2}{\pi}} K_1\left(r_{\perp} \frac{\omega}{\gamma_j v_j}\right)  e^{i\omega t_j } \,.
\end{eqnarray}
We note that the Fourier transform of the single electron decays exponentially for $r_{\perp}\omega/(\gamma_j v_j)\gg 1$ due to the properties of the Bessel function. For angular frequencies $\omega$ such that $r_\perp \omega/\gamma_j v_j \ll 1$, we can approximate $K_1(x)$ as $1/x$ and find
\begin{eqnarray}\label{eq:appspecsingsupp}
	\mathcal{F}_t\left[ B^y_j(\vec{r} = 0, t)\right] & \approx & \frac{ e\mu_0 }{2\pi \,r_{\perp}} \frac{1}{\sqrt{2\pi}}  e^{i\omega t_j} \,.
\end{eqnarray}
This means that the magnetic field of a single electron appears like a delta-peak when seen on time scales much larger than $r_\perp/(\gamma_j v_j)$. In particular, we can conclude that the magnetic field is directly proportional to the current for these time scales.

\begin{figure}[h]
\includegraphics[width=8cm,angle=0]{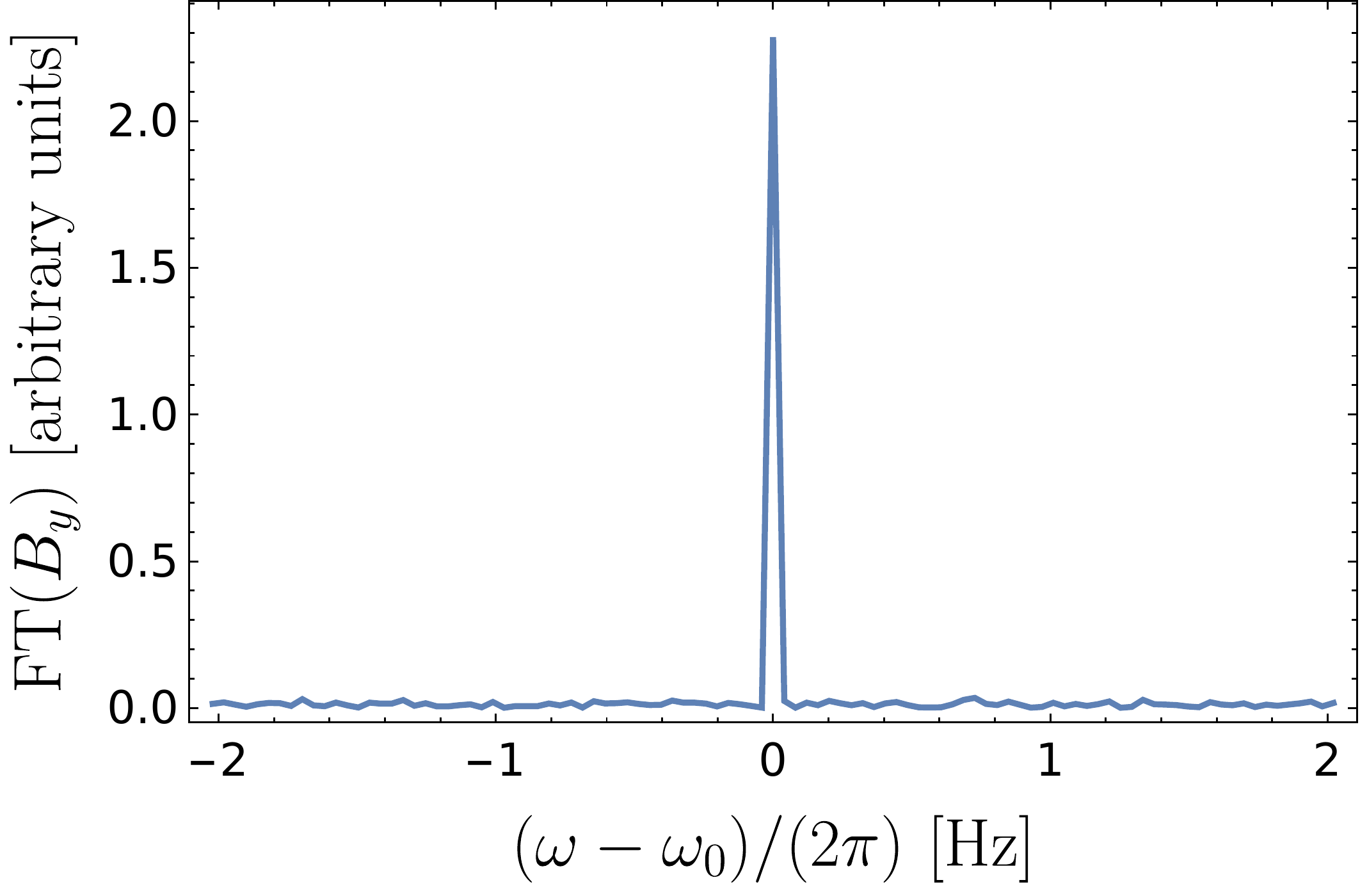}
\caption{\label{fig:discrete1d3dbeamdirectfourier} Spectrum (temporal Fourier transform) of the magnetic field strength $B^y$ due to a current modulated electron beam with $20\,\rm{fA}$ current and uniformly distributed electron positions in the longitudinal direction ($z$-direction) for $10^9$ periods of the base frequency $254\,\rm{MHz}$ (corresponding to a total number of electrons $\sim 500\,000$). The average distance from the beamline is $5\text{w}$, where $\text{w}=1\,\rm{nm}$ is the beam waist. The electrons have a kinetic energy of 18 keV.}
\end{figure}

The single electron Fourier transform can be used to obtain the discrete Fourier transform of the total magnetic field directly from our numerical model above. A plot is given in Fig.~\ref{fig:discrete1d3dbeamdirectfourier}.

To derive the expectation value of the Fourier transform of the magnetic field due to the total beam, we use the probability for an electron to pass the $z=0$ plane at $t=t_j$, given as $\mathcal{P}(t)$. We assume that the beam has a duration $T$ and that $\mathcal{P}(t)$ is normalized on the interval $[-T/2,T/2]$ and vanishes outside of it. Furthermore, we assume that the probability has a Fourier spectrum with distinct lines evenly spaced by $\omega_0$. We restrict our considerations to a one-dimensional model for the beam and set $r_\perp = d$. We write the Fourier decomposition as
\begin{equation}
	\mathcal{P}(t) = \frac{1}{T} \left( a_0 + \sum_{n = -\infty}^{\infty} a_n e^{i n\omega_0 t}  \right)\Theta_{[-\frac{1}{2},\frac{1}{2}]}(t)\,,
\end{equation}
where $\Theta_{[-\frac{1}{2},\frac{1}{2}]}$ is the characteristic function for the interval $[-\frac{1}{2},\frac{1}{2}]$ which takes the value one in the interval and vanishes outside of it.
In particular, $\mathcal{P}(t)$ is periodic with a base frequency of $\omega_0/2\pi$. Using the Poissonian distribution of the electron number in the interval $p_N = \bar{N}^N e^{-\bar{N}}/N!$ where $\bar{N} = E[N]$ is the expected value \cite{vanKampen1992stochastic}, we find
\begin{eqnarray}\label{eq:expfour}
	E\left[\mathcal{F}_t\left[B^y(\vec{r} = 0, t)\right]\right] &=& \frac{ e\mu_0}{2\pi } \frac{1}{\sqrt{2\pi}} e^{-\bar{N}} \sum_{N=1}^{\infty}\frac{\bar{N}^N }{N!}   \int \prod_{j=1}^{N} \left( dt_j \, \mathcal{P}(t_j) \right) \, \sum_k   \frac{\omega}{\gamma_k v_k} K_1\left(d \frac{\omega}{\gamma_k v_k}\right)  e^{i\omega t_k}  \,,
\end{eqnarray}
where the integrals are taken over the whole sum at the end of the equation. For small bunching parameters $r_b < 1$, electrons do not overtake each other and there is a one to one correspondence of the arrival time of an electron and its velocity, which implies $v_k = v(t_k)$ and $\gamma_k = \gamma(t_k)$. The temporally periodic modulation of the electron velocity means that $v(t_k)$ and $\gamma(t_k)$ must be periodic as well. The period is given by the base frequency $\omega_0/2\pi$. We find
\begin{eqnarray}\label{eq:expfour2}
	\nonumber E\left[\mathcal{F}_t\left[B^y(\vec{r} = 0, t)\right]\right]&=& \frac{ e\mu_0}{2\pi } \frac{1}{\sqrt{2\pi}} e^{-\bar{N}} \sum_{N=1}^{\infty}\frac{\bar{N}^N }{(N-1)!} \int_{-T/2}^{T/2} dt \, \mathcal{P}(t)    \frac{\omega}{\gamma(t) v(t)} K_1\left(d\frac{\omega}{\gamma(t) v(t)}\right)  e^{i\omega t} \\
	&=& \frac{ e\mu_0}{2\pi } \frac{\bar{N}}{\sqrt{2\pi}} \int_{-T/2}^{T/2} dt \, \mathcal{P}(t)    \frac{\omega}{\gamma(t) v(t)} K_1\left(d \frac{\omega}{\gamma(t) v(t)}\right)  e^{i\omega t} \,.
\end{eqnarray}
Due to the Bessel function $K_1$ decaying exponentially for arguments larger than one and $\gamma(t)$ and $v(t)$ only varying slightly in time, the spectrum of the beam will not contain frequencies much larger than $\gamma_0 v_0/d$, where $v_0$ and $\gamma_0$ are the average quantities. The integral in equation (\ref{eq:expfour2}) is the Fourier transform of a $2\pi/\omega_0$-periodic function which implies that $E\left[\mathcal{F}_t\left[B^y(\vec{r} = 0, t)\right]\right]$ consists of distinct Fourier-limited spikes at multiples of $\omega_0$. More precisely, there exist coefficients $b_n$ such that
\begin{eqnarray}
	E\left[\mathcal{F}_t\left[B^y(\vec{r} = 0, t)\right]\right] &=& \frac{  I_0 \mu_0  }{(2 \pi)^2 } \left( b_0\, \frac{\sin(\omega T/2)}{\omega} + \sum_{n = -\infty}^{\infty} b_n\, \frac{\sin\left((\omega + n \omega_0)T/2\right)}{\omega + n\omega_0} \right)\,,
\end{eqnarray}
where $I_0 = e \bar{N}/T$ is the average current. For $T\rightarrow \infty$, we have
\begin{eqnarray}
	E\left[\mathcal{F}_t\left[B^y(\vec{r} = 0, t)\right]\right] & \xrightarrow{T\rightarrow \infty} & \frac{  I_0 \mu_0 }{2 \pi } \left( b_0\,\delta(\omega) + \sum_{n = -\infty}^{\infty} b_n\, \delta(\omega + n \omega_0) \right)\,.
\end{eqnarray}
The coefficients can in principle be directly calculated from equation (\ref{eq:expfour}). If $d\,\omega/(\gamma_0 v_0) \ll 1$, we can approximate the Bessel function and find
\begin{eqnarray}\label{eq:expfourapp}
    E\left[\mathcal{F}_t\left[B^y(\vec{r} = 0, t)\right]\right] = \frac{ \mu_0  I_0 }{2 \pi d} \frac{1}{\sqrt{2\pi}}\left( a_0\, \frac{\sin(\omega T/2)}{\omega} + \sum_{n = -\infty}^{\infty} a_n\, \frac{\sin\left((\omega + n \omega_0 )T/2\right)}{\omega+ n\omega_0} \right)\,,
\end{eqnarray}
which is proportional to the spectrum of the probability function $\mathcal{P}(t)$ that is given by the beam bunching. The average magnetic field becomes
\begin{eqnarray}\label{eq:avmagfield}
    E\left[B^y(\vec{r} = 0, t)\right] &=& \mathcal{F}_t^{-1}\left[E\left[\mathcal{F}_t\left[B^y(\vec{r} = 0, t)\right]\right]\right] = \frac{  \mu_0  I(t) }{2 \pi d}\,,
\end{eqnarray}
where $I(t) = I_0 T \mathcal{P}(t)$, which is the magnetic field of a slowly modulated one-dimensional current.

\section{Auto-covariance and noise}\label{sec:singpartauto}

In this section, we derive the covariance matrix of the magnetic field of a one-dimensional modulated beam
\begin{eqnarray}
	\nonumber \rm{Cov}(B_y(t),B_y(t')) &=& E[B_y(t) B^*_y(t')] - E[B_y(t)]E[B^*_y(t')] \\
	\nonumber &=& \frac{1}{2\pi} \int d\,\omega \int d\,\omega' e^{-i\omega t} e^{i\omega' t'} \left(E[\mathcal{F}_t[B_y](\omega) \mathcal{F}^*[B_y](\omega')] - E[\mathcal{F}_t[B_y](\omega)] E[\mathcal{F}^*[B_y](\omega')]\right) \\
	&=:& \frac{1}{2\pi} \int d\,\omega \int d\,\omega' e^{-i\omega t} e^{i\omega' t'} \rm{Cov}(\mathcal{F}_t[B_y](\omega),\mathcal{F}_t[B_y](\omega'))\,,
\end{eqnarray}
which contains all of the information about the noise spectrum of the Gaussian random process $B_y$. 
Again, we consider a beam of $N$ electrons with arrival times at the $z=0$ plane that are randomly distributed over an interval $T$ with the probability distribution $\mathcal{P}(t)$. We find
\begin{eqnarray}
	\nonumber && E[\mathcal{F}_t[B_y](\omega) \mathcal{F}^*[B_y](\omega')] = \left(\frac{ e\mu_0}{2\pi }\right)^2  e^{-\bar{N}} \Bigg[\sum_{N=2}^{\infty}\frac{\bar{N}^N }{N!}\int \prod_{j=1}^{N} \left( dt_j \, \mathcal{P}(t_j) \right) \times \\
	\nonumber &&  \times \sum_{k \neq k'} e^{i\left(\omega t_k - \omega' t_{k'}\right)} \frac{\omega}{\gamma_k v_k} \frac{\omega'}{\gamma_{k'} v_{k'}}\frac{1}{2\pi}  K_1\left(d \frac{\omega}{\gamma_k v_k}\right)K_1\left(d \frac{\omega'}{\gamma_{k'} v_{k'}}\right) \\
	 && + \sum_{N=1}^{\infty}\frac{\bar{N}^N }{N!} \int \prod_{j=1}^{N} \left( dt_j \, \mathcal{P}(t_j) \right)  \sum_k e^{i\left(\omega - \omega'\right)t_k} \frac{\omega\omega'}{2\pi(\gamma_{k} v_{k})^2}  K_1\left(d \frac{\omega}{\gamma_k v_k}\right)K_1\left(d \frac{\omega'}{\gamma_{k} v_{k}}\right)\Bigg]  \,.	
\end{eqnarray}
Again for bunching parameters $r_b < 1$, we can write
\begin{eqnarray}
	\nonumber &&  E[\mathcal{F}_t[B_y](\omega) \mathcal{F}^*[B_y](\omega')]  = \left(\frac{ e\mu_0}{2\pi }\right)^2  \Bigg[ \bar{N}^2 \int_{-T/2}^{T/2} dt \,\mathcal{P}(t) \int_{-T/2}^{T/2} dt' \,\mathcal{P}(t')\times \\
	\nonumber &&   e^{i\left(\omega t - \omega' t'\right)} \frac{\omega}{\gamma(t) v(t)} \frac{\omega'}{\gamma(t') v(t')}\frac{1}{2\pi}  K_1\left(d \frac{\omega}{\gamma v}\right)K_1\left(d \frac{\omega'}{\gamma' v'}\right) \\
	&& + \bar{N} \int_{-T/2}^{T/2} dt  \,\mathcal{P}(t) e^{i\left(\omega - \omega'\right)t} \frac{\omega\omega'}{2\pi(\gamma(t) v(t))^2}  K_1\left(d \frac{\omega}{\gamma(t) v(t)}\right)K_1\left(d \frac{\omega'}{\gamma(t) v(t)}\right)\Bigg] \,.
\end{eqnarray}
Subtracting the square of the expectation value of the spectrum, we obtain
\begin{eqnarray}\label{eq:autocovfreq}
	&& \nonumber \rm{Cov}(\mathcal{F}_t[B_y](\omega),\mathcal{F}_t[B_y](\omega')) =\\
	&& \left(\frac{ e \mu_0}{2\pi}\right)^2  \bar{N} \int_{-T/2}^{T/2} dt  \,\mathcal{P}(t) e^{i\left(\omega - \omega'\right)t} \frac{\omega\omega'}{2\pi(\gamma(t) v(t))^2}  K_1\left(d \frac{\omega}{\gamma(t) v(t)}\right)K_1\left(d \frac{\omega'}{\gamma(t) v(t)}\right)\,.
\end{eqnarray}
This implies that the variance of the spectrum (for $T\gg 1/\omega_0$) is
\begin{eqnarray}
	\rm{Var}_{B_y}(\omega) &=& \rm{Cov}(B_y(\omega),B_y(\omega))  =  \left(\frac{ e \mu_0}{2\pi}\right)^2  \bar{N} \int_{-T/2}^{T/2} dt  \,\mathcal{P}(t) \frac{\omega^2}{2\pi(\gamma(t) v(t))^2}  \left(K_1\left(d \frac{\omega}{\gamma(t) v(t)}\right)\right)^2 \,.
\end{eqnarray}
In particular,
\begin{equation}
	S_{B_y B_y} (\omega) = \frac{1}{T}\left(E\left[\mathcal{F}_t\left[B_y\right](\omega)\right]^2 + \rm{Var}_{B_y}(\omega)\right)
\end{equation}
is the power spectral density. Our result shows that the average spectrum is not changed by the fluctuation, but that there is just a noise floor that is homogeneous for frequencies $\omega \ll \gamma v /d$ and falls off quickly for $\omega \gg \gamma v / d$. For $\omega \ll \gamma v /d$, we find
\begin{eqnarray}
	\rm{Var}_{B_y}(\omega) &=& \left(\frac{ e \mu_0}{2\pi d}\right)^2  \frac{\bar{N}}{2\pi}  \,.
\end{eqnarray}
Furthermore, from equation (\ref{eq:autocovfreq}) we obtain
\begin{eqnarray}
	\rm{Cov}(\mathcal{F}_t[B_y](\omega),\mathcal{F}_t[B_y](\omega')) &=& \left(\frac{ e \mu_0}{2\pi d}\right)^2  \frac{\bar{N}}{2\pi} \int_{-T/2}^{T/2} dt  \,\mathcal{P}(t) e^{i\left(\omega - \omega'\right)t}\,.
\end{eqnarray}
We can calculate the temporal auto covariance with the two point Fourier transform of (\ref{eq:autocovfreq}), with which we find
\begin{eqnarray}\label{eq:varBydelta}
	\rm{Cov}(B_y(t),B_y(t')) &=& \left(\frac{e\mu_0}{2\pi d}\right)^2 \frac{I(t)}{e} \delta(t-t') \Theta_{[-\frac{1}{2},\frac{1}{2}]}\left(\frac{t}{T}\right) \,.
\end{eqnarray}
We find that we are dealing with delta correlated noise.
Introducing a frequency cut-off such that  $\delta(t-t') \rightarrow \Delta f$, we find
\begin{equation}\label{eq:varBy}
	\rm{Var}(B_y(t)) = \left(\frac{\mu_0}{2\pi d}\right)^2 eI(t)\Delta f \,.
\end{equation}
This implies $\rm{Var}(B_y(t))/{E[B_y(t)]}^2 = e\Delta f / I(t)$ which is the signature of shot-noise. Equating $1/\Delta f$ with the time scale of 
the interaction of electron and quantum system $d / \gamma v$, we find that the noise-to-signal ratio $\rm{Var}_{B_y}(t)/E[B_y(t)]^2$ is small if 
\begin{equation}\label{eq:condcont}
     \frac{\gamma v e }{I(t) d} \ll 1 \,.
\end{equation}
The condition in equation (\ref{eq:condcont}) must be fulfilled to obtain a continuous driving signal with small noise. The factor $d / \gamma v$ can be identified as the inverse auto-correlation time $\tau_c$ as it is the inverse of the frequency scale on which the auto-covariance in equation (\ref{eq:autocovfreq}) decays. Thus, the condition in equation (\ref{eq:condcont}) simplifies to $I_\rm{min}\tau_c/e \gg 1$, where $I_\rm{min}$ is the minimal value of the modulated current. This implies that a continuous signal with small noise is ensured if there are many electrons passing the quantum system per auto-correlation time. Assuming a kinetic energy of $18\,\rm{keV}$ and a minimal current of $I_\rm{min} = 20\,\rm{\mu A}$, we obtain a required distance of the quantum system to the center of the beamline of $d \gg 600\,\rm{nm}$. Note that this condition is not a general limit of the mechanism we propose but only of the applicability of the approximation of a continuous driving signal with small noise. If the condition is not fulfilled, then a different model must to be chosen. In the case of quantum systems being driven by the magnetic field of single well separated electrons, we simulate the time evolution of the state of the quantum system for each electron separately; details can be found below in section \ref{sec:nvcenter}.

\section{The averaged magnetic field seen by the quantum system}
\label{sec:dens}

In the following section, we assume that the distance of the quantum system to the center of the beam $d$ is much smaller than the wavelength of the beam modulation $\lambda_0 = 2\pi v/\omega_0$ and that the waist $\text{w}$ of the beam is much smaller than $d$. Based on the magnetic field due to a single electron in equation (\ref{eq:magsingrelsup}), the magnetic field induced by an infinitesimal segment of a one-dimensional beam can be written as
\begin{eqnarray}
	E[d\vec B_j(0,t)] &=& \frac{\mu_0 \gamma I(z,t) dz'}{4\pi} \left[\begin{array}{c} y \\  -x \\ 0 \end{array}\right] \frac{1}{(r_\perp^2 + \gamma^2 z^2)^{3/2}}\,.
\end{eqnarray}
Then, for the whole beam, we write
\begin{eqnarray}
	E[\vec B_j(0,t)] &=& \int E[d\vec B_j(0,t)] = \int_{-\infty}^{\infty} dz\, \frac{\mu_0 I(z,t) }{4\pi \gamma^2} \left[\begin{array}{c} y \\  -x \\ 0 \end{array}\right] \frac{1}{(r_\perp^2/\gamma^2 +  z^2)^{3/2}}\,.
\end{eqnarray}
The integrand is localized to a region of length $r_\perp/\gamma$. If $I(z,t)$ does not change significantly on this length scale, we can approximate
\begin{eqnarray}\label{eq:magclosesub}
	E[\vec B_j(0,t)] & \rightarrow  & \int_{-\infty}^{\infty} dz\, \frac{\mu_0 I(z,t) }{4\pi \gamma^2} \left[\begin{array}{c} y \\  -x \\ 0 \end{array}\right] \frac{2\gamma^2}{r_\perp^2} \delta(z) = \frac{\mu_0 I(0,t) }{2\pi r_\perp^2} \left[\begin{array}{c} y \\  -x \\ 0 \end{array}\right]  \,.
\end{eqnarray}
For the current of the specific case of the Klystron beam, we have the Fourier decomposition
\begin{equation}
	I(z,t) = \left[ I_0 +  \sum_{n = 1}^{\infty}  I_{n\omega_0}(z)    \cos\left(n\omega_0\left( t - \frac{(z-z_0)}{v_0}\right)\right) \right]\,,
\end{equation}
where $I_{n\omega_0}(z)=2I_0 J_n(n\,r_b(z))$ was defined above.
Restricting our considerations to the Fourier component at the base frequency $\omega_0/2\pi$, from the condition that $I(z,t)$ does not change significantly on the length scale $r_\perp/\gamma$, we obtain the conditions $r_\perp \ll \gamma \lambda_0 = 2\pi \gamma v_0 /\omega_0 $ and
\begin{equation}
	r_\perp/\gamma \ll \frac{J_1(r_b(z))}{dJ_1(r_b(z))/dz} = \frac{J_1(r_b(z))}{J_1'(r_b(z)) dr_b(z)/dz}\,.
\end{equation}
For the bunching parameter $r_b(z)\le 1$, we have $J_1(r_b)/J_1'(r_b) \sim r_b $ and the second condition becomes $r_\perp \ll \gamma l $.
\begin{figure}[h]
\includegraphics[width=7cm,angle=0]{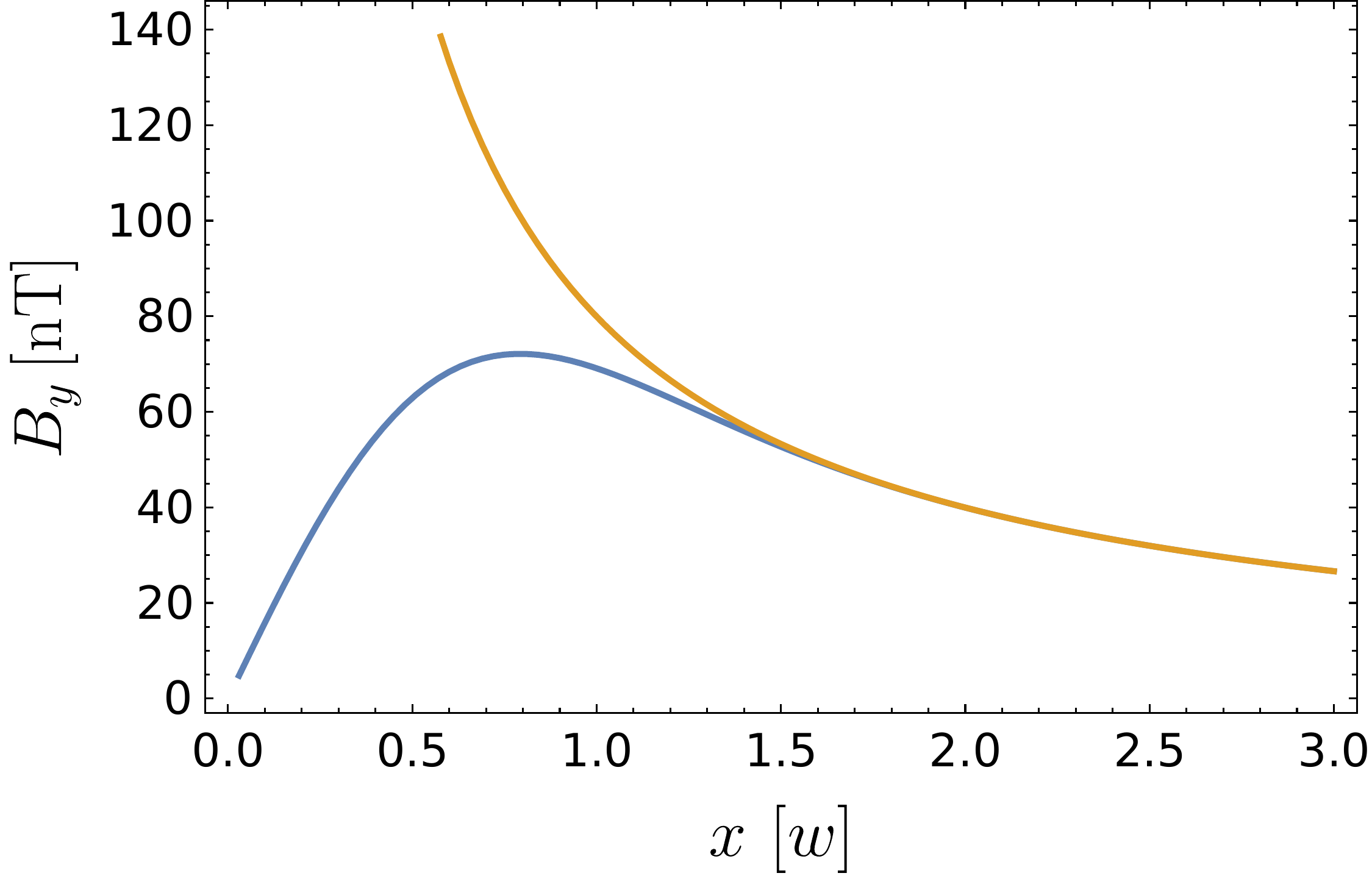}
\caption{\label{fig:plotsgaussian} The time-averaged magnetic field strength due to an electron beam with an averaged current of 20 $\mu$A as a function of the distance $x$ in units of the waist $\text{w}$ from the beam's center ($y=0$) for a Gaussian (blue) and an infinitely thin (orange) electron beam.}
\end{figure}

Therefore, we consider a non-divergent beam with a Gaussian profile $j_e(\vec{r},t) = \hat{z} 2 I(t)\exp(-2|\vec{r}_\perp - \vec{r}_{\perp,0}|^2/\text{w}^2)/(\pi \text{w}^2)$. $j_e(\vec{r})$ is normalized such that $I(t)$ is the average current obtained by integrating the charge density over the $x$-$y$ plane, and $\vec{r}_{\perp,0}$ is the position of the beam's center, i.e. $|\vec{r}_{\perp,0}| = d$. The parameter $\text{w}$ is the waist radius of the beam such that the beam has a $1/e^2$-diameter of $2\text{w}$. Under the conditions above, we obtain
\begin{eqnarray}\label{eq:gauss}
 E[\vec B(t)] & = & \frac{\mu_0 I(t)}{\pi^2 \text{w}^2}  \int d^2r_\perp  \left[\begin{array}{c} y \\  -x \\ 0 \end{array}\right]  \frac{e^{-\frac{2|\vec{r}_\perp - \vec{r}_{\perp,0}|^2}{\text{w}^2}} }{r_\perp^2} \,.
\end{eqnarray}
Equation (\ref{eq:gauss}) gives the general result for distances to the beam $d \ll \lambda$. If additionally $d > 2\text{w}$ holds, one can easily verify that the magnetic field strength of a Gaussian beam can be well approximated by that of an infinitely thin beam \begin{eqnarray}\label{eq:infthin}
 E[\vec B(t)] & = & \frac{\mu_0 I(t)}{2\pi d^2}    \left[\begin{array}{c} y \\  -x \\ 0 \end{array}\right]  \,.
\end{eqnarray}
A plot demonstrating this is shown in Fig.~\ref{fig:plotsgaussian}.

\section{The optical Bloch equations}
\label{sec:Bloch}

To derive the description of the quantum system, we start from the interaction Hamiltonian in the Coulomb gauge
\begin{equation}
	\hat H_\rm{int} = -\vec\mu \cdot \vec{B}\,,
\end{equation}
where $\vec{B}(t) = (B_x(t) , B_y(t) , B_z(t))$ and $ \vec\mu = \vec\mu_L + \vec\mu_S + \vec\mu_I $, where $\vec\mu_L= -\mu_B g_L \vec{L}/\hbar$, $\vec\mu_S= -\mu_B g_S \vec{S}/\hbar$ and $\vec\mu_I = \mu_N g_I \vec{I}/\hbar$. $\mu_B = e\hbar/(2m_e)$ is the Bohr magneton, $g_L = 1 - m_e/m_n \approx 1 $ is the orbital gyromagnetic ratio, $g_S \approx 2 $ is the spin gyromagnetic ratio, $\mu_N = e\hbar/(2m_p)$ is the nuclear magneton and $g_I$ is the total nuclear gyromagnetic ratio. 

The quantum system is considered as a two level system with the free Hamiltonian $H_0 = \hbar\omega_0\sigma_z/2$. For the time evolution of the density matrix $\rho$, we consider the Lindblad equation
\begin{equation}
	\partial_t \rho = \frac{1}{i\hbar} [ \hat H_0 + \hat H_\rm{int}, \rho] + \Gamma \left(\hat{L} \rho \hat{L}^\dagger - \frac{1}{2} \hat{L}^\dagger  \hat{L} \rho - \frac{1}{2} \rho \hat{L}^\dagger  \hat{L}\right)	\,,
\end{equation}
with the Lindblad operator $	\hat{L} = |g\rangle \langle e|$  representing the spontaneous emission into the radiation field corresponding to the natural linewidth of the transition. We obtain
\begin{eqnarray}\label{eq:opticalblochsupp}
	\left( i\hbar \frac{d}{dt} -
	\left[ \begin{array}{cccc} 
			\hbar \omega_0 - i \hbar \frac{\Gamma}{2}  & 0 & -T_{ge}^* & T_{ge}^* \\
			0 & -\hbar\omega_0 - i \hbar \frac{\Gamma}{2} &  T_{ge} & 	-T_{ge} \\
			 -T_{ge} & T_{ge}^* & -i \hbar \Gamma  & 0 \\
			  T_{ge} & - T_{ge}^* &  i \hbar \Gamma  & 0  	
	\end{array}  \right] \right)
	\left[\begin{array}{c}
		\rho_{eg} \\ \rho_{ge} \\ \rho_{ee} \\ \rho_{gg}
	\end{array}\right]   = 0 \,,
\end{eqnarray}
where $T_{ge} = \langle g | H_\rm{int} | e \rangle$ is the time dependent transition moment from the excited to the ground state.  With the transformation $\tilde \rho_{eg} = \rho_{eg} e^{i\omega_0 t}$, $\tilde \rho_{ge} = \rho_{ge} e^{-i\omega_0 t}$, $\tilde\rho_{ee}=\rho_{ee}$
and $\tilde\rho_{gg} = \rho_{gg}$, we find
\begin{eqnarray}
	\left( i\hbar \frac{d}{dt} -
	\left[ \begin{array}{cccc} 
			- i \hbar \frac{\Gamma}{2} & 0 & -T_{ge}^*e^{i\omega_0 t} & T_{ge}^*e^{i\omega_0 t} \\
			0 & - i \hbar \frac{\Gamma}{2} &  T_{ge}e^{-i\omega_0 t} & 	-T_{ge}e^{-i\omega_0 t} \\
			 -T_{ge}e^{-i\omega_0 t} & T_{ge}^*e^{i\omega_0 t} & -i \hbar \Gamma  & 0 \\
			  T_{ge}e^{-i\omega_0 t} & - T_{ge}^*e^{i\omega_0 t} & i \hbar \Gamma & 0  	
	\end{array}  \right] \right)
	\left[\begin{array}{c}
		\tilde\rho_{eg} \\ \tilde\rho_{ge} \\ \tilde\rho_{ee} \\ \tilde\rho_{gg}
	\end{array}\right]   = 0 \,.
\end{eqnarray}
This equation corresponds to the case that the coherence time is given as $T_2 = 2T_1 = 2/\Gamma$ which can be seen by the term $\Gamma/2$ appearing in the components of the matrix operator governing the decay of the off-diagonal elements of the density matrix. The generalization to general $T_2$ becomes (see Sec.4 of \cite{meystre2007elements} and set $\gamma_a = \Gamma_1$, $\gamma_b = 0$ and $\gamma_\rm{ph} = \Gamma_2 - \Gamma_1/2$ and taking into account that the upper level decays into the lower level leading to equation (4.31))
\begin{eqnarray}
	\left( i\hbar \frac{d}{dt} -
	\left[ \begin{array}{cccc} 
			- i \hbar \Gamma_2 & 0 & -T_{ge}^*e^{i\omega_0 t} & T_{ge}^*e^{i\omega_0 t} \\
			0 & - i \hbar \Gamma_2 &  T_{ge}e^{-i\omega_0 t} & 	-T_{ge}e^{-i\omega_0 t} \\
			 -T_{ge}e^{-i\omega_0 t} & T_{ge}^*e^{i\omega_0 t} & -i \hbar \Gamma_1  & 0 \\
			  T_{ge}e^{-i\omega_0 t} & - T_{ge}^*e^{i\omega_0 t} & i \hbar \Gamma_1 & 0  	
	\end{array}  \right] \right)
	\left[\begin{array}{c}
		\tilde\rho_{eg} \\ \tilde\rho_{ge} \\ \tilde\rho_{ee} \\ \tilde\rho_{gg}
	\end{array}\right]   = 0 \,,
\end{eqnarray}
where $\Gamma_j=1/T_j$ and $j\in\{1,2\}$.
With $B_i(t) = \sum_n \mathcal{B}_{i,n\omega_0} \cos(n\omega_0 t + \phi_M)$, we can consider on-resonance driving, and use the rotating wave approximation to find
\begin{eqnarray}
	\left(  \frac{d}{dt} -
	\left[ \begin{array}{cccc} 
			-\Gamma_2  & 0 & -i\Omega/2 & i\Omega/2 \\
			0 & -\Gamma_2 &  i\Omega/2 & 	-i\Omega \\
			 -i\Omega/2 & i\Omega/2 & -\Gamma_1  & 0 \\
			  i\Omega/2 & - i\Omega/2 & \Gamma_1 & 0  	
	\end{array}  \right] \right)
	\left[\begin{array}{c}
		\tilde\rho_{eg} \\ \tilde\rho_{ge} \\ \tilde\rho_{ee} \\ \tilde\rho_{gg}
	\end{array}\right]   = 0 \,,
\end{eqnarray}
where we define the Rabi frequency as
\begin{equation}
	\Omega =  -\langle e | \vec\mu  | g \rangle \cdot \vec{\mathcal{B}}_{\omega_0}/\hbar \,,
\end{equation}
where $\vec{\mathcal{B}}_{\omega_0} = (\mathcal{B}_{x,\omega_0} , \mathcal{B}_{y,\omega_0} , \mathcal{B}_{z,\omega_0})$. 

The finite spectral linewidth of the electromagnetic near-field created by the electron beam will influence the evolution of the quantum system. For the interaction of atoms with laser light, this has been investigated in \cite{Zoller1978:ato,eberly1976atomic,knight1980rabi,kimble1977resonance,agarwal1976exact}. For the case of a modulated current with small fluctuations relative to the mean (e.g. shot noise, modulation phase noise), the result can be applied immediately due to the equivalence of the interaction Hamiltonian. With the resonance condition, the rotating wave approximation leads to the modified optical Bloch equations 
\begin{eqnarray}\label{eq:blochflucsupp}
\hspace*{-4mm}	0 = \left( \frac{d}{dt} -
	\left[ \begin{array}{cccc} 
			-\Gamma_2 - b  & 0 & -i\Omega/2 & 	i\Omega/2 \\
			0 & -\Gamma_2 - b & i\Omega/2 & -i\Omega/2 \\
			-i\Omega/2 & i\Omega/2 & -\Gamma_1  & 0  \\
			i\Omega/2 & -i\Omega/2 & \Gamma_1 & 0 	
	\end{array}  \right] \right)
	\left[\begin{array}{c}
		\tilde\rho_{eg}^1 \\ \tilde\rho_{ge}^{-1} \\ \tilde\rho_{ee}^0 \\ \tilde\rho_{gg}^0
	\end{array}\right]  
\end{eqnarray}
for $\tilde\rho_{ij}^k(t):=\langle \tilde\rho_{ij}(t)\exp(ik\phi(t))\rangle_\rm{ph}$, which are the components of the quantum system's density matrix averaged over the phase noise $\phi(t)$, representing the finite spectral linewidth. The decoherence rate $b$ in equation (\ref{eq:blochflucsupp}) (which is assumed to be much smaller than $\Omega$) is related to the Full Width at Half Maximum (FWHM) linewidth $\delta\omega$ of the modulation as $b=\delta \omega/2$ as we will show in the following.

\subsection{FWHM linewidth and coherence length}
\label{sec:FWHM}

In \cite{Zoller1978:ato}, the inverse coherence length $b$ is defined via $\langle \exp(i\phi(t))\exp(-i\phi(t'))\rangle = \exp(-b|t-t'|)$. 
The magnetic field is given by its Fourier transform as
\begin{equation}
	B(t) = \int \frac{d\,\omega}{2\pi} e^{i\omega t} \hat B(\omega)\,.
\end{equation}
Therefore, the cross correlation becomes
\begin{eqnarray}
	\nonumber E[B^*(t) B(t+\tau)] &=& \int_{-\infty}^\infty dt\, B^*(t) B(t+\tau)\\
	&=& \int_{-\infty}^\infty dt\, \int \frac{d\,\omega}{2\pi} e^{-i\omega t} \hat B(\omega)^* \int \frac{d\,\omega'}{2\pi} e^{i\omega' (t+\tau)} \hat B(\omega')\\
	\nonumber &=& \int \frac{d\,\omega}{2\pi} e^{i\omega \tau}  |\hat B(\omega)|^2\,.
\end{eqnarray}
If a Lorentzian line shape is given, and
\begin{eqnarray}
	|\hat B(\omega)|^2 \propto \frac{(\delta\omega/2)^2}{(\omega_0-\omega)^2+(\delta\omega/2)^2}\,,
\end{eqnarray}
where $\delta\omega$ is the FHWM linewidth in rad/s, we find
\begin{eqnarray}
	E[B^*(t) B(t+\tau)] &\propto & e^{-\frac{\delta\omega}{2}\tau} e^{i\omega_0\tau} \,.
\end{eqnarray}
This implies that the inverse coherence length is $b = \delta\omega/2$.

\section{Transition matrix elements for $^{41}K$}\label{sec:matrixelem41K}

For potassium-41, we have $g_S\approx 2$ and  $g_I \approx -0.000078$. Therefore, we will only consider the coupling of the external magnetic field to the electron spin in the following section. For a hyperfine transition, we can simplify $T_{ge}$ as
\begin{equation}
	T_{ge} =  \mu_B g_S \langle e |  \vec{S}  | g \rangle \cdot \vec{B} /\hbar\,.
\end{equation}
For the case that the magnetic field has only one non-zero component, e.g. the $B_y$-component, we can write
\begin{equation}
	T_{ge} =  \mu_B g_S\langle e | S_y  | g \rangle B_y /\hbar\,.
\end{equation}
Using the $y$-direction as the quantization direction and taking into account that, for the $F=2$ hyperfine state and $m_F=0$ Zeeman sub-level,
\begin{eqnarray}
	|F=2,m_F=0\rangle &=& \sqrt{\frac{1}{2}} \left|\frac{3}{2},\frac{1}{2}\right\rangle \left|\frac{1}{2},-\frac{1}{2}\right\rangle + \sqrt{\frac{1}{2}} \left|\frac{3}{2},-\frac{1}{2}\right\rangle \left|\frac{1}{2},\frac{1}{2}\right\rangle
\end{eqnarray}
and for the $F=1$ and $m_F = 0$,
\begin{eqnarray}
	|F=1,m_F=0\rangle &=& \sqrt{\frac{1}{2}} \left|\frac{3}{2},\frac{1}{2}\right\rangle \left|\frac{1}{2},-\frac{1}{2}\right\rangle - \sqrt{\frac{1}{2}} \left|\frac{3}{2},-\frac{1}{2}\right\rangle \left|\frac{1}{2},\frac{1}{2}\right\rangle\,,
\end{eqnarray}
we obtain 
\begin{eqnarray}\label{eq:transmat00}
\nonumber	T_{ge} &=&   \frac{\mu_B B_y}{\hbar} \langle F=2,m_F=0 | g_L L_y + g_S S_y |F=1,m_F=0\rangle \\
	&=& \frac{\mu_B g_S B_y}{\hbar} \langle F=2,m_F=0 | S_y \left( \sqrt{\frac{1}{2}} \left|\frac{3}{2},\frac{1}{2}\right\rangle \left|\frac{1}{2},-\frac{1}{2}\right\rangle - \sqrt{\frac{1}{2}} \left|\frac{3}{2},-\frac{1}{2}\right\rangle \left|\frac{1}{2},\frac{1}{2}\right\rangle \right) = -\frac{\mu_B g_S B_y}{2}\,.
\end{eqnarray}
and $\tilde T_{ge}/\hbar = -e^{-i\omega_0 t} g_S \mu_B  B_y/(2\hbar) \approx -e^{-i\omega_0 t} e B_y/(2m_e)$, where $e$ is the unit charge and $m_e$ is the electron mass.
For the Rabi frequency, we thus find 
\begin{equation}\label{eq:defrabi}
	\Omega \approx  \frac{g_S \mu_B \mathcal{B}_y}{2\hbar} \approx \frac{e\mathcal{B}_y}{2m_e}\,.
\end{equation}
For the case of $d>2\text{w}$, the magnetic field can be approximated as that of an infinitely thin beam, and 
with equations (\ref{eq:klycurrfour}), (\ref{eq:infthin}) and (\ref{eq:transprobmethods}), we obtain
\begin{equation}\label{eq:defrabiinf}
	\Omega \approx  \sqrt{P_{e\leftrightarrow g}} \frac{I_{\omega_0}}{e} \,.
\end{equation}
where $\sqrt{P_{e\leftrightarrow g}}= r_e/d$ and $r_e=\mu_0 e^2/(4\pi m_e)$ is the classical electron radius.

\section{Transition matrix elements for NV$^-$ centers}
\label{sec:matrixelemNV}

The transition between the $^3A_2$ ground state magnetic sublevels $m_s=0$ and $m_s=\pm 1$ is a mutual polarization of two half-filled molecular orbitals. However, the spin triplet $m_s=0$ and $m_s=\pm 1$ can be described as an effective single spin 1 system with the coupling to an external magnetic field \cite{Wood2016wide,Zheng2019zero}
\begin{equation}
    H_\rm{int} = g_S \mu_B  \vec{S} \cdot \vec{B} /\hbar\,.
\end{equation}
We consider the $x$-direction as the quantization direction. The spin operator $\vec{S} = (S_x,S_y,S_z)$ acts on the fine-structure sub-levels such that $S_x |m_s \rangle  =\hbar m_s |m_s\rangle$, and for $S_\pm = S_y \pm iS_z$ we have
\begin{eqnarray}
    S_\pm |m_s\rangle = \hbar \sqrt{2 - m_s(m_s \pm 1)} |m_s \pm 1\rangle \,.
\end{eqnarray}
We consider the magnetic field oriented in the $y$-direction. Then,
\begin{eqnarray}
	T_{ge} &=&   \frac{g_S \mu_B B_y}{2\hbar} \langle \pm 1 | (S_+ + S_-) | 0 \rangle = \frac{g_S \mu_B B_y}{\sqrt{2}}  \,.
\end{eqnarray}
and $\tilde T_{ge} = e^{-i\omega_0 t} g_S \mu_B B_y / \sqrt{2}$.

\section{The optical Bloch equations with shot noise}
\label{sec:blochshot}

We start with the optical Bloch equations before the rotating frame transformation and neglecting damping:
\begin{eqnarray}\label{eq:blochdampless}
	  \frac{d}{dt} \left[\begin{array}{c}
		\rho_{eg} \\ \rho_{ge} \\ \rho_{ee} \\ \rho_{gg}
	\end{array}\right] &=& 
	-i\left[ \begin{array}{cccc} 
			 \omega_0  & 0 & -T_{ge}/\hbar & T_{ge}/\hbar \\
			0 & -\omega_0  &  T_{ge}/\hbar & 	-T_{ge}/\hbar \\
			 -T_{ge}/\hbar & T_{ge}/\hbar & 0  & 0 \\
			  T_{ge}/\hbar & - T_{ge}/\hbar &  0  & 0  	
	\end{array}  \right] 
\left[\begin{array}{c}
		\rho_{eg} \\ \rho_{ge} \\ \rho_{ee} \\ \rho_{gg}
	\end{array}\right] \,.
\end{eqnarray}
For the sake of simplicity, we assume real transition matrix elements, that is $T_{ge}/\hbar = T_{ge}^*/\hbar$  (phases can be absorbed into the components of the density matrix). We restrict our considerations to the case of $d>2\text{w}$ such that the magnetic field affecting the quantum system can be approximated as that of a one-dimensional beam as given in equation (\ref{eq:infthin}) and we assume $y=0$ and $x=-d$. Furthermore, we consider the situation where $d /\gamma v$ is much smaller than all time scales under consideration, such that we can approximate the average field as equation (\ref{eq:avmagfield}) and its covariance as equation (\ref{eq:varBy}). Here, we only consider the example of the potassium atoms as this corresponds to small fluctuations around the mean. Then, from equation (\ref{eq:transmat00}), we have $T_{ge}/\hbar = -eB_y/(2m_e)$ and $E[T_{ge}]/\hbar = -e\mu_0 I(t)/(4\pi m_e r_\perp) = - r_e I(t)/(e r_\perp) $.

We can write equation (\ref{eq:blochdampless}) as
\begin{equation}\label{eq:stochdiffstrong}
	\dot u = (\mathbf{B}_0 + \alpha \mathbf{B}_1(t,\xi)) u \,,
\end{equation}
where  $u= (\rho_{eg}(t) , \rho_{ge}(t) , \rho_{ee}(t), \rho_{gg}(t))$ and 
\begin{eqnarray}
	\mathbf{B}_0 = i\omega_0\left[ \begin{array}{cccc} 
			-1 & 0 & 0 & 0 \\
			0 & 1 & 0 & 0 \\
			0 & 0 & 0  & 0 \\
			0 & 0 &  0  & 0  	
	\end{array}  \right]  \,,\quad \mathbf{B}_1 = i \frac{T_{ge}}{\hbar\alpha} \mathbf{M} \,\quad \rm{and}\quad \mathbf{M} = \left[ \begin{array}{cccc} 
			0  & 0 & 1 & -1 \\
			0 & 0  &  -1 & 	1 \\
			 1 & -1 & 0  & 0 \\
			 -1 & 1 &  0  & 0  	
	\end{array}  \right] \,.
\end{eqnarray}
$\alpha$ is a parameter estimating the magnitude of fluctuations, which in our case implies
\begin{equation}
	\alpha \sim T_{ge}(t,\xi)/\hbar - E[ T_{ge}(t,\xi)]/\hbar \,.
\end{equation}
With equation (\ref{eq:varBy}) and the frequency scale $\Delta f \sim \gamma v / d$, this equation implies
\begin{equation}
	\alpha \sim \frac{e}{2m_e}\sqrt{\rm{max}_t\left(\rm{Var}(B_y(t,\xi))\right)} \sim \frac{ e^2 \mu_0 }{4\pi m_e d } \sqrt{\frac{I_\rm{max}}{e}  \frac{\gamma v }{ d}}\,.
\end{equation}
In the following, we will derive an ordinary linear differential equation for the expectation value
of the vector of components of the density matrix $u$ given in \cite{vankampen1976}. 
To this end, we must assume that $\alpha \tau_c \ll 1$, where $\tau_c$ is the auto-correlation time
of the magnetic field. If this condition is fulfilled, then the ensemble average of $u$ fulfills the integro-differential equation (see Sec.12 of \cite{vankampen1976})
\begin{equation}
	\frac{d}{dt}{E[u(t)]} = \left(\mathbf{K}_0 + \alpha \mathbf{K}_1(t) + \alpha^2 \mathbf{K}_2(t)\right) E[u(t)] 
\end{equation}
where $\mathbf{K}_0 = \mathbf{B}_0$,
\begin{equation}
	\mathbf{K}_1(t) =  E[\mathbf{B}_1(t,\xi)] = i \frac{E[T_{ge}]}{\hbar\alpha} \mathbf{M}, 
\end{equation}
\begin{equation}
	\mathbf{K}_2(t) = \int_0^t dt' \, \langle\langle \mathbf{B}_1(t,\xi) \mathbf{Y}(t|t') \mathbf{B}_1(t',\xi) \rangle\rangle \mathbf{Y}(t'|t)
\end{equation}
where we consider times much larger than the correlation time $\tau_c$, and the matrix $\mathbf{Y}(t,t')$ is the time-evolution operator for the differential equation
\begin{eqnarray}\label{eq:firstorderdgl}
	\nonumber \frac{d}{dt}{E[u(t)]} &=& \left(\mathbf{K}_0 + \alpha \mathbf{K}_1(t)\right) E[u(t)] \\
	&=& \left(\mathbf{B}_0 +  \alpha E[\mathbf{B}_1(t,\xi)]\right) E[u(t)] \,.
\end{eqnarray}
$\langle\langle ... \rangle\rangle$ denotes the cumulant, and therefore
\begin{eqnarray}
	\nonumber \langle\langle \mathbf{B}_1(t,\xi) \mathbf{Y}(t|t') \mathbf{B}_1(t',\xi) \rangle\rangle  &=& E[\mathbf{B}_1(t,\xi) \mathbf{Y}(t|t') \mathbf{B}_1(t',\xi)] - E[\mathbf{B}_1(t,\xi) ] \mathbf{Y}(t|t') E[ \mathbf{B}_1(t',\xi) ]\\
	\nonumber &=& -\frac{1}{(\hbar\alpha)^2} \left( E[T_{ge}(t)T_{ge}(t')] -  E[T_{ge}(t)]E[T_{ge}(t')] \right)\mathbf{M}\cdot \mathbf{Y}(t|t')\cdot \mathbf{M}\\
	&=& -\frac{1}{(\hbar\alpha)^2} \rm{Cov}\left(T_{ge}(t),T_{ge}(t')\right)\mathbf{M}\cdot \mathbf{Y}(t|t')\cdot \mathbf{M}\,.
\end{eqnarray}
On the time scale of $\omega_0$, the driving $T_{ge}$ can be assumed to be delta-correlated. In particular, for $1/\omega_0 \gg \tau_c$, we can set $\rm{Cov}\left(T_{ge}(t),T_{ge}(t')\right)=\hbar^2 a I(t)I_0^{-1}\delta(t-t')$ based on equation (\ref{eq:varBydelta}), where $a = (e^2\mu_0)^2 I_0/((4\pi m_e d)^2 e)=P I_0/e$ for the case of potassium atoms.
As $\mathbf{Y}(t|t) = \mathbb{I}$, we obtain
\begin{eqnarray}
	\langle\langle \mathbf{B}_1(t,\xi)  \mathbf{Y}(t|t') \mathbf{B}_1(t',\xi) \rangle\rangle  &=&  -\frac{a I(t)}{\alpha^2 I_0}\delta(t-t')\, \mathbf{M}\cdot \mathbf{M}\,,
\end{eqnarray}
and 
\begin{eqnarray}
	\alpha^2\mathbf{K}_2(t) = - \frac{ a I(t)}{I_0}  \left[ \begin{array}{cccc} 
			1  & -1 & 0 & 0 \\
			-1 & 1  &  0 & 	0 \\
			 0 & 0 & 1  & -1 \\
			0 & 0 &  -1  & 1  	
	\end{array}  \right]\,.
\end{eqnarray}
Finally, 
\begin{eqnarray}
	  \frac{d}{dt} E \left[\begin{array}{c}
		\rho_{eg} \\ \rho_{ge} \\ \rho_{ee} \\ \rho_{gg}
	\end{array}\right] &=& 
	\left[ \begin{array}{cccc} 
			 -i\omega_0 - \frac{a I(t)}{I_0}  & \frac{a I(t)}{I_0}  &	  i E[T_{ge}]/\hbar & -i E[T_{ge}]/\hbar \\
			\frac{a I(t)}{I_0} & i\omega_0 - \frac{a I(t)}{I_0} & -i E[T_{ge}]/\hbar & 	i E[T_{ge}]/\hbar \\
			 i E[T_{ge}]/\hbar & -i E[T_{ge}]/\hbar &  - \frac{a I(t)}{I_0}  & \frac{a I(t)}{I_0} \\
			  -i E[T_{ge}]/\hbar & i E[T_{ge}]/\hbar & \frac{a I(t)}{I_0}  & - \frac{a I(t)}{I_0} 	
	\end{array}  \right] 
E \left[\begin{array}{c}
		\rho_{eg} \\ \rho_{ge} \\ \rho_{ee} \\ \rho_{gg}
	\end{array}\right] \,.
\end{eqnarray}
Taking into account that $I(t)$ has a spectrum with distinct lines and a large spacing between spectral lines $\omega_0$, the transformation to the rotating frame, the rotating wave approximation (RWA) and averaging the damping lead to
\begin{eqnarray}\label{eq:blochshot}
	  \frac{d}{dt} E \left[\begin{array}{c}
		\tilde\rho_{eg} \\ \tilde\rho_{ge} \\ \tilde\rho_{ee} \\ \tilde\rho_{gg}
	\end{array}\right] &=& 
	\left[ \begin{array}{cccc} 
			 -a & a I_{2\omega_0}/(2I_0)  & -i \Omega/2 & i \Omega/2 \\
		a I_{2\omega_0}/(2I_0)  &  -a  &  i \Omega/2 & -i \Omega/2 \\
			 -i \Omega/2 & i \Omega/2 & -a  & a \\
			  i \Omega/2 & -i \Omega/2 &  a  & -a  	
	\end{array}  \right] 
E \left[\begin{array}{c}
		\tilde\rho_{eg} \\ \tilde\rho_{ge} \\ \tilde\rho_{ee} \\ \tilde\rho_{gg}
	\end{array}\right] \\
   \nonumber &=& 
	\frac{1}{e}\left[ \begin{array}{cccc} 
			 - P I_0 &   P I_{2\omega_0}/2  & -i \sqrt{P}I_{\omega_0}/2 & i \sqrt{P}I_{\omega_0}/2 \\
			 P I_{2\omega_0}/2  &  - P I_0  &  i \sqrt{P}I_{\omega_0}/2 & -i \sqrt{P}I_{\omega_0}/2 \\
			 -i \sqrt{P}I_{\omega_0}/2 & i \sqrt{P}I_{\omega_0}/2 & -P  & P \\
			  i \sqrt{P}I_{\omega_0}/2 & -i \sqrt{P}I_{\omega_0}/2 &  P  & -P  	
	\end{array}  \right] 
E \left[\begin{array}{c}
		\tilde\rho_{eg} \\ \tilde\rho_{ge} \\ \tilde\rho_{ee} \\ \tilde\rho_{gg}
	\end{array}\right]\,.
\end{eqnarray}
where the Rabi frequency $\Omega = r_e I_{\omega_0}/(e d)$ as defined in equation (\ref{eq:defrabi}), and $I_{n\omega_0} = 2I_0 J_n(nr_b(z))$  is the Fourier coefficient of the modulation at the base frequency (see equation (\ref{eq:klycurrfour})).
We find that shot noise leads to additional decoherence terms and an additional damping term proportional to $a$ in the optical Bloch equations. If we want to ignore this damping, then we have to fulfill the condition
\begin{equation}
    2 a =   \left(\frac{r_e}{d}\right)^2  \frac{2 I_0}{e} \ll \Omega = \frac{r_e}{d}  \frac{I_{\omega_0}}{e}\,,
\end{equation}
which leads to the general condition
\begin{equation}\label{eq:condnegdamping}
     d \gg  \frac{2r_e I_0}{I_{\omega_0}} \,.
\end{equation}
Since $r_e \sim 10^{-15}\,\rm{m}$, the above condition is always fulfilled in the context of this article.

For the method we used above to be applicable, we had to assume that $\alpha\tau_c \ll 1$, where we identified $\tau_c \sim d/(\gamma v)$. 
We obtain
\begin{equation}
	\alpha_c \tau_c \sim \frac{\lambda_e \alpha_\rm{FS}}{2\pi} \sqrt{\frac{I_\rm{max}}{e \,d \gamma v } }   \,,
\end{equation}
from which we find the condition for the distance
\begin{equation}
    d \gg \left(\frac{\lambda_e \alpha_\rm{FS}}{2\pi} \right)^2  \frac{I_\rm{max}}{e\gamma v} \,.
\end{equation}
For a kinetic energy of $18\,\rm{keV}$ and an average current of $I_0=100\,\rm{\mu A}$, we find the right hand side of this condition
to be on the order of $10^{-22}\,\rm{m}$. Therefore, this condition can be fulfilled for the situation that we consider.
Since a larger current leads to a
reduced noise to signal ratio of the magnetic field, it seems counter-intuitive that the minimal distance between beam line and quantum system grows with the current. However, the above condition only applies to the
method presented in \cite{vankampen1976}.

\section{Driving ground state hyperfine transitions in alkali atoms} 
\label{sec:driving41K}

\begin{figure}[h]
\includegraphics[width=8cm,angle=0]{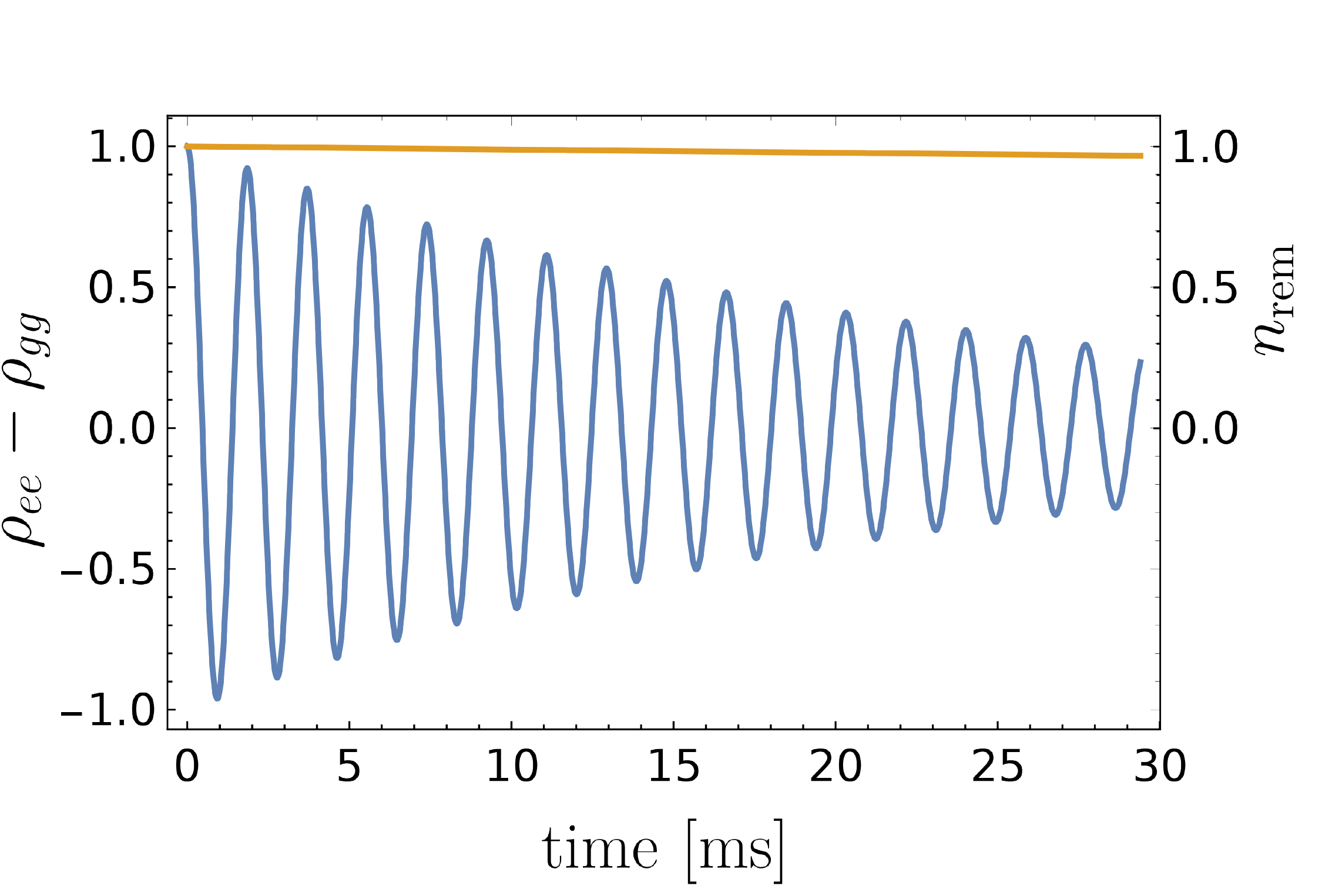}
\caption{\label{fig:blochevolutionsup} The blue curve shows the time-evolution of the inversion $\rho_{ee} - \rho_{gg}$. The center of the Gaussian electron beam of waist radius $\text{w}=50\,\rm{\mu m}$ is at a distance of $5\text{w}$ from the atom. We consider an average current $I_0 = 100\,\mu$A and bunching parameter $r_b= 0.5$ corresponding to a resonant current modulation at the base frequency $\omega_0/2\pi$ of amplitude $2I_0 J_1(r_b)\sim 50\,\mu$A. The resulting Rabi frequency is $\Omega \approx 2\pi \times 540 \,\rm{Hz}$. The decay of coherence is due to the assumed FWHM linewidth of the electron beam modulation of 25 Hz. We calculated (orange line) the ratio of atoms $n_\rm{rem}$ whose state is altered by incoherent scattering of single electrons assuming a remaining current density of $0.1\%$ of the peak value at the position of the atoms as a conservative estimate. After 20ms, less than $1\%$ of the atoms undergo an incoherent scattering interaction with the electrons.
}
\end{figure}

We assume that the atom is in the hyperfine ground state $F\!=\!1$, $m_F\!=\!0$ and is not spatially moving on the time scale of the proposed experiment. Fig.~\ref{fig:blochevolutionsup} plots the simulated hyperfine state response while applying an intensity modulated electron beam with a waist radius of $\text{w}\,=\, 50\,\mathrm{\mu m}$ and a current of $I=100\,\mathrm{\mu A}$ at a distance of $d\,=\,5\text{w}=\,250\,\mathrm{\mu m}$ (beam center to atom), which is modulated on resonance with the hyperfine frequency and bunching parameter $r_b=0.5$ and $\delta E_\rm{kin}/E_\rm{kin}=1/20$ corresponding to $l\sim 1\,$m. At a kinetic energy of 18 keV, a frequency of $254$~MHz corresponds to $\lambda_0$ of about $30\,\rm{cm}$. We find that the wave packet is much smaller than $\lambda_0$ in the interaction region if the initial size of the wave packet is much larger than $\hbar l / (2m_e v\lambda_0 ) \sim 10^{-12}\,$m.
\newline
We evaluate the time evolution of the atomic state based on the modified optical Bloch equations (\ref{eq:blochflucsupp}) in Mathematica using NDSolve.  Furthermore, the condition $d \ll \lambda_0$ that we introduced above is fulfilled and we can use equation (\ref{eq:gauss}) for the description of the expected field strength. Several Rabi oscillations of the hyperfine states are clearly visible in the plot. The source of the decay of coherence is technical noise in the electron beam source which we assumed to lead to a spectral linewidth of the beam modulation of about $25\,\rm{Hz}$. We set $b = \delta\omega/2 = \pi \times 25\,\rm{Hz}$. Furthermore, for the transition under consideration, we have $\Gamma_1 =2\Gamma_2 \ll b$ and we neglect $\Gamma_1$ and $\Gamma_2$ in the simulation. Systematic effects such as transition changes due to inelastic single electron atom interactions, which could also change the electronic state of the atoms, happen on a much longer time scale (see Fig.~\ref{fig:blochevolutionsup}, orange line). The total scattering cross section (causing ionization, elastic and inelastic scattering) for potassium atoms exposed to an 18 keV electron beam is $\sigma_{tot}\approx 1.5 \cdot 10^{-17} \text{cm}^2 $, extrapolated from \cite{Inokuti1971Inelastic}. 
\newline
We now estimate the Doppler shift experienced by atoms in a normal cold atom experiment. If the velocity of the atoms $\Delta v_a$ is small compared to the speed of the electron beam modulation $v$, the observed frequency shift of a moving atom compared to a non-moving atom can be  approximated by $\Delta f=  \Delta v_a f_0/v_e$.
Potassium atoms at a temperature of 40 $\mu$K move with a most probable velocity of 0.12 m/s and will experience an intensity modulated electron beam (velocity of the electron $v=c/4$ ) of frequency 254 MHz in the lab frame with a detuning of around 0.02 Hz, which is negligible.

\section{NV centers in nano-diamonds} 
\label{sec:nvcenter}

In the following, we will consider Nitrogen Vacancy (NV) centers in nano-diamonds as an example. In this situation, the magnetic near field consists of distinct spikes due to the well separated electrons. Therefore, using the expected value for the magnetic field in the optical Bloch equations would not be appropriate and we simulate the effect of the magnetic field of each electron separately. In particular, we will focus on the transition between the $^3A_2$ ground state magnetic sublevels $m_s=0$ and $m_s= 1$ of the NV$^-$ charge state, which are split by $\omega_0 = 2.87\,$GHz \cite{Doherty_2011}. The $m_s = -1$ sub-level is well separated from the $m_s = 1$ sublevel by $\sim 4\,$MHz \cite{Zheng2019zero} such that the transition from $m_s=0$ to $m_s= 1$ can be individually addressed. We consider the $z$-direction as the quantization direction, with the magnetic field oriented in the $x$-direction. Then, we find for the transition matrix elements $T_{0,1} = g_S \mu_B B_x /\sqrt{2}$  (see Appendix \ref{sec:matrixelemNV}). 
This transition exhibits coherence times $T_2$ from $600\,\rm{\mu s}$ \cite{Stanwix2010} up to $600\,\rm{ms}$ (\cite{Bar-Gill2013} using a dynamical decoupling pulse sequence), which is the main decay channel for the Rabi oscillations of the NV$^-$ center. We set $T_1 = 6\,$ms, $T_2 = 3\,$ms here following \cite{Bar-Gill2013}.  As the electron beam source, we consider a standard scanning electron microscope generating a beam waist of $\text{w} = 10\,$nm, beam energy of $2\,$keV, a probe current of $50\,\rm{nA}$ (corresponding to $\sim 100$ electrons per modulation period) and a bunching parameter $r_b\approx 0.5$ ($l \sim 3$cm and $\delta E_\rm{kin}/E_\rm{kin} = 1/20$) directed next to an NV$^-$ center at a distance of $d = 7\text{w} = \,70\,\rm{nm}$, for example embedded in a free standing nanostructure \cite{Batzer2020}.  
We assume that the beam is modulated by velocity modulation and bunching with a spectral modulation linewidth of $10^{-7}\omega_0/(2\pi) \sim 300\,\rm{Hz}$.
Note that, for the above parameters, the wave packet is much smaller than $\lambda_0$ in the interaction region if the initial size of the wave packet is much larger than $2m_e v\lambda_0 /(\hbar l) \sim 6\times 10^{-12}\,$m.
\begin{figure}[h]
\includegraphics[width=7.5cm,angle=0]{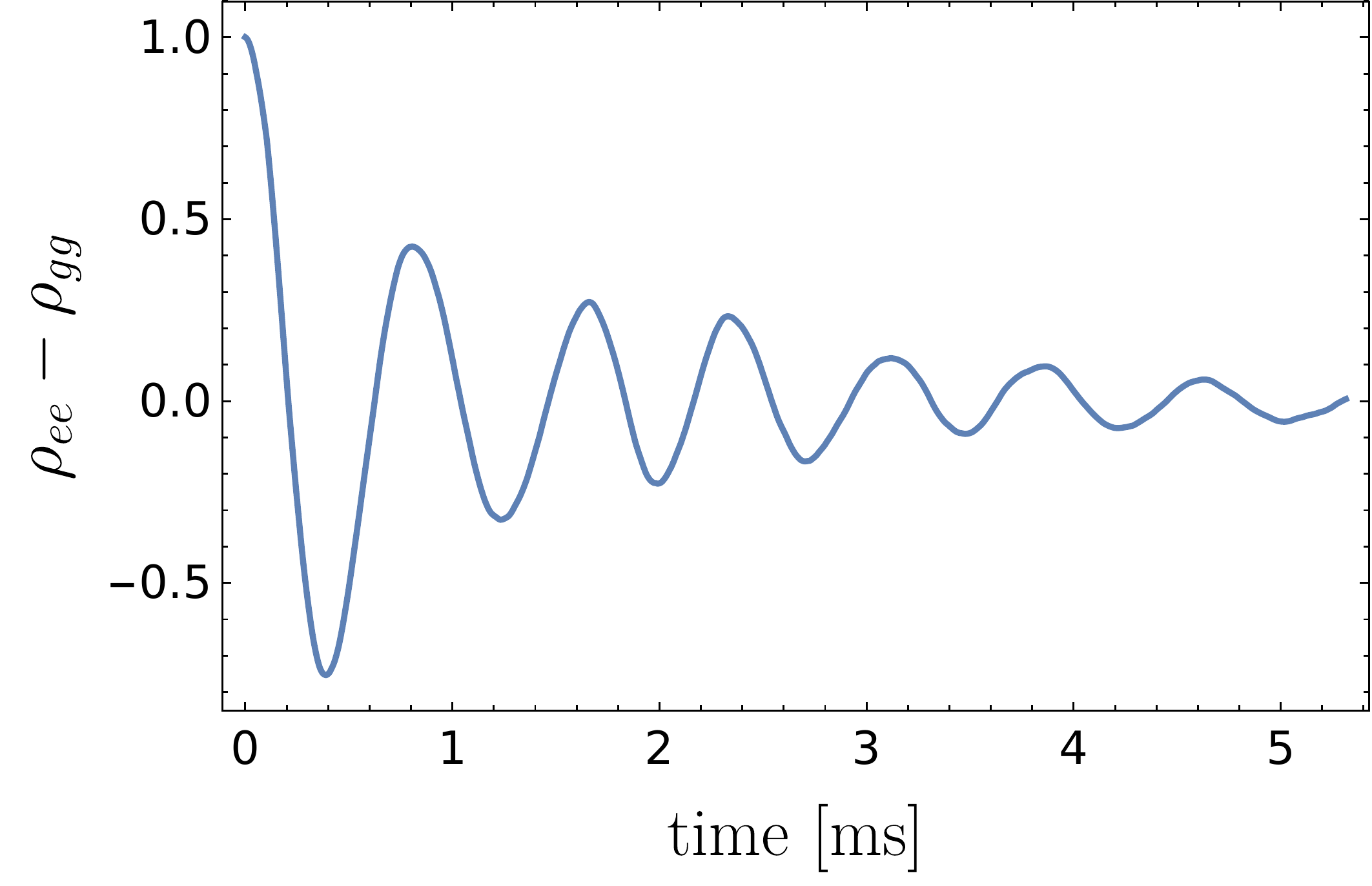}
\caption{\label{fig:blochevolutionsing} Time-evolution of the inversion for the transition $m_s = 0 \rightarrow m_s = 1 $ in the $^3A_2$-state of a NV$^-$ center at a distance of $d = 70\,$nm from a beam of waist $10\,$nm, current $50\,$nA, kinetic energy $2\,$keV and bunching parameter $r_b = 0.5$ ($l = 3$mm, and $\delta E_\rm{kin}/E_\rm{kin} = 1/20$). We set $T_1 = 6\,$ms, $T_2 = 3\,$ms and the FWHM linewidth of the electron beam modulation $b/\pi = 300\,\rm{Hz}$. }
\end{figure}
\newline
The result of a simulation of the expected level response is given in Fig.~\ref{fig:blochevolutionsing}. 
To reduce unwanted systematic effects due to electron scattering on the diamond structure \cite{Tanuma2011} we need to ensure that the electron beam intensity at the position of the NV center is reduced by a factor of $10^{-6}$ compared to its maximum. At this intensity, which is easily fulfilled for a Gaussian beam at 5w, on average less than one electron scatters within a radius of 1 nm next to the NV center every inverse Rabi frequency.

To produce the data for Fig.~\ref{fig:blochevolutionsing}, electrons are generated consecutively by a random process. We modulate the kinetic energy of the particles and calculate the propagation over the drift distance $l$ to obtain the current modulation. The modulation of kinetic energy is sinusoidal $E_\rm{kin}(t) = E_{\rm{kin},0} + \Delta E_\rm{kin} \sin(\omega_0 t + \phi(\xi,t))$, where $\phi(\xi,t)$ is a random process incorporating the finite linewidth of the driving. In particular, $d\phi(\xi,t)/dt = F(\xi,t)$, where $E[F(\xi,t),F(\xi,t')]=2b\delta(t-t')$ which implies that for each temporal interval $[a,b]$, we have
\begin{equation}
    \rm{Var}(\phi(\xi,a)-\phi(\xi,b)) = 2b|a-b|\,.
\end{equation}
Based on this variance and a vanishing average, the phase noise was implemented as a random Gaussian process. 

An interaction region of length $5d/(\gamma v)$ was associated with each electron. Then, the optical Bloch equations in the rotating frame were solved consecutively for each interaction of electrons with the NV$^-$-center over the interaction period using the Python ODE solver solve\_ivp. For electrons with overlapping interaction regions (a rare case for these parameters), the optical Bloch equations were solved together. Between the interaction regions, the analytical solution for the free time evolution was applied.  Finally, the results of 12 runs were averaged to obtain an average over different realizations of phase noise.

\subsection{Possible path to nano-scale resolution}

Electric dipole transitions from the $^3A_2$ ground state of the NV$^-$ center have a transition energy $\ge 1.945$ eV. To ensure a coherent evolution of the quantum system, we want to keep the probability for such transitions during the time needed for one Rabi transition suppressed. At the same time, we want to approach the NV center as closely as possible to maximize the spatial resolution. One option to approach both requirements is to decrease the kinetic energy of the beam electrons while keeping the current constant (see Appendix \ref{sec:electrictransitions} for details). Assuming a kinetic energy of 200 eV and a beam waist radius of w$=$5 nm (corresponding to $\sim 6\,$nm beam diameter FWHM)  \cite{altman2010trends,frank2011very}, a distance of $d=15\,$nm to the NV$^-$ could be achievable. This value for the kinetic energy implies an average velocity of the beam electrons of $v\sim 8.4\times 10^6\,$m/s which is more than one order below the speed of light in diamond at optical frequencies ($n\sim 2.4$ for the refractive index of diamond in the optical regime). 
Due to the distance dependence of electron energy loss to the diamond $\propto K_0(2\omega r_\perp/v)$ \cite{deAbajo2010optical}, we find that the energy transfer to the diamond is strongly suppressed for all transition frequencies near and above the fundamental absorption edge of diamond (at $\sim 5\,$eV) for distances to the diamond surface $r_\perp\gtrsim 5\,$nm \cite{papadopoulos1991optical}. 

\begin{figure*}[h]
\includegraphics[width=\textwidth]{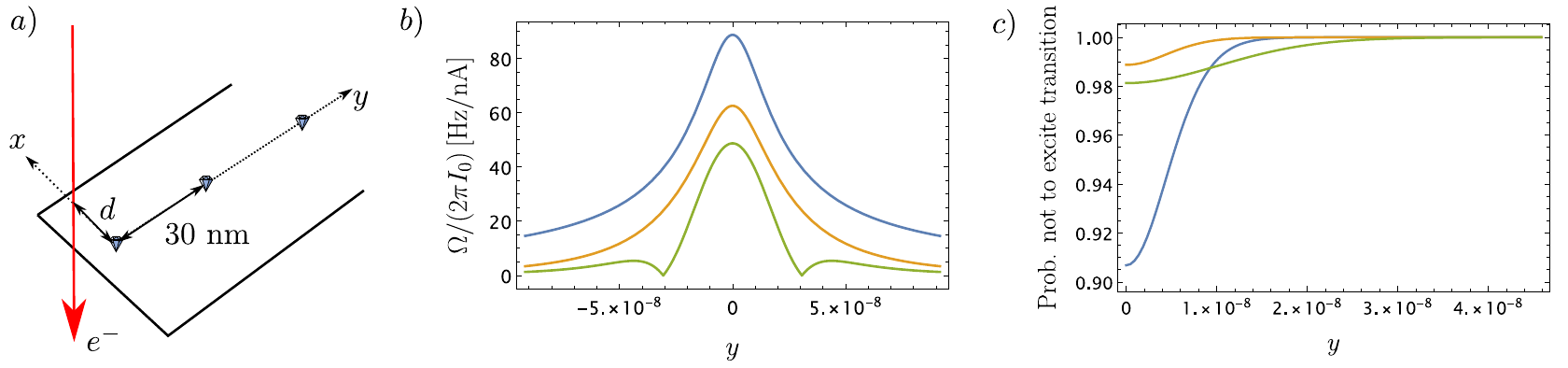}
\caption{\label{fig:rabidistance}  
 {\bf a)} Electron beam (red arrow) close to a 1d arrangement of NV$^-$ centers (blue diamonds) which could be created from nm-sized diamonds with single NV centers \cite{Alkahtani2019}, e.g., deposited on a graphene sheet (black rectangle). 
{\bf b)} Rabi frequency per beam current as a function of position along a the 1d arrangement of NV$^-$ centers for a temporally modulated beam with bunching parameter $r_b=0.5$ (blue curve) and two beams whose distance to the 1d arrangement varies with the resonance frequency, transversely to the 1-d arrangement as $(x,y)=(d[3+2\cos(\omega_0 t)],0)$ (orange curve) and on a circle section as $(x,y)=(d[1 + \sin^2(\omega_0 t)/2] , 2d\sin(\omega_0 t))$ (green). The minimal distance to the arrangement is $d = 15\,$nm. We assumed $w=5\,$nm and a kinetic energy of 200 eV. 
{\bf c)} Probability for the electron beam not to excite the NV$^-$ center at the electric-dipole transition line of $1.945$ eV during one Rabi flop for the three different beam configurations. The effect was averaged over the Gaussian profile of the electron beam. 
} 
\end{figure*}

Due to the non-relativistic velocity of the electrons and the negligibly small imaginary part for the dielectric function of diamond in the optical regime \cite{papadopoulos1991optical}, we approximate the relation between the external electric field and the effective matter-assisted electric field in the diamond by using the relation for the case of a static charge in front of a dielectric \cite{jackson}
$E^\rm{int}=2 E^\rm{ext}/(n^2 + 1)\sim  E^\rm{ext}/3 $ \footnote{We verified the factor $\sim 1/3$ by numerical calculations based on the results presented in Sec. III.7. of \cite{bolotovskii1962theory}. Even higher reduction factors can be achieved in different geometries, for example, a hollow cylinder (see Sec. III.2. of \cite{bolotovskii1962theory}).}. The probability for an incoherent transition of the electric dipole moment will be reduced accordingly by the factor $(2/(n^2 + 1))^2$ in comparison to the probability for a transition with the same electric dipole moment of a quantum system located in vacuum. 

We find that the probability to excite the energetically lowest electric dipole transition at $1.945$ eV during one Rabi transition would be less than $2\%$ for a position-modulated beam (see Fig. \ref{fig:rabidistance}c). The transition probability decays exponentially with increasing transition energy in this regime and higher energetic transitions would be even stronger suppressed. 

Then, the distance dependence of the Rabi frequency provides a pathway towards nano-scale spatial resolution as in aloof EELS \cite{EGERTON201595}. While a temporal modulation of the beam leads to a decay as $d^{-1}$, a much stronger decay can be achieved by employing oscillations of the beam position to generate a driving signal.
Then the oscillating near-field of a moving beam at the first harmonic and the second harmonic (twice the modulation frequency) scale effectively as $d^{-2}$ and $d^{-3}$, respectively. At a distance of $d=15\,$nm to a 1-dimensional array of NV$^-$ centers (see Fig. \ref{fig:rabidistance}a), adjacent NV$^-$ centers with a distance of $\sim 30\,$nm could be resolved, in principle (the spatial dependence of the Rabi frequency shows peaks of width $\sim 40\,$nm (FWHM)), see Fig. \ref{fig:rabidistance}b).

\end{document}